\documentclass{article}
\newlength{\abstractwidth}
\usepackage{cancel}
\usepackage{hyperref}
\usepackage{color}
\usepackage{graphicx}
\usepackage{cite}

\usepackage{tikz, pgf}
\usepackage{tkz-fct}
\usepackage{pgfplots}
\usetikzlibrary{shapes.misc}
\usetikzlibrary{shapes,snakes}
\usetikzlibrary{decorations.pathmorphing}	% For Feynman Diagrams
\usetikzlibrary{decorations.markings}

\usetikzlibrary{arrows.meta}

\newcommand{\backmidarrow}{\tikz  \draw[thick,-{Straight Barb[angle'=60,scale=1.5]}]  ((.1,0)-- (0,0);}
\newcommand{\midarrow}{\tikz  \draw[thick,-{Straight Barb[angle'=60,scale=1.5]}]  (0,0) -- (.1,0);}

\usepackage{amsmath}
\usepackage{amssymb}
\usepackage{latexsym}
\usepackage{physics}
 \usepackage{setspace}
 
 \usepackage{caption}
\usepackage{subcaption}

\numberwithin{equation}{section}

\renewcommand{\thefootnote}{\fnsymbol{footnote}}
\renewcommand{\thanks}[1]{\footnote{#1}}
\newcommand{\starttext}{
\setcounter{footnote}{0}
\renewcommand{\thefootnote}{\arabic{footnote}}}
\newcommand{\bea}{\begin{eqnarray}}
\newcommand{\eea}{\end{eqnarray}}
\newcommand{\be}{\begin{eqnarray}}
\newcommand{\ee}{\end{eqnarray}}

\def\ie{\begin{equation}\begin{aligned}}
\def\fe{\end{aligned}\end{equation}}

\def\ie{\begin{equation}\begin{aligned}}
\def\fe{\end{aligned}\end{equation}}

\def\cC{{\cal C}}

\def\cH{{\cal H}}

\def\nn{\nonumber}

\def\Tr{{\rm Tr}}

\begin{document}

\starttext

\setcounter{footnote}{0}

\begin{flushright}
%\scriptsize 
{\small QMUL-PH-23-03}
\end{flushright}

\vskip 0.3in

\begin{center}

\centerline{\large \bf  Generating functions and large-charge expansion of integrated correlators}
\centerline{\large \bf  in $\mathcal{N}=4$ supersymmetric Yang--Mills theory} 

\vskip 0.2in

{Augustus Brown, Congkao Wen, and Haitian Xie} 
   
\vskip 0.15in

{\small   Centre for Theoretical Physics, Department of Physics and Astronomy,  }\\ 
{\small Queen Mary University of London,  London, E1 4NS, UK}

\vskip 0.15in

{\tt \small a.a.x.brown@qmul.ac.uk,   c.wen@qmul.ac.uk, h.xie@se21.qmul.ac.uk}

\vskip 0.5in

\begin{abstract}
 
 \vskip 0.1in

We recently proved that, when integrating out the spacetime dependence with a certain integration measure, four-point correlators $\langle \mathcal{O}_2\mathcal{O}_2\mathcal{O}^{(i)}_p \mathcal{O}^{(j)}_p  \rangle$ in $\mathcal{N}=4$ supersymmetric Yang-Mills theory with $SU(N)$ gauge group are governed by a universal Laplace-difference equation. Here $\mathcal{O}^{(i)}_p$ is a superconformal primary with charge $p$ and degeneracy $i$. These physical observables, called integrated correlators, are modular-invariant functions of Yang-Mills coupling $\tau$. The Laplace-difference equation is a recursion relation that relates integrated correlators of operators with different charges. In this paper, we introduce the generating functions for these integrated correlators that sum over the charge. By utilising the Laplace-difference equation, we determine the generating functions for all the integrated correlators, in terms of the initial data of the recursion relation. We show that the transseries of the integrated correlators in the large-$p$ (i.e. large-charge) expansion for a fixed $N$ consists of three parts: 1) is independent of $\tau$, which behaves as a power series in $1/p$,  plus an additional $\log(p)$ term when $i=j$; 2) is a power series in $1/p$, with coefficients given by a sum of the non-holomorphic Eisenstein series; 3) is a sum of exponentially decayed modular functions  in the large-$p$ limit, which can be viewed as a generalisation of the non-holomorphic Eisenstein series. When $i=j$, there is an additional modular function of $\tau$ that is independent of $p$ and is fully determined in terms of the integrated correlator with $p=2$. The Laplace-difference equation was obtained with a reorganisation of the operators that means the large-charge limit is taken in a particular way here. From these $SL(2,\mathbb{Z})$-invariant results, we also determine the generalised 't Hooft genus expansion and the associated large-$p$ non-perturbative corrections of the integrated correlators by introducing $\lambda = p\, g^2_{_{YM}}$. The generating functions have subtle differences between even and odd $N$, which have important consequences in the large-charge expansion and resurgence analysis.  We also consider the generating functions of the integrated correlators for some fixed $p$ by summing over $N$, and we study their large-$N$ behaviour, as well as comment on the similarities and differences between the large-$p$ expansion and the large-$N$ expansion. 

\end{abstract}                                            
\end{center}

\baselineskip=15pt
\setcounter{footnote}{0}

\newpage

\setcounter{page}{1}
\tableofcontents

\newpage

%%%%%%%%%%
\section{Introduction}
%%%%%%%%%

It was proposed in \cite{Dorigoni:2021bvj,Dorigoni:2021guq, Dorigoni:2022zcr}  and recently proved in \cite{Dorigoni:2022cua} that the four-point correlation function of superconformal primary operators of the stress tensor multiplet  in $\mathcal{N}=4$ supersymmetric Yang-Mills (SYM) theory, once the spacetime dependence has been integrated out with a certain measure, can be expressed in a lattice sum. These observables are often called integrated correlators.\footnote{The integrated correlator can be related to the partition function of $\mathcal{N}=2^*$ SYM on $S^4$ \cite{Binder:2019jwn}, which can computed using supersymmetric localisation \cite{Pestun:2007rz}. See also \cite{Chester:2019pvm, Chester:2019jas, Chester:2020dja, Chester:2020vyz} on other aspects of the integrated correlator especially  its connections with the four-point amplitude in the type IIB superstring theory on AdS$_5 \times $S$^5$.} For example, in the case of the $SU(N)$ gauge group, the integrated correlator can be expressed as \cite{Dorigoni:2021bvj,Dorigoni:2021guq}
\ie \label{eq:int-SUN}
\cC_{N}(\tau, \bar{\tau}) = \sum_{(m,n)\in \mathbb{Z}^2} \int_0^{\infty} e^{-t Y_{m,n}(\tau, \bar \tau) } B_N(t) dt \, ,
\fe
where the coupling $\tau$ dependence only appears through the quantity $Y_{m,n}(\tau, \bar \tau)$, which is defined as
\ie \label{eq:Ymn}
Y_{m,n}(\tau, \bar \tau) := \pi \frac{|m+n\tau|^2}{\tau_2} \, , \quad {\rm with} \quad \tau = \tau_1+ i \tau_2 = {\theta \over 2\pi} + i {4\pi \over g_{_{YM}}^2} \, , 
\fe
and $B_N(t)$ is a rational function that encodes all the dynamics of the integrated correlator. In \cite{Dorigoni:2022cua}, the generating function that sums over the $N$ dependence for the integrated correlator $\cC_{N}(\tau, \bar{\tau})$ was introduced, and takes the following form, 
\ie \label{eq:int-gen}
\cC(z; \tau, \bar{\tau}) = \sum_{N=2}^{\infty} \cC_{N}(\tau, \bar{\tau}) \,z^N = \sum_{(m,n)\in \mathbb{Z}^2} \int_0^{\infty} e^{-t Y_{m,n}(\tau, \bar \tau) } B(z; t) \, dt \, ,
\fe
where $B(z; t)=\sum_N B_N(t) z^N$ and its explicit expression is given in \eqref{eq:Bztp2}. 
We see that all the non-trivial information about the integrated correlator for any $N$ is contained in a simple function $B(z; t)$. Analogous expressions can be found in \cite{Dorigoni:2022cua, Dorigoni:2022iem} for other classical gauge groups, built on earlier results \cite{Alday:2021vfb}. These expressions make manifest the Montonen–Olive duality \cite{Montonen:1977sn}, as well as its generalisation, the Goddard–Nuyts–Olive (GNO)  duality \cite{Goddard:1976qe}.

In this paper we generalise the lattice-sum representation \eqref{eq:int-SUN} and the generating function \eqref{eq:int-gen}  to integrated correlators that are associated with four-point functions of the form $\langle \mathcal{O}_2 \mathcal{O}_2 \mathcal{O}^{(i)}_{p} \mathcal{O}^{(j)}_{p}  \rangle$, where the operator $\mathcal{O}_{p}$ is a superconformal primary of charge (or dimension) $p$. There are multiple operators which have the same dimension $p$ (when $p>3$), and the index $i$ is to distinguish such degeneracy. When $p=2$,  it reduces to the integrated correlator $\cC_{N}(\tau, \bar{\tau})$ that we discussed above. Because the operators $\mathcal{O}_p^{(i)}$ are the bottom component of supersymmetry transformation, they are not charged under the bonus $U(1)_Y$ symmetry \cite{Intriligator:1998ig, Intriligator:1999ff}. This means that these more general integrated correlators are also $SL(2, \mathbb{Z})$ modular invariant due to Montonen–Olive duality of $\mathcal{N}=4$ SYM. These integrated correlators were recently studied in \cite{Paul:2022piq}, following the earlier work \cite{Binder:2019jwn}. In \cite{Paul:2022piq}, the $SL(2, \mathbb{Z})$ spectral decomposition analysis was applied to these integrated correlators that generalises the earlier work of \cite{Collier:2022emf}. It was assumed in \cite{Paul:2022piq} that only the continuous spectrum (i.e. non-holomorphic Eisenstein series) contributes to these integrated correlators.\footnote{In general a $SL(2, \mathbb{Z})$-invariant quantity can be decomposed into the continuous spectrum of non-holomorphic Eisenstein series and discrete spectrum of Maass cusp forms.} This is the same property that was used in \cite{Collier:2022emf} for the integrated correlator $\cC_{N}(\tau, \bar{\tau})$. This proposal was verified by explicit perturbative and non-perturbative instanton computations \cite{Paul:2022piq}. However, as already pointed out in \cite{Collier:2022emf, Dorigoni:2022zcr} for $\cC_{N}(\tau, \bar{\tau})$, this particular form of the $SL(2, \mathbb{Z})$ spectral decomposition (i.e. only non-holomorphic Eisenstein series contribute) is mathematically equivalent to the lattice-sum representation in \eqref{eq:int-SUN}. We will therefore assume that these more general integrated correlators associated with $\langle \mathcal{O}_2 \mathcal{O}_2 \mathcal{O}^{(i)}_{p} \mathcal{O}^{(j)}_{p}  \rangle$ also have a lattice-sum representation, analogous to \eqref{eq:int-SUN}.

As we showed recently in \cite{recp}, instead of simply organising the operators according to their charges (or equivalently dimensions), it is vital to reorganise operators into different towers and subtowers. We define the operators as
\ie
\mathcal{O}^{(i)}_{p|M}= (\mathcal{O}_2)^p \mathcal{O}^{(i)}_{0|M} \, ,
\fe
where  the operator $\mathcal{O}^{(i)}_{0|M}$ can be  defined recursively as given in \eqref{eq:defOm}. The operator $\mathcal{O}^{(i)}_{0|M}$ has charge $M$, so the total charge of $\mathcal{O}^{(i)}_{p|M}$ is $2p+M$. We denote $\mathcal{O}^{(i)}_{p|M}$ with any $p$ as an operator in the $i$-th subtower of the $M$-th tower. By this construction, two operators will have a vanishing two-point function on $S^4$, where the localisation computation is performed, unless they have the same $M$ and $i$ (i.e. they are in the same subtower of a given tower). The integrated correlators $\cC^{(M,M'|i,i')}_{N,p}(\tau, \bar \tau)$ that we will consider are associated with the four-point correlation functions of the following form, 
\ie \label{eq:cla}
\langle \mathcal{O}_2 \,  \mathcal{O}_2 \, \mathcal{O}^{(i)}_{p|M} \, \mathcal{O}^{(i')}_{p'|M'} \rangle  \,.
\fe
 We have removed the explicit dependence on $p'$ because of the relation $2p+M=2p'+M'$. This is because the four-point correlation functions are non-trivial only when the two higher-dimensional operators have the same dimension (see e.g. \cite{DHoker:1999jke, DHoker:2000xhf, Rayson:2008uje}). For the special case of $M=M'=0$ and $p=1$, $\cC^{(M,M'|i,i')}_{N,p}(\tau, \bar \tau)$ reduces to $\cC_{N}(\tau, \bar{\tau})$ as given in \eqref{eq:int-SUN}.

Furthermore, for  the integrated correlators $\cC^{(M,M'|i,i')}_{N,p}(\tau, \bar \tau)$ to obey a universal Laplace-difference equation, it is critical to normalise them properly \cite{recp}, 
\ie
\cC^{(M,M'|i,i')}_{N,p}(\tau, \bar \tau) = {\widetilde{R}_{N,p}^{(M,M'|i,i')} \over 4} \widehat{\cC}^{(M,M'|i,i')}_{N,p}(\tau, \bar \tau)\, , 
\fe
where the normalisation factor  $\widetilde{R}_{N,p}^{(M,M'|i,i')}$ (given in \eqref{eq:Rt}) is independent of coupling $\tau$. The quantity $\widehat{\cC}^{(M,M'|i,i')}_{N,p}(\tau, \bar \tau)$, which in general is a highly non-trivial modular function of $(\tau, \bar \tau)$, will be the main focus of the paper. In this paper, we will simply refer to $\widehat{\cC}^{(M,M'|i,i')}_{N,p}(\tau, \bar \tau)$ as the integrated correlator. It was shown in \cite{recp} that the integrated correlators $\widehat{\cC}^{(M,M'|i,i')}_{N,p}(\tau, \bar \tau)$ obey a universal Laplace-difference equation on the $\tau$ plane, that relates integrated correlators with different charges $p$. In this paper, we will generalise the lattice-sum representation \eqref{eq:int-SUN} for $\widehat{\cC}^{(M,M'|i,i')}_{N,p}(\tau, \bar \tau)$, and propose the same representation for these more general integrated correlators,
\ie \label{eq:proNp}
\widehat{\cC}^{(M,M'|i,i')}_{N,p}(\tau, \bar \tau) =
\sum_{(m,n)\in \mathbb{Z}^2} \int_0^{\infty} e^{-t Y_{m,n}(\tau, \bar \tau) } \widehat{B}^{(M,M'|i,i')}_{N,p}(t)\, dt \, .
\fe
Once again, all the non-trivial information of the integrated correlators is encoded in the rational functions $\widehat{B}^{(M,M'|i,i')}_{N,p}(t)$.  We will mostly be interested in the integrated correlators for a fixed $N$ but arbitrary $p$. Following the ideas of \cite{Dorigoni:2022cua}, it is very convenient to further introduce generating functions by summing over the charge $p$ dependencies,\footnote{To distinguish the resummation parameter $z$ in summing over $N$ as given in \eqref{eq:int-SUN}, we will use the parameter $w$ for summing over the charge $p$.} 
\ie
\widehat{\cC}^{(M,M'|i,i')}_{N}(w; \tau, \bar \tau) =\sum_p \widehat{\cC}^{(M,M'|i,i')}_{N,p}(\tau, \bar \tau) \, w^p = \sum_{(m,n)\in \mathbb{Z}^2} \int_0^{\infty} e^{-t Y_{m,n}(\tau, \bar \tau) } \widehat{B}^{(M,M'|i,i')}_{N}(w; t)\, dt \, ,
\fe
where $\widehat{B}^{(M,M'|i,i')}_{N}(w; t) = \sum_{p} \widehat{B}^{(M,M'|i,i')}_{N,p}(t)\, w^p$. We will determine the generating functions, and use them to study the properties of the integrated correlators, especially in the large-charge limit. The Laplace-difference equation that is satisfied by the integrated correlators becomes a differential equation for $\widehat{B}^{(M,M'|i,i')}_{N}(w; t)$. 
We will determine the generating functions by solving the differential equation, and the generating functions will enable us to study the charge dependence of the integrated correlators in a straightforward manner.  

 We will be particularly interested in the large-charge (i.e. large-$p$) limit of the integrated correlators. Understanding the behaviour of quantum field theories at the large global charge limit has been of great interest \cite{Hellerman:2015nra,Monin:2016jmo,Alvarez-Gaume:2016vff,Hellerman:2017veg,Jafferis:2017zna,Hellerman:2018xpi} (see \cite{Gaume:2020bmp} for a recent review). Particularly related to our study, the large-charge limit has been studied in the context of extremal correlators in $\mathcal{N}=2$ supersymmetric theories, which can also be computed using supersymmetric localisation, see e.g. \cite{Hellerman:2017sur, Grassi:2019txd, Bourget:2018obm, Beccaria:2018xxl, Beccaria:2018owt, Hellerman:2020sqj,Hellerman:2021yqz,Hellerman:2021duh,Cuomo:2022kio}. 
We find that, for the integrated correlators in $\mathcal{N}=4$ SYM that are considered in this paper, the large-$p$ expansion takes a universal form, summarised in \eqref{eq:uni}, and contains three parts. The first part is independent of $\tau$. It behaves as $\log(p)$ plus a power series in $1/p$ when $M=M', i=i'$ (i.e. when two higher-dimensional operators are the same), otherwise we only have a power series in $1/p$. The second part is a power series in $1/p$, with coefficients given by a sum of the non-holomorphic Eisenstein series with half-integer indices. Finally, the third part is a sum of exponentially decayed modular functions in the large-$p$ limit. These modular functions can be viewed as a generalisation of the non-holomorphic Eisenstein series. When $M=M', i=i'$, there is also an additional $p$-independent modular function that is completely determined in terms of $\cC_{N}(\tau, \bar{\tau})$.  The second and third parts are remarkably similar to the large-$N$ expansion of the integrated correlator $\cC_{N}(\tau, \bar{\tau})$ in \eqref{eq:int-SUN} that was recently obtained in \cite{Dorigoni:2022cua}, with $N$ and $p$ exchanged. However, it should be stressed that even though the large-$p$ and large-$N$ expansions look similar, there are interesting and subtle differences. In particular, when $N$ is even, the number of non-holomorphic Eisenstein series that appear in the large-$p$ expansion do not grow indefinitely as we consider higher orders in the expansion. This is very different from what has been seen in \cite{Chester:2019jas, Dorigoni:2021guq, Dorigoni:2022cua}. This fact also implies that in the generalised 't Hooft genus expansion, at a given order in $1/p$, the large-$\lambda$  expansion truncates.\footnote{In the large-$p$ expansion, we will also consider the generalised 't Hooft genus expansion by introducing $\lambda = p\, g^2_{_{YM}}$.} These structures have interesting consequences in understanding the connections between the power series terms and exponentially decayed terms in the large-$p$ limit from the point of view of resurgence.

The rest of the paper is organised as follows. Section \ref{sec:rev} mainly summarises the results of \cite{recp}. We will begin by briefly reviewing the definition of the integrated correlators and their connection with the partition function of $\mathcal{N}=2^*$ SYM on $S^4$. We will particularly emphasise the importance of reorganising the operators in the localisation computation of the integrated correlators. With an important appropriate normalisation factor,  the integrated correlators obey a universal Laplace-difference equation, as proved in \cite{recp}. In section  \ref{sec:genf},  we will introduce the lattice-sum representation and generating functions for the integrated correlators, following the ideas of \cite{Dorigoni:2022cua}. We will determine the generating functions (in terms of some initial conditions) by utilising the Laplace-difference equation, which becomes a differential equation of the generating functions.  The generating functions have interesting differences depending on whether $N$ is even or odd. We will further determine the singularity structures (discontinuities and poles) of the generating functions, since they are most relevant in the study of the large-charge expansion of the integrated correlators.  We will give several explicit examples to illustrate the general structures. In section \ref{sec:large-p}, we will discuss how to obtain the large-charge expansion of the integrated correlators by utilising the generating functions derived in the previous section. By a simple contour deformation argument, it is easy to show that the $p$-dependence  properties of integrated correlators are determined by the singularities of the generating functions. We will find that the large-charge expansion of the integrated correlators takes a universal form for any $N$, up to a subtle and interesting difference depending  on whether $N$ is even or odd. This difference is sharpened in the generalised `t Hooft limit, and makes connections with resurgence analysis. In section \ref{sec:large-p-ex}, we will give several non-trivial examples of the large-$p$ expansion of the integrated correlators. We will conclude and discuss future research directions in section \ref{sec:confut}. The paper also includes three appendices. 
In appendix \ref{app:genfun}, we will provide more examples of generating functions for the integrated correlators and their singularity properties. Appendix \ref{app:lattice} briefly reviews the connections between the lattice-sum representaion and the $SL(2,\mathbb{Z})$ spectral decomposition of integrated correlators. Appendix \ref{app:largeN} concerns the large-$N$ expansion of the integrated correlators with fixed charge $p$, generalising the results of \cite{Dorigoni:2022cua}.

\vspace{0.3cm}

{\bf Note added}: While we were finalising this paper, \cite{Paul:2023rka} appeared on arXiv, which also studied some large-charge properties of the integrated correlators $\widehat{\cC}^{(M,M'|i,i')}_{N,p}(\tau, \bar \tau)$ for the cases with $M=M'=0$ (the corresponding operators are called maximal-trace operators in \cite{Paul:2023rka}). We found perfect agreement between our results where there are overlaps, up to an overall factor of $4$ due to different normalisations.  

%%%%%%%%%%
\section{Integrated correlators and Laplace-difference equation} \label{sec:rev}
%%%%%%%%%

In this section, we will review the construction of  \cite{recp}, where the superconformal primary operators are reorganised into different towers and subtowers, which greatly simplifies the Gram-Schmidt procedure in the localisation computation of the integrated correlators. The integrated correlators organised in this particular way, with some crucial normalisation factor, were shown to obey a universal Laplace-difference equation that relates integrated correlators with different charges. 

\subsection{Review of integrated correlators}
\label{sec:rev-local}

The operators $\mathcal{O}^{(i)}_p$ that appear in the four-point correlation functions that we will study are half-BPS superconformal primary operators. They are in the $[0,p,0]$ representation of the $SU(4)$ R-symmetry, and have scaling dimension $p$, which is protected by supersymmetry. There are in general more than one distinguishable operator for a given charge (or dimension) $p$, and so the superscript $i$ denotes such degeneracy. In terms of the fundamental scalar fields $\Phi^I$ (with $I=1,\cdots, 6$), they take the following form: 
\ie  \label{eq:mul-tr}
T_{p_1, \cdots, p_n}(x, Y) = {p_1 \cdots p_n\over p} T_{p_1}(x, Y) \cdots T_{p_n}(x, Y) \, ,
\fe
where $p = \sum p_i$, and each $T_p(x, Y)$ is a single-trace operator, 
\ie  \label{eq:sin-tr}
T_p(x, Y) = {1\over p} Y_{I_1} \cdots Y_{I_p} \Tr\left( \Phi^{I_1}(x) \cdots \Phi^{I_p}(x) \right) \, ,
\fe
and $Y_{I}$ is a null $SO(6)$ vector that conveniently sums over the R-symmetry index $I$. Note for a finite $N$, not all the $T_{p_1, \cdots, p_n}(x, Y)$ are independent. For example, $T_4= \frac{1}{2} T_{2,2}$ when $N=2,3$. In fact, for $N=2$ the only independent operator at dimension $2p$ is $T_{2,2,\ldots,2}$ (with $p$ copies of $T_2$); they are often called maximal-trace operators. We will consider a special class of correlators of the form
$\langle \mathcal{O}_2\, \mathcal{O}_2\, \mathcal{O}^{(i)}_p\, \mathcal{O}^{(j)}_p  \rangle$.
Because of the constraints from superconformal
symmetry, the correlators can be decomposed as \cite{Eden:2000bk, Nirschl:2004pa}
\ie \label{eq:22pp}
    \langle \mathcal{O}_2(x_1, Y_1)\mathcal{O}_2(x_2, Y_2)\mathcal{O}^{(i)}_p(x_3, Y_3)\mathcal{O}^{(j)}_p(x_4, Y_4) \rangle = \mathcal{G}_{\rm free}^{(i,j)}(x, Y) + \mathcal{I}_4 (x, Y)\mathcal{H}_{N, p}^{(i,j)} (U, V; \tau, \bar \tau) \, ,
\fe
where we separate the correlator into a free part, $\mathcal{G}_{\rm free}^{(i,j)}$, which may be simply computed by free-field Wick contractions, and a part that non-trivially depends on the dynamics of the theory. For the dynamic part, one can further factor out all the R-symmetry dependence, denoted as $\mathcal{I}_4 (x, Y)$ in \eqref{eq:22pp}. The factor $\mathcal{I}_4 (x, Y)$ is completely fixed by the symmetry; we will follow the convention of \cite{Binder:2019jwn}, and the expression for $\mathcal{I}_4 (x, Y)$ can be found in that reference. The function $\mathcal{H}_{N, p}^{(i,j)} (U, V; \tau, \bar \tau)$ is our main focus, which  is a function of the Yang-Mills coupling $\tau$ and the cross ratios, 
\ie
U= {x_{12}^2 x_{34}^2 \over x_{13}^2 x_{24}^2} \, , \quad \qquad V= {x_{14}^2 x_{23}^2 \over x_{13}^2 x_{24}^2} \, .
\fe
As argued in \cite{Binder:2019jwn} (see also \cite{Paul:2022piq}), this particular type of correlator, once the spacetime dependence has been integrated out with a certain measure, can be computed by supersymmetric localisation. More precisely, we define the integrated correlators as \cite{Binder:2019jwn} 
\ie \label{eq:measure}
\cC_{N, p}^{(i,j)}(\tau, \bar \tau) = -{2\over \pi } \int_0^{\infty} dr \int^{\pi}_0 d \theta {r^3 \sin^2 \theta \over U^2} \mathcal{H}_{N, p}^{(i,j)} (U, V; \tau, \bar \tau) \, , 
\fe
with $U=1-2r \cos\theta  +r^2$ and $V=r^2$. Importantly, as shown in \cite{Binder:2019jwn}, the integrated correlators $\cC_{N, p}^{(i,j)}(\tau, \bar \tau)$ are related to the partition function $\mathcal{Z}_N(\tau, \tau_A'; m)$ of $\mathcal{N}=2^{\star}$ SYM on $S^4$, deformed by the higher-dimensional operators $\mathcal{O}^{(i)}_p$. The partition function, and thus the integrated correlators,  can be computed by supersymmetric localisation \cite{Pestun:2007rz,Nekrasov:2002qd}. The integrated correlators $\cC_{N, p}^{(i,j)}(\tau, \bar \tau)$ are modular functions of $(\tau, \bar \tau)$, and of course also have non-trivial dependence on the parameters $N, p$ and $i, j$. 

\subsection{Organisation of operators and Laplace-difference equation}
\label{sec:OLDE}

Due to the dimensionful radius of $S^4$, operators with different dimensions will mix. In general, a dimension-$p$ operator on $S^4$ could mix with operators of dimensions $(p-2), (p-4)$, etc. This is resolved by a Gram-Schmidt procedure \cite{Gerchkovitz:2016gxx, Binder:2019jwn}. To efficiently perform the procedure, we will reorganise the operators into towers and subtowers, such that operators in different (sub)towers are orthogonal to each other. We will then only need to perform a Gram-Schmidt procedure within the (sub)towers.
We therefore reorganise the operators in the following manner:
\ie \label{eq:def-opreator}
\mathcal{O}^{(i)}_{p|M} = (T_2)^p \mathcal{O}^{(i)}_{0|M}\, ,
\fe
where the operator $\mathcal{O}^{(i)}_{0|M}$ has dimension $\Delta = M$ and consequently $\mathcal{O}^{(i)}_{p|M}$ has dimension $2p+M$. The operator $\mathcal{O}^{(i)}_{0|M}$ is defined as follows. We begin by considering a general operator $T_{p_1, p_2, \ldots, p_n}$ as defined in \eqref{eq:mul-tr} with all $p_i >2$ (namely, $T_2$ is excluded). These operators are ordered according to their dimensions, which are labeled as $B^{(i)}_M$ with dimension $M$ and $i$ denotes possible degeneracy. The first few examples are $B_0 = \mathbb{I}$ (i.e. the identity operator), $B_3 = T_3$, $B_4 = T_4$, and $B_5 = T_5$. We have omitted the degeneracy index $i$, since there is no degeneracy when $M<6$. For $M=0$, we define $\mathcal{O}_{0|0} = B_0 = \mathbb{I}$, and $\mathcal{O}^{(i)}_{0|M}$ with $M \geq 3$ is defined recursively as
\ie \label{eq:defOm}
\mathcal{O}^{(i)}_{0|M}  = B_M^{(i)} - \sum_{M'\leq M, \, j<i} \frac{\langle B_M^{(i)} \, , \mathcal{O}^{(j)}_{0|M'} \rangle_{S^4}}{\langle (T_{2})^{\delta}  \mathcal{O}^{(j)}_{0|M'}\, , \mathcal{O}^{(j)}_{0|M'} \rangle_{S^4}} \,  (T_{2})^{\delta} \mathcal{O}^{(j)}_{0|M'} \, , 
\fe
where $\delta=(M-M')/2$, and
\ie \label{eq:2pt}
\langle \mathcal{O}_A\, , \mathcal{O}_B \rangle_{S^4}
= {\partial_{\tau'_A}\partial_{\tau'_B} \mathcal{Z}_N(\tau, \tau_A'; 0) \big{\vert}_{\tau_A'=0} \over \mathcal{Z}_N(\tau, 0; 0) }\, ,
\fe
where $\mathcal{Z}_N(\tau, \tau_A'; 0)$ is the partition function $\mathcal{Z}_N(\tau, \tau_A'; m)$ with $m=0$. For a general mass $m$,  the partition function $\mathcal{Z}_N(\tau, \tau_A'; m)$ can be expressed as 
\ie \label{eq:parZ}
\mathcal{Z}_N(\tau, \tau_A'; m) =& \int d^{N-1}a  \left \vert \exp\left( 2\pi i \sum_{i=1}^N a_i^2 +i \sum_{p>2} \pi^{p/2} \tau'_p \sum_{i=1}^N a_i^p \right)\right\vert^2    
Z_{\rm 1-loop}(a; m) \left\vert Z_{\rm inst}(\tau, \tau', a;m) \right\vert^2 \, , 
\fe
where the integration variables $a_i$ are constrained by $\sum_{i=1}^N a_i =0$, and $Z_{\rm 1-loop}$ and $Z_{\rm inst}$ give the perturbative and non-perturbative contributions, respectively. When $m=0$, both $Z_{\rm 1-loop}(a;m)$ and $ Z_{\rm inst}(\tau, \tau', a; m)$ simply become $1$, and we have
\ie \label{eq:m0}
\mathcal{Z}_N(\tau, \tau_A'; 0) =& \int d^{N-1}a  \left \vert \exp\left( 2\pi i \sum_{i=1}^N a_i^2 +i \sum_{p>2} \pi^{p/2} \tau'_p \sum_{i=1}^N a_i^p \right)\right\vert^2    
\, .
\fe
One may find more detailed localisation computation of the integrated correlators in \cite{Binder:2019jwn, Paul:2022piq, recp}.

 Using \eqref{eq:defOm} and the above expression for the partition function, it is straightforward to obtain $\mathcal{O}_{0|M}$. For $M<6$, there is no degeneracy, and we find 
\ie \label{eq:exOm}
\mathcal{O}_{0|0} =\mathbb{I}\, , \qquad \mathcal{O}_{0|3} = T_3\, , \qquad \mathcal{O}_{0|4} = T_4- {2N^2-3\over N(N^2+1)} T_{2,2}\, ,
\qquad \mathcal{O}_{0|5} = T_5- {5(N^2-2)\over N(N^2+1)} T_{3,2} \, .
\fe
They coincide with the so-called single-particle operators \cite{Aprile:2020uxk}. 
For $M \ge 6$, in general there is non-trivial degeneracy. For example, when $M=6$, we can choose $B_6^{(1)}= T_{3,3}$ and $B_6^{(2)}= T_{6}$, with which we found 
\ie
\mathcal{O}_{0|6}^{(1)} &= T_{3,3} -{9\over N^2+7} T_{4,2} +\frac{3 \left(5 N^2+1\right)}{N
   \left(N^2+3\right) \left(N^2+7\right)} T_{2,2,2}\, , \cr
\mathcal{O}^{(2)}_{0|6} &= T_6 - \frac{3N^4-11N^2+80}{N(N^4+15N^2+8)}T_{3,3}- \frac{6(N^2-4)(N^2+5)}{N(N^4+15N^2+8)}T_{4,2}+ \frac{7(N^2-7)}{N^4+15N^2+8}T_{2,2,2} \, .
\fe
The operators $\mathcal{O}^{(i)}_{p|M}$ (for any $p$) with different $M$ are denoted as different towers, and within a given $M$-th tower, $i$ labels different subtowers \cite{recp}. One of the crucial properties of $\mathcal{O}^{(i)}_{p|M}$ is that they are mutually orthogonal on $S^4$ \cite{Gerchkovitz:2016gxx}. More precisely, they obey the following conditions, 
\ie \label{eq:orthog}
\langle \mathcal{O}^{(i_1)}_{p_1|M_1} \, , \,\,  \mathcal{O}^{(i_2)}_{p_2|M_2} \rangle_{S^4} = 0 \,, \quad {\rm unless}  \quad M_1 = M_2 \quad {\rm and} \quad i_1 =i_2\, ,
 \fe 
for any $p_1, p_2$. Precisely because of this property, we only need to perform the Gram-Schmidt procedure within a given (sub)tower, as was done in \cite{recp}. We will briefly review the results of \cite{recp} below.

As we commented earlier, the integrated correlators are related to the partition function of $\mathcal{N}=2^*$ SYM on $S^4$. Following \cite{recp}, they can be expressed as, 
\ie \label{eq:newC}
{\cC}^{(M, M'|i,i')}_{N,p}(\tau, \bar \tau) = {\widetilde{R}^{(M, M'|i,i')}_{N,p}  \over 4 } \, {\widehat \cC_{N,p}}^{(M, M'|i,i')}(\tau, \bar \tau) \, ,
\fe
where
\ie \label{eq:newCh}
\widehat{\cC}^{(M, M'|i,i')}_{N,p}(\tau, \bar \tau) = \frac{v_{p|M}^{ \mu}\, \bar{v}_{p'|M'}^{ \nu}\partial_{{\tau'}^{(i)}_{\mu|M}}\partial_{{\bar{\tau}}^{{}_\prime (i')}_{\nu|M'}} \partial_m^2  \log \mathcal{Z}_N(\tau, \tau_A'; m) \big{|}_{m=\tau_A'=0}  }{ \mathcal{N}^{(i)}_{M} \,  D^{(M, M')}_{N,p} }\, ,
\fe
with $\mathcal{Z}_N(\tau, \tau_A'; m)$ being the partition function of $\mathcal{N}=2^*$ SYM on $S^4$ as given in \eqref{eq:parZ}. Here $\partial_{{\tau'}^{(i)}_{\mu|M}}$ (and $\partial_{{\bar{\tau}}^{{}_\prime (i')}_{\nu|M'}}$) corresponds to inserting the operator $\mathcal{O}^{(i)}_{\mu|M}$ (and $\mathcal{O}^{(i')}_{\nu|M'}$). The sum over $\mu$ (or $\nu$) in $v_{p|M}^{ \mu}\,\partial_{{\tau'}^{(i)}_{\mu|M}}$ (or $\bar{v}_{p'|M'}^{ \nu} \, \partial_{{\bar{\tau}}^{{}_\prime (i')}_{\nu|M'}}$) is from $0$ to $p$ for $M>0$ (or $0$ to $p'$ for $M'>0$), and from $1$ to $p$ for $M=0$ (or $1$ to $p'$ for $M'=0$)\footnote{This is because for $M=0$, the first non-trivial operator is $T_2$, which, by our definition, is $\mathcal{O}_{1|0}$, i.e. $M=0, p=1$}.
We will now explain all the other ingredients as follows. The mixing coefficients $v_{p|M}^{\mu}$  for the operators $\mathcal{O}_{p|M}^{(i)}$ are found by performing the Gram-Schmidt procedure within a given (sub)tower because of the orthogonality property \eqref{eq:orthog}. The mixing coefficients are therefore defined by requiring 
\ie \label{eq:vMp}
 \left \langle  v^{\mu}_{p|M} \mathcal{O}_{\mu|M}^{(i)} \, ,  \mathcal{O}_{q|M}^{(i)}  \right \rangle_{c} =0 \, , \qquad {\rm with} \qquad q=0, 1, \ldots, p-1 \, ,
\fe
where the connected two-point function is defined as
\ie \label{eq:MMpp}
 \left\langle  \mathcal{O}_{p|M}^{(i)} \, ,  \mathcal{O}_{p'|M'}^{(i')} \right\rangle_{c}  = 
\left\langle  \mathcal{O}_{p|M}^{(i)} \, ,  \mathcal{O}_{p'|M'}^{(i')} \right\rangle_{S^4} - \left\langle  \mathcal{O}_{p|M}^{(i)} \right\rangle_{S^4} \left\langle \mathcal{O}_{p'|M'}^{(i')} \right\rangle_{S^4} \, .
\fe
This uniquely defines the mixing coefficients $v_{p|M}^{\mu}$ as
\ie \label{eq:vgeneral1-app}
v_{p|M}^{\mu} = {p \choose \mu} \frac{\left( \frac{N^2+2M-1}{2}+\mu \right)_{p-\mu}}{(2\, i \, \tau_2)^{p-\mu}} \, .
\fe

Another crucial ingredient of the construction is the normalisation factors in the denominator in \eqref{eq:newCh}, which are given as \cite{recp}
\ie \label{eq:DD}
 \mathcal{N}^{(i)}_{M} &=
 {M^2  \over 4^M} \, \tau_2^{ {N^2-1 \over 2} }\,  {R}^{(M, M|i,i)}_{N,0} \, , \cr
  D_{N,p}^{(M,M')} &= 2^{-2p-\delta}\,\tau_2^{-(a+1+\delta+2p)}  \,(a+1)_{p+\Theta(\delta)}  \, \left(\delta+p-\Theta(\delta) \right)! \, , 
\fe
where $\Theta(x)=0$ if $x \geq 0$ and $\Theta(x)=x$ if $x < 0$, and 
\ie \label{eq:para}
a = \frac{N^2+M+M'-3}{2} \, , \qquad \delta = \frac{M-M'}{2} \, .
\fe 
Without loss of generality, we will assume $M \geq M'$ and we simply have $\Theta(\delta)=0$. 
Finally, the prefactor $\widetilde{R}^{(M, M'|i,i')}_{N,p}$ in \eqref{eq:newC}, which is independent of the coupling $\tau$, is given by (when $M \geq M'$)
\ie \label{eq:Rt}
\widetilde{R}^{(M, M'|i,i')}_{N,p}  = {2^{2p+ \delta} \, M^2  \over (M+2p)^{2} }   \,   (a+1)_{p}  \, \Gamma \left(p+\delta+1 \right)   {R}^{(M, M|i,i)}_{N,0}  \, ,
\fe
with ${R}^{(M, M|i,i)}_{N,p}$ determined by the two-point function, 
\ie \label{eq:RR}
\langle \mathcal{O}^{(i)}_{p|M}(x_1, Y_1) \,,  \mathcal{O}^{(i)}_{p|M}(x_2, Y_2) \rangle = R_{N,p}^{(M,M|i,i)}  \left( \frac{Y_1 \cdot Y_2}{ x_{12}^2} \right)^{2p+M}\,.
\fe
In the special case $M=M', i=i'$, the denominator in \eqref{eq:newCh} reduces to
\ie
\mathcal{N}^{(i)}_{M} D_{N,p}^{(M,M)} = 
v_{p|M}^{ \mu}\, \bar{v}_{p'|M'}^{ \nu}\partial_{{\tau'}^{(i)}_{\mu|M}}\partial_{{\bar{\tau}}^{{}_\prime (i)}_{\nu|M}}  \log \mathcal{Z}_N(\tau, \tau_A'; 0) \big{|}_{\tau_A'=0}  \, , 
\fe
and 
\ie
\widetilde{R}^{(M, M|i,i)}_{N,p} = {R}^{(M, M|i,i)}_{N,p} \, .
\fe

After setting up all the necessary ingredients, we are ready to present the Laplace-difference equation that is satisfied by $\widehat{\cC}^{(M, M'|i,i')}_{N,p}(\tau, \bar \tau)$ \cite{recp}, 
\ie \label{eq:gen-LDu}
\Delta_{\tau}\, {\widehat \cC_{N,p}}^{(M, M'|i, i')} & (\tau, \bar \tau)  =\, \left(p+1+ \delta  \right)\left(p+ a+1 \right) {\widehat \cC_{N,p+1}}^{(M, M'|i,i')}(\tau, \bar \tau) + p\left(p+ a +\delta \right) {\widehat \cC_{N,p-1}}^{(M, M'|i,i')}(\tau, \bar \tau) \cr
& -\, \left[ 2p\left( p+ a\right) + (2p+a+1)(\delta+1) \right] {\widehat \cC_{N,p}}^{(M, M'|i,i')}(\tau, \bar \tau) -4 \, \delta_{M,M'} \delta_{i, i'}\, \cC_{N,1}^{(0,0)}(\tau, \bar \tau) \, ,
\fe
where the laplacian $\Delta_\tau = 4 \tau_2^2 \partial_{\tau}\partial_{\bar \tau}$ and $a$ and $\delta$ are given in 
\eqref{eq:para}. The ``source term" $\cC_{N,1}^{(0,0)}(\tau, \bar \tau)$ denotes the integrated correlator associated with $\langle \mathcal{O}_2 \mathcal{O}_2 \mathcal{O}_2 \mathcal{O}_2 \rangle$, which appeared in \eqref{eq:int-SUN} as $\cC_{N}(\tau, \bar \tau)$ and has been studied in \cite{Dorigoni:2021bvj, Dorigoni:2021guq, Dorigoni:2022cua} (see also \cite{Chester:2019pvm, Chester:2019jas, Chester:2020dja, Chester:2020vyz, Collier:2022emf, Hatsuda:2022enx}). Correspondingly, $B_N(t)$, which appeared in its lattice-sum representation, will be denoted as $B^{(0,0)}_{N,1}(t)$ (note that we also omit $i,i'$ indices since there is no degeneracy in this case). Furthermore, the prefactor ${R}^{(0,0)}_{N,1}$ is simply $(N^2-1)/2$, namely, 
\ie \label{eq:CN-norm}
\cC^{(0,0)}_{N,1}(\tau, \bar \tau)= {N^2-1 \over 8} \widehat{\cC}^{(0,0)}_{N,1}(\tau, \bar \tau) \, .
\fe

The Laplace-difference equation provides a powerful recursion relation, which in principle determines ${\widehat \cC_{N,p}}^{(M, M'|i, i')}(\tau, \bar \tau)$ once the initial condition ${\widehat \cC_{N,0}}^{(M, M'|i, i')}(\tau, \bar \tau)$ is given.  Interestingly, the difference 
\ie \label{eq:ddHu}
\widehat{ \mathcal{H}}_{N,p}^{(M, M'|i, i')} & (\tau, \bar \tau) =  {\widehat \cC_{N,p}}^{(M, M'|i, i')}(\tau, \bar \tau) -  {\widehat \cC_{N,p-1}}^{(M, M'|i, i')}(\tau, \bar \tau)\, ,
\fe
 also satisfies a three-term recursion of a similar manner
\ie \label{eq:LDHu}
 \Delta_{\tau} \widehat{ \mathcal{H}}_{N,p}^{(M, M'|i, i')}(\tau, \bar \tau)=& \, (p+1+\delta) \left( p+a+1 \right) \widehat{ \mathcal{H}}_{N,p+1}^{(M, M'|i, i')}(\tau, \bar \tau)+ \, (p-1) \left(p+a+\delta-1 \right)\widehat{ \mathcal{H}}_{N,p-1}^{(M, M'|i, i')}(\tau, \bar \tau) \cr 
&- \left[(p+a)(p+\delta)+p(p+a+\delta)\right] \widehat{ \mathcal{H}}_{N,p}^{(M, M'|i, i')}(\tau, \bar \tau)\, .
\fe
Note the above recursion relation for $\widehat{\mathcal{H}}_{N,p}^{(M, M'|i, i')}(\tau, \bar \tau)$ is valid for $p\geq 1$. The Laplace-difference equation for $\widehat{\mathcal{H}}_{N,p}^{(M, M'|i, i')}(\tau, \bar \tau)$ is particularly useful since it removes the ``source term" $\cC^{(0,0)}_{N,1}(\tau, \bar \tau)$ when $M=M', i=i'$.

\section{Generating functions of integrated correlators}
\label{sec:genf}

As anticipated in the introducation, we will assume the lattice-sum representation for the integrated correlators, 
\ie \label{eq:proNp2}
\widehat{\cC}^{(M,M'|i,i')}_{N,p}(\tau, \bar \tau) =
\sum_{(m,n)\in \mathbb{Z}^2} \int_0^{\infty} e^{-t Y_{m,n}(\tau, \bar \tau) } \widehat{B}^{(M,M'|i,i')}_{N,p}(t)\, dt \, ,
\fe
where $
Y_{m,n}(\tau, \bar \tau)$ is defined in \eqref{eq:Ymn}. The function $\widehat{B}^{(M,M'|i,i')}_{N,p}(t)$ is a rational function of $t$ that contain all the dynamics of the integrated correlators. It has a Taylor expansion, 
\ie \label{eq:chat}
\widehat{B}^{(M,M'|i,i')}_{N,p}(t)= \sum_{s=2}^{\infty} \widehat{c}^{(M,M'|i,i')}_{N,p; s} \, t^{s-1} \, , 
\fe
and obeys several important properties, such as 
\ie \label{eq:prop}
\widehat{B}_{N,p}^{(M,M'|i,i')} (t) = {1 \over t} \widehat{B}_{N,p}^{(M,M'|i,i')} (1/t) \, , \qquad  \int_0^{\infty} {dt \over \sqrt{t}} \widehat{B}_{N,p}^{(M,M'|i,i')} (t) =0 \, .
\fe
It should be stressed that assumption of the lattice-sum representation is mathematically equivalent to assume that the $SL(2, \mathbb{Z})$ spectral decomposition of the integrated correlators only contains the continuous spectrum (i.e. the non-holomorphic Eisenstein series) \cite{Dorigoni:2022zcr} (see also appendix \ref{app:lattice} for more details). This proposal (in the language of $SL(2, \mathbb{Z})$ spectral decomposition) has been checked in \cite{Paul:2022piq} by explicit perturbative and non-perturbative instanton computations for some non-trivial examples. These checks provide highly non-trivial evidence for this representation.  

We argue that the $SL(2,\mathbb{Z})$-invariant Laplace-difference equation \eqref{eq:gen-LDu}, which  recursively determines ${\widehat \cC_{N,p}}^{(M, M'|i, i')}(\tau, \bar \tau)$ for all $p$ in terms of the initial data ${\widehat \cC_{N,0}}^{(M, M'|i, i')}(\tau, \bar \tau)$ (or ${\widehat \cC_{N,1}}^{(0, 0)}(\tau, \bar \tau)$ for the case of $M=M'=0$), provide further strong evidence for the lattice-sum representation \eqref{eq:proNp2}.  It is easy to see that, following from the Laplace-difference equation, if the initial data ${\widehat \cC_{N,0}}^{(M, M'|i, i')}(\tau, \bar \tau)$ has the lattice-sum representation, all the integrated correlators will enjoy the same representation.\footnote{The general solutions to the Laplace-difference equation that we will present later further confirm this statement.} When $M=M'=0$, it was proved recently in \cite{Dorigoni:2022cua} that indeed ${\widehat \cC_{N,1}}^{(0, 0)}(\tau, \bar \tau)$ can be expressed in the lattice-sum form for any $N$, as in \eqref{eq:int-SUN}. Furthermore, using the results of \cite{Paul:2022piq}, it was shown in \cite{recp} (see appendix A of the reference) that all initial conditions ${\widehat \cC_{N,0}}^{(M, M'|i, i')}(\tau, \bar \tau)$ with $M, M'<6$ can be written as linear combinations of ${\widehat \cC_{N,1}}^{(0, 0)}(\tau, \bar \tau)$, which implies that they also all have the lattice-sum representation.  This shows that ${\widehat \cC_{N,p}}^{(M, M'|i, i')}(\tau, \bar \tau)$ should have the lattice-sum representation for all $N$ and $p$, at least for any $M,M'<6$.

Once  the lattice-sum representation is given, following \cite{Dorigoni:2022cua}, it is natural to further introduce the generating functions for the integrated correlators by summing over the $p$-dependence of the integrated correlators,\footnote{More generally, one may consider the generating functions for summing over both $N$ and $p$, namely $\widehat{\cC}^{(M,M'|i, i')}(w; z; \tau, \bar \tau) =\sum_{N, p}\widehat{\cC}^{(M,M'|i, i')}_{N, p}(\tau, \bar \tau)\,z^N w^p$.} 
\ie \label{eq:gen3}
\widehat{\cC}^{(M,M'|i,i')}_{N}(w; \tau, \bar \tau) =\sum_p \widehat{\cC}^{(M,M'|i,i')}_{N,p}(\tau, \bar \tau) \, w^p = \sum_{(m,n)\in \mathbb{Z}^2} \int_0^{\infty} e^{-t Y_{m,n}(\tau, \bar \tau) } \widehat{B}^{(M,M'|i,i')}_{N}(w; t)\, dt \, ,
\fe
where 
\ie \label{eq:Bgen3}
\widehat{B}^{(M,M'|i,i')}_{N}(w; t) = \sum_{p=0}^{\infty} \widehat{B}^{(M,M'|i,i')}_{N,p}(t)\, w^p =  \sum_{p=0}^{\infty} \sum_{s=2}^{\infty} \widehat{c}^{(M,M'|i,i')}_{N,p; s} \, t^{s-1}\, w^p \, .
\fe 
The integrated correlators can then be expressed in terms of generating functions as a contour integral, 
\ie
\widehat{\cC}^{(M,M'|i,i')}_{N,p}(\tau, \bar \tau) = {1\over 2\pi i} \oint_C {dw \over w^{p+1}} \widehat{\cC}^{(M,M'|i,i')}_{N}(w; \tau, \bar \tau) \, ,
\fe
where the contour $C$ circles the origin $w=0$ clockwise. 
It will prove to be convenient to introduce an intermediate quantity, 
\ie \label{eq:Bsw}
\widehat{B}^{(M,M'|i,i')}_{N; s}(w) = \sum_{p=0}^{\infty} \widehat{c}^{(M,M'|i,i')}_{N,p; s} w^p \, ,
\fe
and then $\widehat{B}^{(M,M'|i,i')}_{N}(w; t)=\sum_{s=2}^\infty \widehat{B}^{(M,M'|i,i')}_{N; s}(w)\, t^{s-1}$.

One can also introduce  the generating function for $\widehat{ \mathcal{H}}_{N,p}^{(M, M'|i,i')}(\tau, \bar \tau)$ in \eqref{eq:ddHu} by defining 
\ie
\widehat{ \mathcal{H}}_{N}^{(M, M'|i,i')}(w;\tau, \bar \tau) = \sum_{p=0}^{\infty} \widehat{ \mathcal{H}}_{N,p}^{(M, M'|i,i')}(\tau, \bar \tau)\, w^p \, , 
\fe
with its lattice-sum representation given by
\ie
\widehat{ \mathcal{H}}_{N}^{(M, M'|i,i')}(w;\tau, \bar \tau) = \sum_{(m,n)\in \mathbb{Z}^2} \int_0^{\infty} e^{-t Y_{m,n}(\tau, \bar \tau) } \,\widehat{H}^{(M, M'|i,i')}_N(w; t)\, dt\, ,
\fe
and 
\ie \label{eq:hhat}
\widehat{H}^{(M, M'|i,i')}_N(w; t) =\sum_{s=2}^{\infty} \widehat{H}^{(M, M'|i,i')}_{N;s}(w)\,t^{s-1} = \sum_{s=2}^{\infty} \sum_{p=0}^{\infty} \hat{h}^{(M, M'|i,i')}_{N,p;s}\, t^{s-1} w^p\, , 
\fe
with $\hat{h}^{(M, M'|i,i')}_{N,p;s}=\hat{c}^{(M, M'|i,i')}_{N,p;s}- \hat{c}^{(M, M'|i,i')}_{N,p-1;s}$. It is easy to see that $\widehat{B}^{(M, M'|i,i')}_{N}(w;t)$ and $\widehat{H}^{(M, M'|i,i')}_{N}(w;t)$ are related by the following relation, 
\ie \label{eq:genC}
\widehat{B}^{(M, M'|i,i')}_{N}(w;t) = \frac{1}{1-w} \widehat{H}^{(M, M'|i,i')}_{N}(w;t) \,. 
\fe

In the next section, we will determine the generating functions using the Laplace-difference equation. To obtain the generating functions, we will first determine $\widehat{B}^{(M,M'|i,i')}_{N;s}(w)$ defined in \eqref{eq:Bsw}, and then obtain $\widehat{B}^{(M,M'|i,i')}_{N}(w;t)$ by summing over the index $s$. The Laplace-difference equations \eqref{eq:gen-LDu} and \eqref{eq:LDHu} become differential equations of $\widehat{B}^{(M,M'|i,i')}_{N;s}(w)$ and $\widehat{H}^{(M,M'|i,i')}_{N;s}(w)$, and we will see that the solutions to these equations are Hypergeometric functions, from which one can perform the summation over the index $s$. We will begin with the generating functions for $M=M'$, $i=i'$. By including some extra differential operators, we then generalise our method to obtain $\widehat{B}^{(M,M'|i,i')}_{N}(w;t)$ for the most general cases.

\subsection{Generating functions for $M=M'$, $i=i'$} \label{sec:M=Mp}

We begin with the case of $M=M'$, $i=i'$, namely the higher-dimensional operators that appear in the definition of the integrated correlators are identical.  When $M=M'$, $i=i'$, there is a non-trivial source term for the Laplace-difference equation \eqref{eq:gen-LDu} of 
${\widehat \cC_{N,p}}^{(M,M|i,i)}(\tau, \bar \tau)$. We will simplify the process by considering the Laplace-difference equation for $\widehat{ \mathcal{H}}_{N,p}^{(M,M|i,i)}  (\tau, \bar \tau)$ as given in \eqref{eq:gen-LDu}, which removes the source term. It is straightforward to see that the Laplace-difference equation \eqref{eq:gen-LDu} implies the following differential equation for $\widehat{H}^{(M,M|i,i)}_{N; s}(w)$,
\ie \label{eq:diffH}
s(s-1) \widehat{H}_{N;s}^{(M,M|i,i)}(w) -{(w-1)^2}  \left[\left(a+1\right) {{d \widehat{H}_{N;s}^{(M,M|i,i)}(w)}\over{d w}}  + w \frac{d^2 \widehat{H}_{N;s}^{(M,M|i,i)}(w)}{d w^2} \right] = 0 \, ,
\fe
where we have set $\delta=0$ in \eqref{eq:gen-LDu} and here $a = ({N^2+2M-3})/{2}$ since $M=M'$. Here we have used the fact that the laplacian $\Delta_{\tau}$, when acting on the lattice-sum representation, becomes $t \partial_t^2 (t \widehat{H}_{N}^{(M,M|i,i)}(w;t))$, which gives the first term in the differential equation \eqref{eq:diffH}.
The solution to this differential equation is given by the standard Hypergeometric function. After taking into account the boundary conditions, we obtain the general formula for $\widehat{H}_{N;s}^{(M,M|i,i)}(w)$,  
\ie \label{eq:solH}
\widehat{H}_{N;s}^{(M,M|i,i)}(w) = \frac{(a+1)\, \hat h_{N,1; s}^{(M,M|i,i)}}{ s(s-1)} \left[
(1-w)^s \, _2F_1\left(s, s+ a ; a+1 ;w\right)- 1 \right] +\, \hat h_{N,0; s}^{(M,M|i,i)}\, .
\fe  
where $\hat h_{N, p; s}^{(M, M'|i,i')}$ are the Taylor coefficients of $\widehat{H}_{N}^{(M, M'|i, i')}(w;t)$, as defined in \eqref{eq:hhat}. 
 We need both $\hat h_{N, 0; s}^{(M, M|i,i)}$ and $\hat h_{N,1;s}^{(M,M|i,i)}$ as the boundary conditions, as the recursion relation \eqref{eq:LDHu} is only valid when $p\geq 1$.  

Using the relation \eqref{eq:genC} and the solution \eqref{eq:solH}, we can express $\widehat{B}^{(M,M|i,i)}_{N;s}(w)$  as, 
\ie \label{eq:genC2}
\widehat{B}^{(M, M|i,i)}_{N;s}(w) =q_{N;s}^{(M, M|i, i)} \, (-1)^s \, (1-w)^{s-1} \, {}_2F_1\left(s, s+ a ; a+1 ;w\right) -\, {4\, {c}^{(0,0)}_{N,1;s} \over s(s-1)(1-w)} \, ,
\fe
with
\ie \label{eq:qns}
q_{N;s}^{(M, M|i,i)} = (-1)^{s} \left({\hat{c}^{(M, M|i,i)}_{N,0;s}}+{4\, {c}^{(0,0)}_{N,1;s} \over s(s-1)} \, \right) \, ,
\fe
where we have used $\hat h_{N,p; s}^{(M, M|i, i)}=\hat{c}^{(M,M|i,i)}_{N,p;s}- \hat{c}^{(M, M|i,i)}_{N,p-1;s}$, and have expressed $\hat{c}^{(M, M|i,i)}_{N,1;s}$ in terms of $\hat{c}^{(M, M|i,i)}_{N,0;s}$ and ${c}^{(0,0)}_{N,1;s}$ using the Laplace-difference equation.\footnote{Note ${c}^{(0,0)}_{N,1;s}={(N^2-1) \over 8}\hat{c}^{(0,0)}_{N,1;s}$ according to \eqref{eq:CN-norm}.}
Importantly, $q_{N;s}^{(M, M|i,i)}$ contains all the initial data, and $\hat{c}^{(M,M|i,i)}_{N,p;s}$ is always proportional to $(-1)^s s (s-1) (2s-1)^2$, and is symmetric under $s \leftrightarrow 1-s$.   Below are a few examples of $\hat{c}^{(M,M|i,i)}_{N,p;s}$, 
\begin{align}
 {\hat c}^{(0,0)}_{2,1;s} &= \frac{2}{3}(-1)^s s(s-1)(2s-1)^2\, , \qquad {\hat c}^{(0,0)}_{2,2;s}=\frac{2}{15}(-1)^s s(s-1)(2s-1)^2 (s^2-s+8) \, , \\
{\hat c}^{(3,3)}_{3,0;s} &= \frac{3}{40}(-1)^s s(s-1)(2s-1)^2 (s^2-s+18) \,, \cr
{\hat c}^{(4,4)}_{4,0;s} &= \frac{(-1)^s}{1028160} s(s-1)(2s-1)^2 (25 s^6-75 s^5+8599 s^4-17073 s^3+291832 s^2-283308 s+2140848) \, , \nonumber
\end{align}
and ${\hat c}^{(0, 0)}_{N,0;s} =0$ for any $N$. Here we have omitted the degeneracy index $i$ since $M\leq 5$ exhibit no degeneracies. These were calculated by using \eqref{eq:newCh} and the representation \eqref{eq:proNpE}. Because of the representation \eqref{eq:proNpE} (or equivalently the lattice-sum representation), we only need to focus perturbative terms to determine these coefficients ${\hat c}^{(M,M'|i,i')}_{N,p;s}$ \cite{Dorigoni:2022zcr}.

As ${c}^{(0,0)}_{N,1;s}$ and ${\hat{c}^{(M, M|i,i)}_{N,0;s}}$ are both proportional to $(-1)^s s(s-1)(2s-1)^2$, it is easy to see that $q_{N;s}^{(M, M|i,i)}$ is a polynomial in $s$, and we find it has degree
\ie \label{eq:degreeQ}
x = 2N + 2 \left\lfloor {M/2} \right\rfloor -2 \, ,
\fe 
where $\left\lfloor y \right\rfloor$ is the floor function.

To complete the final step of obtaining the generating functions for the integrated correlators, we also need to sum over the parameter $s$ in $\widehat{B}^{(M,M|i,i)}_{N;s}(w)$, 
\ie \label{eq:BNzt-ssum}
\widehat{B}_{N}^{(M,M|i,i)}(w; t):=\sum_{s=2}^{\infty} \widehat{B}^{(M,M|i,i)}_{N;s}(w) \, t^{s-1} \, .
\fe
Firstly, for the last term in \eqref{eq:genC2}, which has a pole at $w=1$, we have
\ie \label{eq:pole}
\widehat{B}_{N}^{(M,M|i,i)}(w; t) \big{\vert}_{\rm pole} = -\sum_{s=2}^{\infty} {4 {c}^{(0,0)}_{N,1;s} \over s(s-1)(1-w)} t^{s-1} \, .
\fe
It is easy to see that $\widehat{B}_{N}^{(M,M|i,i)}(w; t) \big{\vert}_{\rm pole}$ is independent of $M$ and $i$, and obeys the following relation, 
\ie \label{eq:pole2}
t \, \partial_t^2  \left( t\, \widehat{B}_{N}^{(M,M|i,i)}(w; t) \big{\vert}_{\rm pole}\right) = {4 \over w-1} \sum_{s=2}^{\infty} {c}^{(0,0)}_{N,1;s} t^{s-1} = {4 \over w-1}  B^{(0,0)}_{N,1}(t) \,,
\fe
where  $B^{(0,0)}_{N,1}(t)$ again is simply $B_N(t)$ that appears in the lattice-sum representation \eqref{eq:int-SUN} (here we write it in our unified notation). The generating function for $B^{(0,0)}_{N,1}(t)$, by summing over the $N$-dependence, was determined recently in \cite{Dorigoni:2022cua}, and is given by
\ie \label{eq:Bztp2}
B^{(0,0)}_{1}(z; t) = \sum_{N=2}^{\infty} B^{(0,0)}_{N,1}(t) \, z^N = \frac{3 t z^2 \left[(t-3) (3 t-1)(t+1)^2 -
   z(t+3) (3 t+1) (t-1)^2 \right]}{2(1-z)^{\frac{3}{2}}
   \left[(t+1)^2-(t-1)^2 z\right]^{\frac{7}{2}}} \, .
\fe
The result \eqref{eq:pole} shows that the generating functions in general have a pole at $w=1$, and from \eqref{eq:pole2}, the residue at this pole obeys the following relation,  
\ie \label{eq:resz1}
t \, \partial^2_t \left(t\, {\rm Res}_{w=1} \widehat{B}^{(M,M|i,i)}_{N}(w;t) \right) = 4 B^{(0,0)}_{N,1}(t)\, ,
\fe
where $B^{(0,0)}_{N,1}(t)$ is given in \eqref{eq:Bztp2}. The differential operator $t \, \partial^2_t (t \ldots)$, when translated into the $\tau$-plane, becomes $\Delta_{\tau}$. Therefore the contribution from the residue at $w=1$ of the generating function cancels precisely the ``source term"  $-4 \, \cC^{(0,0)}_{N,1}(\tau, \bar \tau)$ in the Laplace-difference equation \eqref{eq:gen-LDu}.

For the remaining part of $\widehat{B}^{(M,M|i,i)}_{N;s}(w)$, we see that there is an interesting distinction in $\widehat{B}^{(M,M|i,i)}_{N;s}(w)$ depending on whether $N$ is even or odd, due to the Hypergeometric function containing integer or half-integer arguments. In particular, the parameter $a$ defined in \eqref{eq:para} that appears in the solution \eqref{eq:genC2} is an integer if $N$ is odd and a half integer if $N$ is even.  This distinction leads to different structures of the final expression for the generating function $\widehat{B}_{N}^{(M,M|i,i)}(w; t)$, and consequently the integrated correlators, especially in the large-charge expansion. We will now discuss each case separately.

\subsubsection{Even $N$}

We begin with the case of even $N$, for which $a$ is a half-integer. From \eqref{eq:genC2}, the remaining part of $\sum_s \widehat{B}^{(M,M|i,i)}_{N;s}(w)\, t^{s-1}$ is given by 
\ie \label{eq:remainingB}
\sum_{s=1}^{\infty} q_{N,s}^{(M,M|i,i)} \, (-1)^s \, (1-w)^{s-1} \, {}_2F_1\left(s, s+ a ; a+1 ;w\right) t^{s-1} \, .
\fe
Focusing first on the sum of the Hypergeometric function, we define 
\ie \label{eq:FFa}
\mathcal{F}^{(M)}_N (w,t):=& \sum_{s=1}^\infty {}_2F_1\left(s,s+a ;a+1;w\right)\,t^{s-1} \, ,
\fe
which for the even $N$ case (again in this case the parameter $a$ is a half integer) is given by
\ie \label{eq:FFaeven}
\mathcal{F}^{(M)}_N (w,t)=\, \frac{(-1)^{a+1/2}}{2^{2a}} \binom{a}{\frac{1}{2}} \left(\frac{\left(t^2-2 t (w+1)+(w-1)^2\right)^{a-1/2}}{(w\, t)^{a}}\, G(w, t) -\frac{ f_{N}^{(M)} (w,t)}{(w\, t)^{a-1/2}} \right)\, ,
\fe
where we have defined
\ie \label{eq:coth}
G(w,t):=\coth ^{-1}\left(\frac{\sqrt{w}-1}{\sqrt{t}}\right)+\coth
   ^{-1}\left(\frac{\sqrt{w}+1}{\sqrt{t}}\right) \, .
\fe
The function $f_N^{(M)}(w,t)$ is a polynomial in $w$ and $t$, which can be written as 
\ie \label{eq:fa}
f_N^{(M)}(w,t)=g_N^{(M)}(w,t)+\frac{t^{2a-2}}{(w-1)^{2a-2}} \, g_N^{(M)} \left(w,\frac{(w-1)^2}{t} \right) \, ,
\fe
and $g_N^{(M)}(w,t)$ is a polynomial obtained by expanding  $\frac{\left(t^2-2 t (w+1)+(w-1)^2\right)^{a-1/2}}{(w\, t)^{1/2}}\, G(w, t)$ up to $O(t^{a-3/2})$. Said another way, the function $f_N^{(M)}(w,t)$ is essentially determined by $G(w, t)$ by ensuring that the full function $\mathcal{F}^{(M)}_N (w,t)$ has the right properties in the small-$t$ expansion.  When $N=2, M=0$, as an example, we have $a=1/2$ and
\ie \label{eq:hypN2}
\mathcal{F}^{(0)}_{2}(w,t) = \sum_{s=1}^{\infty} {}_2F_1 \left(s, s+\frac{1}{2}; \frac{3}{2}, w\right)\, t^{s-1} =
-\frac{G(w,t) }{2 \sqrt{w\, t}}\, .
\fe
In this particular case, $f_{2}^{(0)}(w,t)=0$ according to \eqref{eq:fa}, since $g_{2}^{(0)}(w,t)=0$. 

We now take into account the prefactors $(-1)^s$ and $(1-w)^{s-1}$, giving
\ie \label{eq:mFa}
\sum_{s=1}^\infty (-1)^s (1-w)^{s-1} {}_2F_1\left(s,s+a ;a+1;w\right)\,t^{s-1}= -\mathcal{F}_N^{(M)}(w, t(w-1))\,,
\fe
which can be written as 
\ie \label{eq:FFaevenw-1}
\mathcal{F}_N^{(M)}(w, t(w-1)) = {1\over 2^{2a} } \binom{a}{ {1\over 2} } \left[ - \frac{(t+1)^{2a-1} (1-w/w_1)^{a-1/2}}{(w \, t )^a \sqrt{w-1}} G(w,t(w-1)) + \frac{ f_N^{(M)}(w,t(w-1))}{(w \, t)^{a-1/2} (1-w)^{a-1/2}} \right] \, ,
\fe
where $w_1={(1+t)^2}/{(1-t)^2}$. 
%\GB{
%\ie \label{eq:FFaevenw-1}
%\mathcal{F}_N^{(M)}(w, t(w-1)) = -2^{-2a} \binom{a}{1/2} \left( \frac{(t+1)^{2a-1} (1-w/w_1)^{a-1/2}}{(w \, t )^a \sqrt{w-1}} G(w,t(w-1)) - \frac{f_N^{(M)}(w,t(w-1))}{(-w \, t)^{a-1/2} (w-1)^{a-1/2}} \right) \, ,
%\fe}
Finally, we take into account the prefactor $q_{N;s}^{(M,M|i,i)}$. From \eqref{eq:remainingB}, one can see that each $s$ in $q_{N;s}^{(M,M|i,i)}$ becomes a differential operator in $t$, defined by 
\ie \label{eq:defS}
S\cdot f(t):=  \partial_t \left( t\, f(t) \right) \, ,
\fe
where $f(t)$ is a general polynomial. As  $q_{N;s}^{(M,M|i,i)}$ is a polynomial in $s$, it gets promoted to the differential operator $q_{N;S}^{(M,M|i,i)}$ acting on the function $\mathcal{F}_N^{(M)}(w, t(w-1))$.
%\GB{and $(S+k):= t^{-k} \partial_t \, t^{k+1}$. Not really necessary for the differential operator, but could be useful when introducing the integral operator, as $(S+k)^{-1}$ is the inverse of $(S+k)$}
Putting all the terms together, the generating function is given by 
\ie \label{eq:Btotal}
\widehat{B}^{(M,M|i,i)}_{N}(w;t) = -q_{N;S}^{(M,M|i,i)} \cdot \mathcal{F}^{(M)}_N(w, t(w-1)) + \widehat{B}_{N}^{(M,M|i,i)}(w; t) \big{\vert}_{\rm pole} \, ,
\fe
where $q_{N;s}^{(M,M|i,i)}$ is given in \eqref{eq:qns}. This equation is valid for both even and odd $N$.

For the even $N$ case, we can use the expression of $\mathcal{F}^{(M)}_N(w, t(w-1))$ in \eqref{eq:FFaevenw-1}, and the simple structure of $q_{N;S}^{(M,M|i,i)}$, to write down the general form of  $\widehat{B}^{(M,M|i,i)}_{N}(w;t)$,\footnote{The formula does not apply to the special case $N=2, M=0$. For this particular case, the generating function is given in \eqref{eq:BtzN2}. }
\ie \label{eq:Bgen}
\widehat{B}^{(M,M|i,i)}_{N}(w;t) &=\frac{ (t+1)^{2b} \left(1-{w / w_1} \right)^{b} \sqrt{w-1}  P_{N}^{(M,M|i,i)}(w,t)}{\pi (w\,t)^{a}} \, G(w, t(w-1))\cr
&+\, \frac{(t+1)\, Q^{(M,M|i,i)}_N(w,t)}{(w\, t)^{a-1/2}} + \widehat{B}_{N}^{(M,M|i,i)}(w; t) \big{\vert}_{\rm pole} \, ,
\fe
where $P_{N}^{(M,M|i,i)}(w,t)$ (or $Q^{(M,M|i,i)}_N(w,t)$) is a polynomial of degree-$2x$  (or degree-$2(x-1)$)  in $t$ and degree-$(x-1)$  in $w$, with the parameter $a$ given in \eqref{eq:para}, $x$ (the degree  of $q_{N;s}^{(M,M|i,i)}$ as a polynomial in $s$) given in \eqref{eq:degreeQ}, and $b$ given as
\ie \label{eq:parb}
b =a - 1/2 - x = \frac{N^2+2M}{2}-2N -2 \left\lfloor \frac{M}{2} \right\rfloor \, .
\fe
Furthermore, the polynomials $P_{N}^{(M,M|i,i)}(w,t)$ and $Q^{(M,M|i,i)}_N(w,t)$ obey some interesting symmetry properties. For instance, under $t \rightarrow 1/t$, 
\ie
Q_N^{(M,M|i,i)}(w,t) = t^{2(x-1)} Q_N^{(M,M|i,i)}(w,1/t)\, , \qquad P_N^{(M,M|i,i)}(w,t) = t^{2x} P_N^{(M,M|i,i)}(w,1/t) \, ,
\fe
and correspondingly the generating function has the property $\widehat{B}^{(M,M|i,i)}_{N}(w;t)=1/t \, \widehat{B}^{(M,M|i,i)}_{N}(w;1/t)$ as stated in \eqref{eq:prop}. 
 We further find that $P_N^{(M,M|i,i)}(w,t)$ satisfies an analogous symmetry in $w$ as  
 \ie
 P_N^{(M,M|i,i)}(w,t) = w^{x-1} P_N^{(M,M|i,i)}(1/w,\,-t) \, .
 \fe
 
The general structure \eqref{eq:Bgen} can be understood from \eqref{eq:Btotal} and using the fact that $q_{N;S}^{(M,M|i,i)}$ is an order-$x$ differential operator in $t$ acting on $\mathcal{F}_N^{(M)}(w, t(w-1))$. The first term of $\widehat{B}^{(M,M|i,i)}_{N}(w;t)$ in \eqref{eq:Bgen} arises from  the first term given in \eqref{eq:FFaevenw-1}, specifically when the derivatives act on the term in front of the $G(w,t(w-1))$. The pre-factor $q_{N;S}^{(M,M|i,i)}$ has a factor of $(2S-1)$, which leads to an extra factor of $(w-1)$ when acting on $(t+1)^{2a-1} (1-w/w_1)^{a-1/2}/(w \, t )^a$.   One can also see that ${f^{(M)}_N(w,t(w-1))}/{(w\, t)^{a-1/2}}$ in \eqref{eq:FFaevenw-1} (and the contribution from the first term when the derivatives act on $G(w,t(w-1))$, which becomes a rational function) leads to the second term in $\widehat{B}^{(M,M|i,i)}_{N}(w;t)$. This term naively contains some higher-order pole $1/(w-1)^{a-1/2}$ due to $t \rightarrow t(w-1)$. However this is canceled out by the numerator $f^{(M)}_N(w,t(w-1))$, which behaves as  $(w-1)^{a-1/2}$, as can be seen from the definition \eqref{eq:fa}. Finally,  the last term $\widehat{B}_{N}^{(M,M|i,i)}(w; t) \big{\vert}_{\rm pole}$ is given in \eqref{eq:pole}, which obeys the relation \eqref{eq:pole2}.

As we will see, by a simple application of the residue theorem, what is important for understanding the charge dependence of integrated correlators is the singularity structure of the generating functions on the $w$-plane. As shown in \eqref{eq:Bgen} explicitly, the generating functions have a pole at $w=1$, which is determined by \eqref{eq:pole2}. 
 The generating functions in general also have branch cuts along $(1, w_1)$ (again $w_1= (t+1)^2/(t-1)^2$), and the discontinuities  can be read off from \eqref{eq:Bgen}, and take the following general form, 
\ie \label{eq:dis}
{\rm disc} \widehat{B}^{(M,M|i,i)}_{N}(w;t) = -i\frac{ (t+1)^{2b} \left(1-{w / w_1} \right)^{b}\sqrt{w-1}\,  P_{N}^{(M,M|i,i)}(w,t)}{ (w\,t)^{a}} \, ,
\fe
where we have also used that $G(w, t (w-1)) \sqrt{w-1}$ has branch cuts along $(1, w_1)$ with discontinuity given by
\ie \label{eq:discG}
{\rm disc}\left[ {G(w, t (w-1))  \sqrt{w-1} }\right] = - {i\, \pi  \sqrt{w-1} } \, . 
\fe
It is of importance to note that the parameter $b$, as defined in \eqref{eq:parb}, is an integer when $N$ is even. Therefore, in this case, $(t+1)^{2b} \left(1-{w / w_1} \right)^{b}$ is simply a polynomial in $t$ and $w$. This will not be the case for odd $N$, and we will discuss the implications of this later.

The discontinuity can also be found directly from $\mathcal{F}^{(M)}_N(w, t(w-1))$. As $(a-1/2)$ is an integer, the discontinuity in \eqref{eq:FFaevenw-1} arises solely from the first term, and so 
\ie \label{eq:dcFeven}
{\rm disc} \mathcal{F}^{(M)}_N(w, t(w-1)) = {i\, \pi \over   2^{2a}} \, \binom{a}{{1\over 2}} \frac{(t+1)^{2a-1} (1-w/w_1)^{a-1/2}}{(w \, t )^a (w-1)^{1/2}} \, .
\fe
As $q_{N;S}^{(M,M|i,i)}$ will not affect the discontinuity on the $w$-plane, we find 
\ie
{\rm disc} \widehat{B}^{(M,M|i,i)}_{N}(w;t) =  -q_{N;S}^{(M,M|i,i)} \cdot \left( {i\, \pi \over   2^{2a}}  \binom{a}{{1\over 2}} \frac{(t+1)^{2a-1} (1-w/w_1)^{a-1/2}}{(w \, t )^a (w-1)^{1/2}} \right) \, .
\fe
Then the general structure of the discontinuity given in \eqref{eq:dis} follows using the fact that $q_{N;S}^{(M,M|i,i)}$ is an order-$x$ differential operator in $t$ with a factor of $(2S-1)$, with the operation $S$ given in \eqref{eq:defS}.

\subsubsection{Odd $N$}
\label{eq:oddN}

We now consider the case of odd $N$. In this case, the parameter $a$ (as given in \eqref{eq:para}) is an integer, and so the Hypergeometric function in \eqref{eq:remainingB} has integer arguments. Using the Pfaff transformation
\ie
{}_2F_1\left(a,b;c;w\right) = (1-w)^{-b} {}_2F_1\left(b,c-a;c;\frac{w}{w-1}\right)\, ,
\fe
and the definition of Jacobi polynomials
\ie
P_n^{(\alpha, \beta)}(w) = \frac{(\alpha + 1)_n}{n!} {}_2F_1 \left(-n, 1+\alpha + \beta +n; \alpha +1; \frac{1-w}{2}\right) \, ,
\fe 
we can express $\mathcal{F}^{(M)}_N (w,t)$ defined in \eqref{eq:FFa} in terms of Jacobi polynomials, 
\ie
\mathcal{F}_N^{(M)}(w,t) = \sum_{s=1}^\infty (1-w)^{-s} \frac{\Gamma(a+1)}{(s)_a} P_{s-1}^{(a,-a)} \left(\frac{1+w}{1-w} \right) t^{s-1} \, .
\fe
We can then apply the generating function of the Jacobi polynomials 
\ie \label{eq:Jac}
\sum_{n=0}^{\infty} P_n^{(\alpha, \beta)}(z) t^n=2^{\alpha+\beta} R^{-1}(1-t+R)^{-\alpha}(1+t+R)^{-\beta} \, , 
\fe
where $R=\left(1-2 z t+t^2\right)^{\frac{1}{2}}$, and find
\ie \label{eq:FFaodd}
\mathcal{F}_N^{(M)}(w,t) = \frac{\Gamma(a+1)}{(1-w)} \, (S)_{a}^{-1} \cdot \tilde{R}^{-1} \left(1-\frac{t}{1-w}+\tilde{R}\right)^{-a} \left(1+\frac{t}{1-w}+\tilde{R}\right)^{a} \, ,
\fe
where $\tilde{R} = \left(1 - 2t \frac{1+w}{(1-w)^2} +\frac{t^2}{(1-w)^2}\right)^{1/2}$. As before, $S$ should be understood as a differential operator, whereas its inverse from the factor $(S)_{a}^{-1}$ should be understood as an integral on any function $f(t)$ it acts on,
 \ie
(S+k)^{-1} \cdot\, f(t):= t^{-k-1} \int dt \, t^{k} \, f(t) \, ,
 \fe
which is the inverse of $(S+k)$.

Evaluating \eqref{eq:FFaodd}, we find, similarly to \eqref{eq:FFaeven}, 
\ie \label{eq:FFaodd1}
\mathcal{F}_N^{(M)}(w,t) = {(-1)^{a+1}\pi  \over  2^{2a+1}  }\binom{a}{{1\over 2}} \left( \frac{(t^2-2t(w+1)+(w-1)^2)^{a-1/2}}{(w t)^a} - \frac{f_N^{(M)}(w,t)}{(w t)^a} \right) \, ,
\fe
where, similarly to the even $N$ case \eqref{eq:fa}, $f_N^{(M)}(w,t)$ is a polynomial in $w$ and $t$, defined by
\ie
f_N^{(M)}(w,t) = g_N^{(M)}(w,t) + \frac{t^{2a-1}}{(w-1)^{2a-1}} g_N^{(M)}\left(w,\frac{(w-1)^2}{t}\right) \, ,
\fe
and $g_N^{(M)}(w,t)$ is a polynomial obtained by expanding $(t^2-2t(w+1)+(w-1)^2)^{a-1/2}$ up to $O(t^{a-1})$.

To obtain the total generating function, we rescale $t \rightarrow t(w-1)$, and obtain
\ie \label{eq:FFaoddw-1}
\mathcal{F}_N^{(M)}(w, t(w-1)) =  - \frac{\pi} {2^{2a+1}} \binom{a}{{1\over 2}} \left( \frac{(t+1)^{2a-1} (1 - w/w_1)^{a-1/2}}{(w t)^a \sqrt{1-w}} - \frac{f_N^{(M)}(w,t(w-1)}{(w t)^a (1-w)^a} \right) \, .
\fe
Finally, from \eqref{eq:Btotal}, we find the generating functions take the following general form,\footnote{Just as in the case of the even $N$, where $N=2$ is a special case, here $N=3$ is special, and the corresponding generating functions for $N=3$  do not follow the same structure as \eqref{eq:genoddN}. When $N=3$, we can have $M=0$ or $M=3$, which are given in \eqref{eq:B3zt} and \eqref{eq:B33}, respectively.} 
\ie \label{eq:genoddN}
\widehat{B}^{(M,M|i,i)}_{N}(w;t) &= \frac{ (t+1)^{2b} \left(1-{w / w_1} \right)^{b} \sqrt{1-w} \, P_{N}^{(M,M|i,i)}(w,t)}{2 \, (w\,t)^{a}}+\, \frac{(t+1)\, Q^{(M,M|i,i)}_{N}(w,t)}{(w\, t)^{a}} + \widehat{B}_{N}^{(M,M|i,i)}(w; t) \big{\vert}_{\rm pole} \, ,
\fe
where $\widehat{B}_{N}^{(M,M|i,i)}(w; t) \big{\vert}_{\rm pole}$ is again determined in \eqref{eq:resz1}, with the parameters $a$ and $b$ are given in \eqref{eq:para} and \eqref{eq:parb}, respectively. The function $P_{N}^{(M,M|i,i)}(w,t)$ is a polynomial of degree-$2x$ in $t$ and degree-$(x-1)$ in $w$, just as in the even $N$ case, and the polynomial $Q^{(M,M|i,i)}_{N}(w,t)$ has degree $2(a-1)$ in $t$ and $(a-1)$ in $w$.  They are very similar to the polynomials that appeared in the even $N$ case, and also satisfy similar symmetries, 
\ie
P_{N}^{(M,M|i,i)}(w,t) = t^{2x}P_{N}^{(M,M|i,i)}(w,1/t)\, , \qquad   
Q^{(M,M|i,i)}_{N}(w,t) &=t^{2(a-1)} Q^{(M,M|i,i)}_{N}(w,1/t) \, ,
\fe
and
\ie
 P_{N}^{(M,M|i,i)}(w,t) =w^{x-1}P_{N}^{(M,M|i,i)}(1/w,-t)\,  . 
\fe
The generating function again has a pole at $w=1$ due to the last term $\widehat{B}_{N}^{(M,M|i,i)}(w; t) \big{\vert}_{\rm pole}$. It is easy to see that $\widehat{B}^{(M,M|i,i)}_{N}(w;t)$ also has  branch cuts along $(1, w_1)$ with discontinuity of the form,  
\ie \label{eq:dis2} 
{\rm disc} \widehat{B}^{(M,M|i,i)}_{N}(w;t) = -i \frac{ (t+1)^{2b} \left(1-{w / w_1} \right)^{b} \sqrt{w-1}\, P_{N}^{(M,M|i,i)}(w,t)}{ (w\,t)^{a} } \, .
\fe
Similarly to the even-$N$ case, this discontinuity can be obtained directly from \eqref{eq:FFaoddw-1}. The discontinuity of \eqref{eq:FFaoddw-1} is 
\ie \label{eq:dFoddN}
{\rm disc} \mathcal{F}^{(M)}_{N}(w,t(w-1)) =  \frac{i \, \pi}{ 2^{2a}} \binom{a}{\frac{1}{2}}\frac{ (t+1)^{2a-1}\left(1-w/w_1\right)^{a-1/2}}{(w \, t)^a \sqrt{w-1}} \, ,
\fe
and so the discontinuity of $\widehat{B}^{(M,M|i,i)}_{N}(w;t)$ is given by 
\ie
{ \rm disc} \widehat{B}^{(M,M|i,i)}_{N}(w;t) = -q_{N,S}^{(M,M|i,i)} \cdot \left(  \frac{i\, \pi}{ 2^{2a}} \binom{a}{{1\over 2}} \frac{ (t+1)^{2a-1}\left(1-w/w_1\right)^{a-1/2}}{(w \, t)^a \sqrt{w-1}} \right) \, , 
\fe
 which again, after using the definition of $q_{N,S}^{(M, M|i,i)}$ as in \eqref{eq:qns}, leads to the general expression of the discontinuity  ${ \rm disc} \widehat{B}^{(M,M|i,i)}_{N}(w;t)$ as given in \eqref{eq:dis2}.

We see that, very interestingly, even though the generating functions for even $N$ and odd $N$ take different forms, they have identical singularity structures. However, when $N$ is odd, the parameter $b$ as given in \eqref{eq:parb} is a half-integer, and so unlike the even $N$ cases, the combination $(t+1)^{2b} \left(1-{w / w_1} \right)^{b}$ in ${ \rm disc} \widehat{B}^{(M,M|i,i)}_{N}(w;t)$ is not a polynomial of $t$ and $w$ anymore. In particular, it has branch cuts along $(w_1, \infty)$. This will lead to an important difference in the large-$p$ expansion of the integrated correlators for even and odd $N$.

\subsubsection{Generating functions for $M \neq M'$ or $M=M'$ with $i \neq i'$}

We now consider the generating functions for integrated correlators with $M \neq M'$ or $M = M'$ but with $i \neq i'$. Without loss of generality, we will assume $M \geq M'$ so that $\delta = (M-M')/2 \geq 0$. 
In these cases, because of the lack of the source term $-4 \, \cC^{(0,0)}_{N,1}(\tau, \bar \tau)$, it is not necessary to use the difference $ {\widehat \cH_{N,p}}^{(M, M'|i, i')} (\tau, \bar \tau) $ to remove the source term. The Laplace-difference equation is now
\ie \label{eq:gen-MneqM'}
\Delta_{\tau}\, {\widehat \cC_{N,p}}^{(M, M'|i, i')} & (\tau, \bar \tau)  =\, \left(p+1+ \delta  \right)\left(p+ a+1 \right) {\widehat \cC_{N,p+1}}^{(M, M'|i,i')}(\tau, \bar \tau) + p\left(p+ a +\delta \right) {\widehat \cC_{N,p-1}}^{(M, M'|i,i')}(\tau, \bar \tau) \cr
& -\, \left[ 2p\left( p+ a\right) + (2p+a+1)(\delta+1) \right] {\widehat \cC_{N,p}}^{(M, M'|i,i')}(\tau, \bar \tau) \, .
\fe
In terms of $\widehat{B}^{(M,M'|i,i')}_{N; s}(w)$, which was introduced in \eqref{eq:Bsw}, the Laplace-difference equation becomes
\ie \label{eq:gen-MneqM'csumw}
(w-1)^2  \partial_w^2 & \widehat{B}^{(M,M'|i,i')}_{N; s}(w)  + (w-1) \left[ (a+\delta+3)w-(a+\delta+1) \right] \partial_w \widehat{B}^{(M,M'|i,i')}_{N; s}(w) \cr
& - \left[s(s-1)-(w-1) \left(a+\delta +1 - \frac{a \, \delta}{w} \right) \right]\widehat{B}^{(M,M'|i,i')}_{N; s}(w) = 0 \, . 
\fe
The solution to this differential equation is again given in terms of Hypergeometric function, and takes the following form, 
\ie
\widehat{B}^{(M,M'|i,i')}_{N;s}(w) = \frac{\hat{c}^{(M,M'|i,i')}_{N,0;s}(a-\delta+1)_\delta \, \delta!}{(s-\delta)_{2\delta}  \, w^{\delta}}  \left[ (1-w)^{s-1} \, _2F_1 (s-\delta, s+a, a-\delta+1, w) \right. \cr 
\left. - \sum_{r=0}^{\delta-1} \sum_{l=0}^\delta \binom{s-1}{l} \frac{(s-\delta)_{r-l} (a+s)_{r -l}}{(a-\delta+1)_{ r-l} (r -l)!} (-1)^{l} w^r  \right] \, .
\fe
 We therefore see that the generating functions in general can be written as 
\ie \label{eq:BMneqM's}
\widehat{B}^{(M,M'|i,i')}_{N;s}(w) =(-1)^s q_{N; s}^{(M,M'|i,i')} \left(  (1-w)^{s-1} \, _2F_1 (s-\delta, s+a, a-\delta+1, w) -   K_{N; s}^{(M,M')}(w)  \right) \, , 
\fe
where $K_{N;s}^{(M,M')}(w)=0$ when $M=M'$ (with $\delta=0$), otherwise, 
\ie
K_{N;s}^{(M,M')}(w) = 
\sum_{r=0}^{\delta-1} \sum_{l=0}^\delta \binom{s-1}{l} \frac{(s-\delta)_{r-l} (a+s)_{r -l}}{(a-\delta+1)_{ r-l} (r -l)!} (-1)^{l} w^r  \, . 
\fe
It is easy to see that the role of $K_{N;s}^{(M,M')}(w)$ is to ensure that, by definition, $\widehat{B}^{(M,M'|i,i')}_{N;s}(w)$ behaves as $w^0$ in the small-$w$ expansion. Finally, $q_{N;s}^{(M,M'|i,i')}(w)$, which again contains all the initial data, is given by 
\ie \label{eq:qMMp}
q_{N;s}^{(M,M'|i,i')}(w) = (-1)^{s} \left(\frac{\hat{c}^{(M,M'|i,i')}_{N,0;s}(a-\delta+1)_\delta \, \delta!}{(s-\delta)_{2\delta}  \, w^{\delta}} + \delta_{M,M'} \delta_{i,i'} \frac{4 \, c_{N,1; s}^{(0,0)}}{s(s-1)} \right)\, .
\fe
We see that, when $M=M'$ and $i=i'$, $q_{N;s}^{(M,M'|i,i')}$ reduces to $q_{N;s}^{(M,M|i,i)}$ given in \eqref{eq:qns}. As we are only focusing on the cases with $M \neq M'$ or $i \neq i'$, we will ignore the term with $c_{N,1; s}^{(0,0)}$. When $M \neq M'$, $\hat{c}^{(M,M'|i,i')}_{N,0;s}$ always has a factor of $(-1)^s (s-\delta)_{2 \delta} (2s-1)^2$ (and when $M=M'$, $\hat{c}^{(M,M'|i,i')}_{N,0;s}$ vanishes at least at $s=0, 1$), and is symmetric under $s \rightarrow 1-s$. For example, 
\begin{align}
    \hat{c}^{(4,0)}_{4,0;s} &=\frac{(-1)^s }{7560}(s-2) (s-1) s (s+1) (2 s-1)^2 \left(5 s^4-10 s^3+251 s^2-246 s+360\right)\, , \\
\hat{c}^{(5,3)}_{5,0;s} &=\frac{(-1)^s }{60480}  (s-2) (s-1) s (s+1) (2 s-1)^2 \left(s^6-3 s^5+115 s^4-225 s^3+3604 s^2-3492 s+5040\right) \, . \nonumber
\end{align}
It is therefore clear that $q_{N,s}^{(M,M'|i,i')}$ is a polynomial in s, and its degree is
\ie \label{eq:degQQ}
x=2N+2 \left\lfloor \frac{M}{2} \right\rfloor -2 -2\delta\, ,
\fe
which generalises \eqref{eq:degreeQ} for $\delta\neq 0$.

We now perform the summation over the index $s$. As there is a factor of $(s-\delta)_\delta$ in $\widehat{B}^{(M,M'|i,i')}_{N;s}(w)$, which can be seen from the expansion of the Hypergeometric function, we can start the sum at $s=\delta+1$. We therefore find
\ie \label{eq:Bwtsum}
\widehat{B}_{N}^{(M,M'|i,i')}(w; t) = -\sum^{\infty}_{s=\delta+1} q_{N;s}^{(M,M'|i,i')}(w) \left[ (1-w)^{s-1} \, _2F_1 (s-\delta, s+a, a-\delta+1, w) -  K_{N; s}^{(M,M')}(w)  \right] (-t)^{s-1} \, .
\fe
As $ q_{N;s}^{(M,M'|i,i')}(w)$ and $ K_{N;s}^{(M,M'|i,i')}(w)$ are both polynomials in $s$, the last term in \eqref{eq:Bwtsum} sums to  
\ie
q_{N; S}^{(M,M'|i,i')}(w) \, K_{N;S}^{(M,M')}(w) \cdot \left( \frac{(-t)^{\delta} }{1+t} \right)\, , 
\fe
where $S$ is the differential operator defined in \eqref{eq:defS}. To compute the summation of the first term in \eqref{eq:Bwtsum}, we first find the  generating function for the Hypergeometric function, 
\ie
\mathcal{F}^{(M,M')}_{N}(w,t) := \sum_{s=\delta+1}^\infty \, {}_2F_1 \left(s-\delta, s+a, a-\delta+1, w \right)\, t^{s-1} \, ,
\fe
which reduces to $\mathcal{F}^{(M)}_{N}(w,t)$ in \eqref{eq:FFa} when $M=M'$.

We can shift $s$ so that the sum starts from $s=1$, as in the $M=M'$ case, so we have
\ie
\mathcal{F}^{(M,M')}_{N}(w,t) = \sum_{s=1}^\infty {}_2F_1 \left(s, s+a+\delta, a-\delta+1, w\right) \, t^{s+\delta-1} \, .
\fe
Using the identity 
\ie
\partial_z^n \left( z^{c-1} \, _2F_1 (a,b,c,z) \right) \equiv (c-n)_n \,z^{c-n-1} \, _2F_1 (a,b,c-n,z) \, ,
\fe
$\mathcal{F}^{(M,M')}_{N}(w,t)$ can be written as 
\ie
\mathcal{F}^{(M,M')}_{N}(w,t) = \sum_{s=1}^\infty \frac{1}{w^{a-\delta} \, (a-\delta+1)_{2\delta}}{\partial}^{2\delta}_w  \left(w^{a+\delta} {}_2F_1 (s, s+a+\delta, a+\delta+1, w)\right) \, t^{s+\delta-1} \, .
\fe
As $a+\delta = {(N^2+2M-3)}/{2}$, this is exactly the same Hypergeometric function that appeared in \eqref{eq:FFa}, and so taking the sum inside the differential, we find
\ie
\mathcal{F}^{(M,M')}_{N}(w,t) &=  \frac{t^{\delta}}{w^{a-\delta} \, (a-\delta+1)_{2\delta}} {\partial}^{2\delta}_w  \left(w^{a+\delta} \, \mathcal{F}_{N}^{(M)}(w,t)\right) \, ,
\fe
where $\mathcal{F}_{N}^{(M)}(w,t)$ is given in \eqref{eq:FFaeven} for even $N$ and \eqref{eq:FFaodd1} for odd $N$.  
Therefore, finally, the generating function is given by 
\ie \label{eq:Bwtsum2}
\widehat{B}_{N}^{(M,M'|i,i')}(w; t) =  -q_{N;S}^{(M,M'|i,i')}(w) \cdot \left[\mathcal{F}^{(M,M')}_{N}(w,t(w-1)) - K_{N,S}^{(M,M')}(w) \cdot \left(\frac{(-t)^\delta}{1+t}\right)  \right] \, ,
\fe
where $q_{N;s}^{(M,M'|i,i')}(w)$ is given in \eqref{eq:qMMp} in terms of the initial data of the recursion relation \eqref{eq:gen-LDu} (i.e. $\hat{c}^{(M,M'|i,i')}_{N,0;s}$). 

Unlike the case with $M=M', i=i'$,  the generating functions $\widehat{B}_{N}^{(M,M'|i,i')}(w; t)$ do not  have a pole at $w=1$ due to the  absence of the source term when $M\neq M'$ or $i\neq i'$,  but once again they do have branch cuts along $(1, w_1)$. We will now compute the corresponding discontinuity. From \eqref{eq:FFaeven} and \eqref{eq:FFaodd1}, and using \eqref{eq:discG}, we find  that the discontinuity of $\widehat{B}_{N}^{(M,M'|i,i')}(w; t)$ is given by
\ie \label{eq:discB}
{\rm disc} \widehat{B}_{N}^{(M,M'|i,i')}(w; t) &= q_{N;S}^{(M,M'|i,i')}(w) \cdot \frac{i \pi \, (-1)^{a+\delta+1/2} \, 2^{-2(a+\delta)} }{t^a (w-1)^a w^{a-\delta} (a-\delta+1)_{2\delta}} \binom{a+\delta}{\frac{1}{2}}\cr
&\left[ \partial_w^{2 \delta} \left((t^2-2t(w+1)+(w-1)^2)^{a+\delta-1/2}\right)\big{\vert}_{t \rightarrow t(w-1)} \right] \, .
\fe

To see the form of the discontinuity, we can perform the derivatives 
 $\partial_w^{2 \delta}$ explicitly, and so ${\rm disc} \widehat{B}_{N}^{(M,M'|i,i')}(w; t)$ can then be expressed as
\ie\label{eq:discBintermed}
{\rm disc} \widehat{B}_{N}^{(M,M'|i,i')}(w; t) 
&=-\frac{i \pi \, 2^{-2a} }{ (w-1)^{\delta+1/2} w^{a-\delta} }  \binom{a+\delta}{\frac{1}{2}} \cr
& q_{N,S}^{(M,M'|i,i')}(w) \cdot \left[ t^{-a} (t+1)^{2a-2\delta-1} (1-w/w_1)^{a-\delta-1/2} h(w,t(w-1)) \right] \, ,
\fe
where $h(w,t)$ is a polynomial in $w$ and $t$, explicitly given by
\ie
h(w,t) = \sum_{n=0}^\delta \binom{\delta}{n} \left(\delta+{1\over 2}-n \right)_n \left(a-\delta+{1\over 2}+n \right)_{2\delta-n} (1+t-w)^{2(\delta-n)} \left(t^2-2t(w+1)+(w-1)^2\right)^n \, .
\fe 
A factor of $(w-1)^{a-\delta-1/2}$ has been factored out in \eqref{eq:discBintermed}, and $h(w,t(w-1))$ has an extra factor of $(w-1)^\delta$. Furthermore, the prefactor $q_{N,S}^{(M,M'|i,i')}(w)$ is an order-$x$ (as given in \eqref{eq:degQQ}) differential operator in $t$, and so
\ie \label{eq:diBMMp}
{\rm disc} \widehat{B}_{N}^{(M,M'|i,i')}(w; t) =-i \frac{\sqrt{w-1} \left(1-{w}/{w_1}\right)^b (t+1)^{2b} P_{N}^{(M,M'|i,i')}(w,t)}{(w \, t)^a} \, ,
\fe
where
\ie
b&=a-\delta-1/2-x = \frac{N^2+2M}{2}-2N -2 \left\lfloor \frac{M}{2} \right\rfloor \, ,
\fe
and $P_{N}^{(M,M'|i,i')}(w,t)$ is a polynomial of degree-$(x+\delta-1)$ in $w$ and degree-$2(x+\delta)$ in $t$. % that has absorbed $\pi \, 2^{-2(a+\delta)} \binom{a+\delta}{1/2}$. 
In writing down the general expression, we have used the fact that an extra factor of $(w-1)$ comes from the factor of $(2S-1)$ in $q_{N,S}^{(M,M'|i,i')}(w)$ when acting on $t^{-a} (t+1)^{2a-2\delta-1} (1-w/w_1)^{a-\delta-1/2}$. Once again $P_{N}^{(M,M'|i,i')}(w,t)$ obeys some interesting symmetries, for example, 
\ie
P_{N}^{(M,M'|i,i')}(w,t) = t^{2(x+\delta)} P_{N}^{(M,M'|i,i')}(w,1/t) \, .
\fe
However, it does not have the corresponding symmetry in $w$, and so $P_{N}^{(M,M'|i,i')}(w,t)$ is not palindromic in $w$. Compared with \eqref{eq:dis} and \eqref{eq:dis2}, we find that remarkably the discontinuity ${\rm disc} \widehat{B}_{N}^{(M,M'|i,i')}(w; t)$ for all cases takes a unified form.

\subsection{Examples of generating functions}

We will now give some concrete examples of generating functions, and their singularity structures on the $w$-plane (i.e. their poles and discontinuities). We will begin with the simplest case of $M=M'=0$, then move on to examples with $M=M' > 0$ and $M \neq M'$.

\subsubsection{$M=M'=0$}

Let us begin to consider the operators $\mathcal{O}_{p|0}= (T_2)^p$  with the simplest $SU(2)$ gauge group, namely $M=M'=0, $ and $N=2$. Because in this is the case  there is no degeneracy, we will remove the index $i$. From \eqref{eq:genC2}, we have 
\ie 
\widehat{B}^{(0,0)}_{2;s}(w) = \frac{ {c}^{(0,0)}_{2, 1; s}}{ s\, (s-1) \, (1-w)} \left[  (1-w)^s \, {}_2F_1 \left(s, s+\frac{1}{2}; \frac{3}{2}, w\right) -1\right] \, ,
\fe
where  
\ie
{c}^{(0,0)}_{2, 1; s} = {3 \over 2} \hat{c}^{(0,0)}_{2,1;s}(s)= (-1)^s s(s-1)(2s-1)^2\, .
\fe

Going through the procedure we outlined in the previous section, we find 
 the generating function is given by, 
\ie \label{eq:BtzN2}
\widehat{B}^{(0,0)}_{2}(w; t)=4 t \frac{ (t-1)^2 \left(3
   t^2-2 t+3\right) w- (t+1)^2 \left(3 t^2-10 t+3\right)}{(t+1)^3 (w-1) \left[ (t+1)^2-(t-1)^2
   w\right]^2}\, .
   \fe
For this particular $N=2$ case, the $\coth ^{-1}$ components of $G(w,t(w-1))$  in $\mathcal{F}^{(0)}_2(w,(w-1)t)$ actually cancel out, and the generating function only contains poles but no branch cuts in the $w$-plane.

 In general for $N>2$, the generating functions have both poles and branch cuts and take the general form as given in \eqref{eq:Bgen}. To illustrate the universal structures, we will consider an example of even $N$, namely $N=4$. When $N=4$, we have
\ie \label{eq:BztN4}
\widehat{B}^{(0,0)}_{4;s}(w) = \frac{{c}^{(0,0)}_{4,1;s}}{ s\, (s-1) \, (w-1)} \left[ (1-w)^s \, {}_2F_1 \left(s, s+\frac{13}{2}; \frac{15}{2}, w\right) -1 \right] \, ,
\fe 
where ${c}^{(0,0)}_{4,1;s}$ is given by,
\ie
{c}^{(0,0)}_{4,1;s} = {15 \over 2} \hat{c}^{(0,0)}_{4,1;s} ={1 \over 12}  (-1)^s (s-1) s (2 s-1)^2 \left(s^2-s+4\right) \left(s^2-s+18\right) \, .
\fe

Using the procedure we outlined in the previous paragraphs, it is straightforward to obtain the generating function $\widehat{B}^{(0,0)}_4(w;t)$, which takes exactly the general form given in \eqref{eq:Bgen}, with $a=13/2$ and $b=0$, 
\ie \label{eq:Bgen-example}
\widehat{B}^{(0,0)}_{4}(w;t) &=\frac{ \sqrt{w-1}  P_{4}^{(0,0)}(w,t)}{\pi (w\,t)^{13/2}} \, G(w, t(w-1))+ \frac{(t+1)\, Q^{(0,0)}_4(w,t)}{(w\, t)^{6}} + \widehat{B}_{4}^{(0,0)}(w; t) \big{\vert}_{\rm pole} \, ,
\fe
and the relevant polynomials are given by
\begin{align} \label{eq:P004}
 &    P^{(0,0)}_4(w,t) = {45045 \pi \over 33554432}\left[6837 \,\tilde{t}^{12}_{0,5}\,-32775
   \,\tilde{t}^{12}_{1,4} \left(w+1 \right) +5688 \,\tilde{t}^{12}_{2,3}
   \left(11 w^2+18 w+11\right) \right. \cr
   &  -5457 \,\tilde{t}^{12}_{3,2} \left(11
   w^3+21 w^2+21 w+11\right) 
   +899 \,\tilde{t}^{12}_{4,1} \left(33 w^4+60 w^3+70 w^2+60 w+33\right) \cr
   & \left. -190 \,\tilde{t}^{12}_{5,0} \left(33 w^4+12 w^3+38 w^2+12 w+33\right) (w+1)\right] \, , 
   \end{align}
   and
   \begin{align} \label{eq:Q004}
 &    Q^{(0,0)}_4(w,t) = -{39 \over 16777216}\left[7896735 \,\tilde{t}^{10}_{0,5}-1155 \,\tilde{t}^{10}_{1,4} (37333 w+19101) \right. \cr
    &+ \, 231 \,\tilde{t}^{10}_{2,3} \left(451717 w^2+270630 w+87645\right)  -132 \,\tilde{t}^{10}_{3,2} \left(1182061 w^3+291179 w^2+165375 w+51345\right) \cr
  & +  \, 110  \,\tilde{t}^{10}_{4,1} \left(1610089 w^4-799548 w^3+95802 w^2+10332
   w+4221\right)
   \cr 
    &-   \left. \tilde{t}^{10}_{5,0} \left(89980301 w^5-131315679 w^4+50765066
   w^3-4821894 w^2-181335 w-232155\right)\right] \, ,
\end{align}
 where we denote $ \tilde{t}^{n}_{m,q} = (t^{m}+t^{n-m})(w-1)^q$. Finally, the pole term is 
\ie \label{eq:B4pole}
\widehat{B}^{(0,0)}_4(w;t) \big{\vert}_{\rm pole} = -\frac{2 \left(3 t^6-24 t^5-2 t^4-70 t^3-2
   t^2-24 t+3\right)}{(t+1)^7 (w-1)} \, . 
   \fe

We will eventually use the residue theorem to compute the large-$p$ expansion of the integrated correlator, so what is relevant is the singularity structures on the $w$-plane. From \eqref{eq:Bgen-example}, we see that $\widehat{B}^{(0,0)}_4(w;t)$ has a pole at $w=1$ with residue from \eqref{eq:B4pole}
\ie \label{eq:N4-pole}
{\rm Res}_{w=1} \widehat{B}^{(0,0)}_4(w;t) = -\frac{2(3t^6-24t^5-2t^4-70t^3-2t^2-24t+3)}{(t+1)^7} \, ,
\fe
and it has branch cuts along $(1, w_1)$, with the discontinuity given by, 
\ie \label{eq:N4-disc}
{\rm disc} \widehat{B}^{(0,0)}_4(w;t) = -\frac{i \sqrt{w-1} P^{(0,0)}_4(w,t) }{(t \, w)^{13/2}  } \, ,
\fe
with $P^{(0,0)}_4(w,t)$ given in \eqref{eq:P004}. 

We now consider the cases with odd $N$. For example, for $N=3$ (and $M=0,M'=0$), we find 
 \begin{align} \label{eq:B3zt}
& \widehat{B}^{(0,0)}_{3}(w;t) = \frac{3 t \left(t^4+4 t^3+17 t^2+10 t+12\right)}{(t+1)^5 (w-1)}\\
+ &\, \frac{48 t w \left[ 12 t^2 \left(t+1\right)-t \left(7 t^3-21 t^2+33 t-11\right)
   (w-1)+(t-1)^2 \left(t^3-6 t^2+12 t-5\right) (w-1)^2 \right]}{(t+1)^7 \sqrt{1-w} \left(\sqrt{1-w}+\sqrt{1-{w}/{w_1}} \right)^4 \left(1-{w}/{w_1}\right)^{3\over 2}}  \cr
   +& \, \frac{24 t \left[8 t^2-6 t \left(3 t^2+4 t+5\right) (w-1)+\left(8 t^4-t^3-3 t^2-7
   t+11\right) (w-1)^2-(t-1)^2 \left(t^3-t^2+t+1\right) (w-1)^3 \right]}{(t+1)^6 (1-w) \left(\sqrt{1-w}+\sqrt{1-{w}/{w_1}}\right)^3 \left( 1- w/{w_1}\right)^{3\over 2}} \, .\nonumber
 \end{align}
 As we commented, just like the first even $N$ case (i.e. $N=2$), for this special case, $\widehat{B}^{(0,0)}_{3}(w;t)$ does not follow the general structure \eqref{eq:genoddN}. However, the singularity properties, which we will use for the computation of the large-$p$ expansion of the integrated correlator, are universal. More concretely, 
it is straightfoward to see that $\widehat{B}^{(0,0)}_{3}(w;t)$ has a pole at $w=1$ with residue,  
\ie \label{eq:B30pole}
{\rm Res}_{w=1} \widehat{B}^{(0,0)}_{3}(w;t) = -\frac{3 \left(t^4-7 t^3-7 t+1\right)}{(t+1)^5} \, , 
 \fe
and branch cuts along $(1, w_1)$, with discontinuity
 \ie \label{eq:B30cuts}
{\rm disc} \, \widehat{B}^{(0,0)}_{3}(w;t) &= -\frac{ i \, \sqrt{w-1}\, P^{(0,0)}_3(w,t) }{(w \, t)^3 \, (t+1)^3 
   \left(1-{w}/{w_1} \right)^{3/2}}\, ,
 \fe
where $P^{(0,0)}_3(w,t)$ is given by
\ie
P^{(0,0)}_3(w,t) &= 3\left[5 \tilde{t}^{8}_{0,3}-21
  \tilde{t}^{8}_{1,2}(w+1)+\tilde{t}^{8}_{2,1} \left(34 w^2+37 w+34\right) \right. \cr
  &\left.-9 \tilde{t}^{8}_{3,0} (w+1) \left(3 w^2-2 w+3\right)+9\tilde{t}^{8}_{4,1}\left(w^2+w+1\right) \right]
\, .
\fe

\subsubsection{$M=M'\neq 0$}

Let us now consider the generating functions for $M=M'\neq 0$. We will consider the simplest case with $M=3, M'=3, N=3$ as an example, and other cases take similar forms. Here we will only give the singularity structures of the generating function, i.e. the discontinuity and the pole. The generating function for this particular case can be found in appendix \ref{app:genfun}. As explained earlier, singularities are eventually the relevant information for understanding the charge dependence of integrated correlators.

 The singularity structures of $\widehat{B}_{3}^{(3,3)}(w; t)$ follow the general structure of \eqref{eq:resz1}. As in all the cases, $\widehat{B}_{3}^{(3,3)}(w; t)$ has a pole at $w=1$ and branch cuts along $(1, w_1)$, with the following expressions for the residue and the discontinuity,
\ie 
{\rm Res}_{w=1} \widehat{B}^{(3,3)}_{3}(w;t) = -\frac{3 \left(t^4-7 t^3-7 t+1\right)}{(t+1)^5} \, , 
 \fe
and 
\begin{align} \label{eq:dsB333}
{\rm disc}\, \widehat{B}^{(3,3)}_{3}(w;t) =-\frac{ i
   \sqrt{w-1} \, P^{(3,3)}_3(w,t) }{(tw)^6 (t+1) \sqrt{1-{w/w_1}}}  \, ,
\end{align} 
with the polynomial $P^{(3,3)}_3(w,t)$ given by
\begin{align}
& P^{(3,3)}_3(w,t)  =3 \left[11 \tilde{t}^{12}_{0,5}-56
   \tilde{t}^{12}_{1,4} (w+1) +5 \tilde{t}^{12}_{2,3}
    \left(23 w^2+36 w+23\right) \right. \cr 
    & -12\tilde{t}^{12}_{3,2} (w+1) \left(10
   w^2+7 w+10\right) +\tilde{t}^{12}_{4,1}
   \left(65 w^4+92 w^3+97 w^2+92 w+65\right) \cr
   &\left. -4\tilde{t}^{12}_{5,0} (w+1) \left(w^2+w+1\right)
   \left(4 w^2-5 w+4\right)+\tilde{t}^{12}_{6,1}\left(w^2+w+1\right)^2\right] \,.  
\end{align} 

\subsubsection{$M \neq M'$}

Finally, we consider examples with $M \neq M'$. We will consider the case $M=4, M'=0$ with $N=4$ as an example, and again we will only give the singularity structures. Recall that, for $M \neq M'$ or $i \neq i'$, 
\ie 
{\rm Res}_{w=1} \widehat{B}^{(M,M'|i,i')}_{N}(w;t) = 0 \, , 
\fe
and so we only need to consider the discontinuity. The discontinuity takes the general form of \eqref{eq:diBMMp},
\begin{align} \label{eq:dB404}
& {\rm disc} \widehat{B}^{(4,0)}_{4}(w;t) =- i\frac{\sqrt{w-1}\, P^{(4,0)}_4(w,t) }{ (w\, t)^{17/2} } \, , 
\end{align}
and the polynomial $P^{(4,0)}_4(w,t)$ is given by
\begin{align} 
& P^{(4,0)}_4(w,t)= \frac{36465 \pi }{17179869184  } \left[2386324\tilde{t}^{16}_{0,7}-1025423\tilde{t}^{16}_{1,6}(19w+11)\right. \\
&+213186 \tilde{t}^{16}_{2,5} \left(323 w^2+306 w+99\right)-84545 \tilde{t}^{16}_{3,4} \left(1615 w^3+1785 w^2+945 w+231\right) \cr
&+38932 \tilde{t}^{16}_{4,3} \left(4199 w^4+4420 w^3+2730 w^2+1092 w+231\right)\cr
&-27795 \tilde{t}^{16}_{5,2} \left(4199 w^5+3315 w^4+1950 w^3+910 w^2+315 w+63\right)\cr
&+962 \tilde{t}^{16}_{6,1} \left(46189 w^6+14586 w^5+6435 w^4+2860 w^3+1155 w^2+378 w+77\right) \cr
&\left. -139 \tilde{t}^{16}_{7,0} \left(46189 w^7-17017 w^6-3003 w^5-1001 w^4-385 w^3-147 w^2-49 w-11\right)\right] \, , \nonumber
\end{align}
where we have omitted the degeneracy indices $i,i'$ since there is no degeneracy for $M<6$. 
An analogous expression for ${\rm disc}\, \widehat{B}^{(5,3)}_{5}(w;t)$ can be found in appendix \ref{app:genfun}. 

The computation for the generating functions for other cases is very similar. We will not list results for more examples here. In the next section, we will use these generating functions, especially their singularity structures, to obtain the large-charge expansion of integrated correlators.

\section{Large-charge expansion}
\label{sec:large-p}

{As we commented in the introduction, it has been shown in the literature that the charge of some global symmetry provides a very useful expansion parameter for performing the analytical study of physical observables in QFT. In particular, it allows us to access the strong coupling regime, and can often be described by some effective field construction (see e.g. \cite{Hellerman:2015nra,Monin:2016jmo,Alvarez-Gaume:2016vff,Hellerman:2017veg,Jafferis:2017zna}). In this section, we will utilise the generating functions of integrated correlators we obtained previously to study the large-charge properties of the integrated correlators. It will provide explicit analytical results of the integrated correlators in the large-charge expansion. As we will show, all the integrated correlators behave universally in the large-charge expansion (up to some subtle difference between even and odd $N$), and we will further find that the large-charge expansion of the integrated correlators take a rather similar form to that of the large-$N$ expansion. In particular, the same types of modular functions appear in both cases.

 Let us begin with the general expression for the integrated correlators. From \eqref{eq:gen3}, we have
\ie 
\widehat{\cC}^{(M,M'|i,i')}_{N, p}(\tau, \bar \tau) = \sum_{(m,n)\in \mathbb{Z}^2}\int_0^{\infty} dt \, \frac{1}{2 \pi i} \oint_C {dw \over w^{p+1}} e^{-t Y_{m,n}(\tau, \bar \tau) } \widehat{B}^{(M,M'|i,i')}_N(w; t)\, .
\fe
It is convenient to separate $\widehat{\cC}^{(M,M'|i,i')}_{N, p}(\tau, \bar \tau)$ into several different parts. Firstly, we will separate out the $(m,n)=(0,0)$ case in the lattice-sum, which contributes to the $\tau$-independent part of the integrated correlators since $Y_{0, 0} (\tau, \bar \tau)=0$. In fact, in the large-$p$ expansion, there is additional $\tau$-independent contribution, as we show in \eqref{eq:const}. We will denote all the $\tau$-independent contributions as $ \widehat{\cC}^{(M,M'|i,i')}_{N,p}\big{\vert}_{\rm const}$.  For the remaining $(m,n)\neq (0,0)$ part, we will deform the contour, as shown in figure \ref{fig:contour}. Instead of surrounding $w=0$, it will surround the pole at $w=1$, and the discontinuity along $(1,w_1)$. These contribtuions will be denoted as $\widehat{\cC}^{(M,M'|i,i')}_{N,p}(\tau, \bar \tau) \big{\vert}_{\rm pole}$ and $\widehat{\cC}^{(M,M'|i,i')}_{N,p}(\tau, \bar \tau) \big{\vert}_{\rm disc}$, respectively. 
To summarise, in the large-charge expansion, the integrated correlators can be conveniently decomposed into the following different contributions 
\ie 
\widehat{\cC}^{(M,M'|i,i')}_{N,p}(\tau, \bar \tau) = \widehat{\cC}^{(M,M'|i,i')}_{N,p}\big{\vert}_{\rm const} \, + \, \widehat{\cC}^{(M,M'|i,i')}_{N,p}(\tau, \bar \tau) \big{\vert}_{\rm pole} \, 
+ \widehat{\cC}^{(M,M'|i,i')}_{N,p}(\tau, \bar \tau) \big{\vert}_{\rm disc} \, .
\fe
Below we will discuss each part separately; we will start with the discontinuity part and consider the remaining parts afterwards.

\begin{figure}[t!]
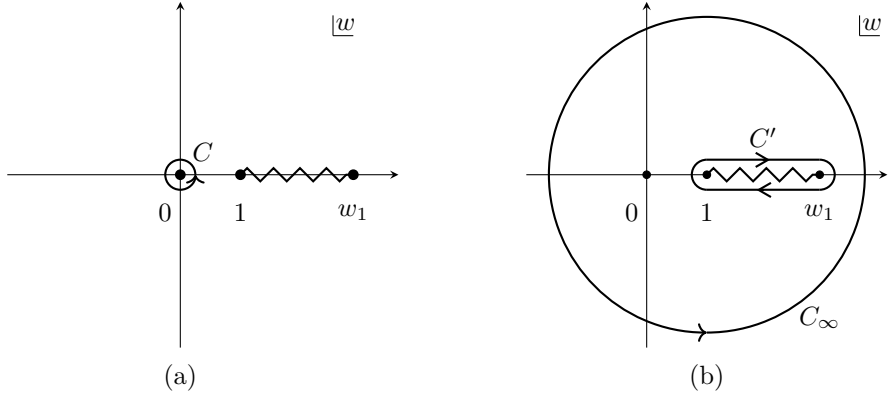

\begin{center}
\tikzpicture[scale=1.0]
\scope [very thick, every node/.style={sloped,allow upside down},xshift=-03.5cm,yshift=0cm]
%
%\draw [draw=black, ]        (0,0) circle (1 and 1);
%
%\draw (0.0,-0.0)  node($a$);
%\draw (0, -2.5) node{$(a)$};

\draw[thin] (-2.3,0) -- (2.3,0) ;
\draw[-stealth,thin] (0,-2.3) -- (0,2.3) ;]
\draw [fill=black] (-0.0,0.0) circle [radius=.05];
\draw[thick, -> ]     (.2,0) arc (0:360:0.2cm);% syntax (starting point coordinates) arc (starting angle:ending angle:radius)
\draw (2.2, 2.0) node{$w$};
\draw[thin] (2.03,1.85) -- (2.3,1.85) ;
\draw[thin] (2.03,1.85) -- (2.03,2.12) ;

%%%%%%%%%%%%%%%%%%%%
\draw [fill=black] (2.3,0.0) circle [radius=.05];
\draw [fill=black] (0.8,0.0) circle [radius=.05];
\draw[snake,thick] (0.8,0) -- (2.3,0) ;  
\draw (-0.2, -0.5) node{$0$};
\draw (0.30, 0.30) node{$C$};
\draw (0.8, -0.5) node{$1$};
\draw (2.3, -0.5) node{$w_1$};
\draw [-stealth,thin](2.3,0) -- (2.9,0.0) ;
%%%%%%%%%%%%%%%
\draw (0, -2.7) node{(a)};
\endscope 
%%%%%%%%%%%%%
%%%%%%% %%%%%%
\scope[xshift=3.5cm,yshift=0cm]
\draw[thick, ->] (0,-2.1) arc (-90:270:2.1cm);% syntax (starting point coordinates) arc (starting angle:ending angle:radius);
% \draw[thick,->](0.0,0.0) circle [radius=2.1cm];
\draw[ thick, ] (0.0,-0.2) arc (270:90:0.2cm);% syntax (starting point coordinates) arc (starting angle:ending angle:radius);
\draw[ thick ] (1.5,0.2) arc (90:-90:0.2cm);% syntax (starting point coordinates) arc (starting angle:ending angle:radius);
\draw [thin](-2.4,0) -- (0,0) ;
\draw [-stealth,thin](1.5,0) -- (2.4,0.0) ;
\draw[-stealth,thin] (-0.8,-2.3) -- (-0.8,2.3) ;]
\draw [fill=black] (1.5,0.0) circle [radius=.05];
\draw [fill=black] (-0.0,0.0) circle [radius=.05];
\draw[snake,thick] (0,0) -- (1.5,0) ;  
\draw[thick] (0,0.2) --   node {\midarrow}  (1.5,0.2)  ; 
\draw[thick] (1.5,-0.2)--   node {\backmidarrow}  (0,-0.2)   ; 
\draw (-1, -0.5) node{$0$};
\draw (0, -0.5) node{$1$};
\draw (1.5, -0.5) node{$ w_1$};
\draw (0.75, 0.5) node{$C'$};
\draw (1.5, -1.9) node{$C_\infty$};
\draw [fill=black] (-0.8,0.0) circle [radius=.05];
\draw (0, -2.7) node{(b)};
\draw (2.2, 2.0) node{$w$};
\draw[thin] (2.03,1.85) -- (2.3,1.85) ;
\draw[thin] (2.03,1.85) -- (2.03,2.12) ;
\endscope
\endtikzpicture
\end{center} 
\caption{We have used $w_1 = {(t+1)^2}/{(t-1)^2}$. (a) The contour $C$ encircling the pole at $w=0$. 
(b) The deformed contour $C'$ encircles the cut, together with the contour at infinity, $C_\infty$, which gives a vanishing contribution. 
}
\label{fig:contour}
\end{figure}

\subsection{Discontinuity part}
\label{sec:dispart}

As we will see, the most interesting part of the large-charge expansion of integrated correlators arises from the discontinuity part, $\widehat{\cC}^{(M,M'|i,i')}_{N,p}(\tau, \bar \tau) \big{\vert}_{\rm disc}$, which can be written as
\ie \label{eq:poledisc}
\widehat{\cC}^{(M,M'|i,i')}_{N,p}(\tau, \bar \tau) \big{\vert}_{\rm disc}:=&\, {1\over 2\pi i}\sum_{(m,n)\neq 0}\int_0^{\infty} e^{-t Y_{m,n}(\tau, \bar \tau) }\int_1^{w_1} {dw \over w^{p+1}} {\rm disc}\widehat{B}^{(M,M'|i,i')}_{N}(w; t) \, dt\, ,
\fe
where the notation $(m,n) \neq (0,0)$ means excluding the case $m=n=0$. 
The analysis is very similar to that of the large-$N$ expansion for the integrated correlators associated with $\langle \mathcal{O}_2\mathcal{O}_2\mathcal{O}_2\mathcal{O}_2 \rangle$  done in \cite{Dorigoni:2022cua}. It is convenient to separate the integration region into $(1, \infty)$ and $(w_1, \infty)$:
\ie
\widehat{\cC}^{(M,M'|i,i')}_{N,p}(\tau, \bar \tau) \big{\vert}_{\rm disc} = \widehat{\cC}^{(M,M'|i,i'), {\rm P}}_{N, p}(\tau, \bar \tau) + \widehat{\cC}^{(M,M'|i,i'), {\rm NP}}_{N, p}(\tau, \bar \tau)\, ,
\fe 
with
\ie \label{eq:integral-form}
\widehat{\cC}^{(M,M'|i,i'), {\rm P}}_{N, p}(\tau, \bar \tau) &= {1\over 2\pi i} \sum_{(m,n) \neq (0,0)} \int_0^\infty e^{-t\, Y_{m n}(\tau, \bar{\tau})}  \int^{\infty\pm i\, \delta_N\, \epsilon }_1  {dw \over w^{p+1}} {\rm disc} \widehat{B}^{(M,M'|i,i')}_N(w;t)\,dt \, ,  \cr
\widehat{\cC}^{(M,M'|i,i'), {\rm NP}}_{N, p}(\tau, \bar \tau) &= -{1\over 2\pi i} \sum_{(m,n) \neq (0,0)} \int_0^{\infty} e^{-t\, Y_{m n}(\tau, \bar{\tau})}  \int^{\infty \pm i\, \delta_N\, \epsilon}_{w_1}  {dw \over w^{p+1}} {\rm disc} \widehat{B}^{(M,M'|i,i')}_N(w;t) \,dt \, ,
\fe
where  $\delta_N=1$ for odd $N$ and $\delta_N=0$ for even $N$. The factor $\pm i \epsilon$ in the integrations given in \eqref{eq:integral-form} is to choose the integration contour slightly above or below the real axis to avoid the cut in the odd-$N$ case. Recall that the exponent $b$ of $\left(1-{w / w_1} \right)^{b}$ in the discontinuity ${\rm disc} \widehat{B}^{(M,M'|i,i')}_N(w;t)$ given in \eqref{eq:dis} and \eqref{eq:dis2} is an integer for even $N$ and a half-integer for odd $N$. Therefore in the odd $N$ case, ${\rm disc} \widehat{B}^{(M,M'|i,i')}_N(w;t)$ itself has branch cuts along $(w_1, \infty)$, but for the even-$N$ case it is a polynomial. For this reason we need to deform the integration slightly by $\pm i \epsilon$ for the cases with odd $N$.

For odd $N$, the factor $\pm i \epsilon$ in the integration has no effect in the case of $\widehat{\cC}^{(M,M'|i,i'), {\rm P}}_{N, p}(\tau, \bar \tau)$, and gives an overall $\pm i$ ``ambiguity" in the case of $\widehat{\cC}^{(M,M'|i,i'), {\rm NP}}_{N, p}(\tau, \bar \tau)$ \cite{Dorigoni:2022cua}. This   ``ambiguity" is in a complete agreement with the resurgent  analysis. More precisely,  as we will see, $\widehat{\cC}^{(M,M'|i,i'), {\rm P}}_{N, p}(\tau, \bar \tau)$ contains power series terms in the large-$p$ expansion, 
and $\widehat{\cC}^{(M,M'|i,i'), {\rm NP}}_{N, p}(\tau, \bar \tau)$ is exponentially decayed in the large-$p$ limit. Furthermore, $\widehat{\cC}^{(M,M'|i,i'), {\rm P}}_{N, p}(\tau, \bar \tau)$ is not Borel summable, and the $\pm i$ reflects the choices in resummation of the perturbative expansion $\widehat{\cC}^{(M,M'|i,i'), {\rm P}}_{N, p}(\tau, \bar \tau)$, and the completion of the asymptotic perturbative series is precisely $\widehat{\cC}^{(M,M'|i,i'), {\rm NP}}_{N, p}(\tau, \bar \tau)$. For even $N$, however, the function $\widehat{B}^{(M,M'|i,i')}_N(w;t)$ has no branch cuts along $(w_1, \infty)$, and so we can simply integrate along the real axis. Therefore, there is no such ambiguity. As we will see, this is also in agreement with resurgence. %\GB{Cut rest of paragraph:}This becomes especially clear in the generalised 't Hooft limit $\lambda=p\, g^2_{_{YM}}$. For odd $N$, we will see that the large-$\lambda$ expansion at each order in the $1/p$ expansion is asymptomatic and not Borel summable; for even $N$, the large-$\lambda$ expansion is in fact a polynomial (i.e. the power series terminates) at each order in $1/p$ expansion, therefore there is no ``ambiguity" in summing the series. 

To perform the computation explicitly, we make a suitable change of variables.  For the perturbative part, we use the substitution  $w=e^{\frac{\mu}{p}}$, and find
\ie
\widehat{\cC}^{(M,M'|i,i'), {\rm P}}_{N, p}(\tau, \bar \tau) = {1\over 2\pi i} \sum_{(m,n) \neq (0,0)} \int_0^\infty dt \, e^{-t\, Y_{m n}(\tau, \bar{\tau})}\frac{1}{p}\int_0^\infty & {\rm disc} \widehat{B}_N^{(M,M'|i,i')}(e^\frac{\mu}{p}; t)\, e^{-\mu} \, d\mu \, .
\fe
By expanding the integrand ${\rm disc} \widehat{B}_N^{(M,M'|i,i')}(e^\frac{\mu}{p}; t)$ in the large-$p$ limit, we can perform the $\mu$ integral, which leads to objects of the form $t^s+t^{1-s}$. Using the definition  of the non-holomorphic Eisensetein series, 
\ie \label{eq:EE}
E(s;  \tau, \bar \tau ) &= \sum_{(m,n)\neq (0,0)} \left(Y_{m,n}(\tau, \bar \tau)\right)^s = {1\over \Gamma(s)}\sum_{(m,n) \neq (0,0)} \int_0^\infty e^{-t\, Y_{m n}(\tau, \bar{\tau})} t^{s-1} \, dt \, ,
\fe
as well as the functional relation $\Gamma(s) E(s; \tau, \bar \tau) = \Gamma(1-s) E(1-s; \tau, \bar \tau)$, we find 
\ie \label{eq:SU3T1}
 \widehat{\cC}^{(M,M'|i,i'), {\rm P}}_{N, p}(\tau, \bar \tau) = \sum_{r=1}^{\infty} p^{-r-1/2} \sum_{m=1}^{r} a^{(M,M'|i,i')}_{N; r,m} E(m+1/2; \tau, \bar \tau) \,,
\fe
where the coefficients $a^{(M,M'|i,i')}_{N; r,m}$ are rational \footnote{To be precise, in some cases the large-$p$ expansion of $\widehat{\cC}^{(M,M'|i,i')}_{N, p}(\tau; \bar{\tau})$ may have some overall factors of $\pi, \sqrt{2}$, etc. }. 

Before we move on to the other contribution $\widehat{\cC}^{(M,M'|i,i'), {\rm NP}}_{N, p}(\tau, \bar \tau)$, let us pause here to comment on an important difference between even and odd $N$ in \eqref{eq:SU3T1} that we have already anticipated when we discussed the generating functions.  For odd $N$,  the coefficients in front of the Eisenstein series $a^{(M,M'|i,i')}_{N; r,m}$ are all non-zero; for even $N$, however, the coefficients vanish when $m$ is too large. More precisely, we have,  
\ie
a^{(M,M'|i,i')}_{N; r,m} =0\, , \qquad {\rm if} \qquad m>(N^2+M+M')/2-2 \,  .
\fe
This is because $(1+t)^{2b}(1-w/w_1)^b$ is a polynomial for even $N$, which truncates in the variable $t$ (and $w$). Therefore, according to the definition of $E(s;  \tau, \bar \tau )$ in \eqref{eq:EE}, the index $s$ cannot be too big if the expansion in $t$ truncates. %\GB{Cut:} When translating into generalised 't Hooft coupling $\lambda = p\, g_{_{YM}}^2$, using the expression \eqref{eq:Es} for the Eisenstein series, this implies that the large-$\lambda$ expansion of $\widehat{\cC}^{(M,M'|i,i')}_{N, p}(\tau; \bar{\tau})$ contains only a finite number of terms (up to the order $O(1/\lambda^{(N^2+M+M'-3)/2}$))  at any given order in the $1/p$ expansion. We will see all these interesting properties in explicit examples. 

For computing the non-perturbative part $\widehat{\cC}^{(M,M'|i,i'), {\rm NP}}_{N, p}(\tau, \bar \tau)$, we make the change of variables $w = w_1 e^{\frac{\mu}{p}}$ to give 
\ie \label{eq:NP-sad}
\widehat{\cC}^{(M,M'|i,i'), {\rm NP}}_{N, p}(\tau, \bar \tau) = 2\sum_{(m,n) \neq (0,0)} \int_1^\infty h_{N,p}^{(M,M'|i,i')}(t)\, e^{-S_p(t)} \, dt \, , 
\fe
where we have used the symmetry property \eqref{eq:prop} to change the integration region $t\in (0, \infty)$ to $t\in (1, \infty)$, which, as we will see, turns out to be convenient; furthermore $h_{N,p}^{(M,M'|i,i')}(t)$  and $S_p(t)$ are defined as follows: 
\ie \label{eq:h}
h_{N,p}^{(M,M'|i,i')}(t) &= -\frac{1}{p} \int_0^{\infty \pm i \, \delta_N \, \epsilon} {\rm disc} \widehat{B}^{(M,M'|i,i')}_N(w_1 e^{\frac{\mu}{p}}, t) \, e^{-\mu} \, \frac{d\mu}{2 \pi i} \, ,\fe
and
\ie \label{eq:action}
S_p(t) &= t \, Y_{m n}(\tau, \bar{\tau}) + 2\, p \log(\frac{t+1}{t-1}) \, .
\fe
Importantly, $h_{N,p}^{(M,M'|i,i')}(t)$ does not contain exponentially behaved terms in the large-$p$ limit, which is in the ``action" $S_p(t)$, more precisely in $e^{-S_p(t)}$. In the integration region, $S_p(t)>0$, therefore in the large-$p$ limit $\widehat{\cC}^{(M,M'|i,i'), {\rm NP}}_{N, p}(\tau, \bar \tau)$ is exponentially decayed. The saddle-point is at $\partial_t S(t)|_{t=t^*} =0$, with
the solution 
\ie
 t^*= \sqrt{ 1 + 4p/Y_{m,n} (\tau, \bar \tau)} \, . 
 \fe
 To obtain the large-$p$ expansion, we substitute $t=  t^* + \alpha$  back into \eqref{eq:NP-sad} and perform Gaussian integrals over $\alpha$. We find the result of this analysis takes the following form,  
 \ie \label{eq:non-pert}
 p^{v_1} \sum_{r=0}^{\infty} \sum_{m=0}^{2r} p^{-r/2} d^{(M,M'|i,i')}_{N; r,m} D_p \left(m{-}{3r\over 2} {+}\, v_2; \tau, \bar \tau \right)\, ,
\fe
where $D_p(s;  \tau, \bar \tau )$ is a generalisation of $E(s;  \tau, \bar \tau )$ with an exponentially decayed piece, 
\ie \label{eq:dD}
D_p(s;  \tau, \bar \tau ) = \sum_{(m,n)\neq (0,0)} \exp(-4 \sqrt{p\, Y_{m,n}(\tau, \bar \tau) }) \left(Y_{m,n}(\tau, \bar \tau)\right)^s \, . 
\fe
When the parameter $p=0$, it reduces to $E(s;  \tau, \bar \tau )$.  
As we explained earlier, there may be an overall $\pm i$ ambiguity in \eqref{eq:non-pert} depending on whether $N$ is odd or even. Furthermore, $v_1, v_2$ are some simple rational numbers (such as $1/4, 1/2$) depending on the parameters $(M, M'|i, i')$ and $N$, and the coefficients $d^{(M,M'|i,i')}_{N; r,m}$ are rational numbers (again, up to some overall factors of $\pi, \sqrt{2}$, etc., depending on particular cases). We note that exactly the same modular function $D_p(s; \tau, \bar \tau)$ appeared in the large-$N$ expansion of $\cC^{(0,0)}_{N,1}(\tau, \bar \tau)$ (again, this is the simplest integrated correlator with $M=M'=0, p=1$) with $N$ and $p$ exchanged \cite{Dorigoni:2022cua}.\footnote{Non-holomorphic modular invariant functions analogous to $D_N(s; \tau,\bar\tau)$ and the generalisations have recently appeared in another context \cite{Luo:2022tqy}}.  %In that case, there is a natural holographic interpretation in terms of string world-sheet instantons, when we translate gauge theory parameters into string theory ones using the AdS/CFT dictionary.  It would be of great interest to understand the holographic description of our result. 

\subsection{Remaining parts}
\label{sec:rem-part}

We will now consider the additional contributions to the integrated correlators that are beyond what has been considered in subsection \ref{sec:dispart}. They are relatively simpler to compute. The remaining parts can be separated into two parts: one is  independent of the charge $p$, which arises from the pole at $w=1$ of the integrated correlators, as given in \eqref{eq:poleint}; the other is independent of the coupling $\tau$, as explained in appendix \ref{app:lattice} (more precisely it is given in\eqref{eq:const}).   

\subsubsection{The $p$-independent term from $w=1$ pole}
\label{sec:p-indep}

We begin by considering the contribution from the pole at $w=1$, which is given by 
\ie \label{eq:poleint}
\widehat{\cC}^{(M,M'|i,i')}_{N,p}(\tau, \bar \tau) \big{\vert}_{\rm pole}:=&\, {1\over 2\pi i}\sum_{(m,n)\neq 0}\int_0^{\infty} e^{-t\, Y_{m,n}(\tau, \bar \tau) }\int_1^{w_1}  {\rm Res}_{w=1}\widehat{B}^{(M,M'|i,i')}_{N}(w; t) \, dt\, .
\fe
As we commented earlier, this contribution is associated with the source term $\cC^{(0,0)}_{N,1}(\tau, \bar \tau)$, which is only relevant when $M=M', i=i'$. Furthermore, this contribution is independent of $M, i$ and $p$. The residue at the $w=1$ pole,  ${\rm Res}_{w=1} \widehat{B}^{(M, M|i, i)}_{N}( w;t)$, is a rational function of $t$, which obeys
\ie \label{eq:st-pole}
t \, \partial^2_t \left(t\, {\rm Res}_{w=1} \widehat{B}^{(M, M|i, i)}_{N}( w;t) \right) = 4 B^{(0,0)}_{N,1}(t) \, .
\fe
One can in fact write down the generating function by summing over $N$.  Using the result \eqref{eq:Bztp2} and appropriate boundary conditions, we find,
\ie \label{eq:gen-pole}
\sum_{N=2}^{\infty} {\rm Res}_{w=1} \widehat{B}^{(M, M|i, i)}_{N}( w;t)\, z^N =  \frac{z}{2t} \left[\frac{t+1}{(1-z)^2}+\frac{(t-1)^4
   z-(t+1)^4}{(1-z)^{3/2} \left[(t+1)^2-(t-1)^2
   z\right]^{3/2}}\right] \, .
   \fe
According to \eqref{eq:poleint}, the contribution to the integrated correlators arising from the $w=1$ pole can be expressed as 
\ie
\widehat{\cC}^{(M, M|i, i)}_{N,p}(\tau, \bar \tau) \big{\vert}_{\rm pole} = - \sum_{(m,n) \neq (0,0) } \int^{\infty}_0 e^{- t Y_{m,n} (\tau, \bar \tau)}  {\rm Res}_{w=1} \widehat{B}^{(M, M|i, i)}_{N}( w;t)  \,  dt \, ,
\fe
where we have excluded $(m,n)=(0,0)$ here in the lattice sum, as it will be considered in the next subsection. Explicitly, ${\rm Res}_{w=1} \widehat{B}^{(M, M|i, i)}_{N}( w;t)$ is determined by its generating function as given in \eqref{eq:gen-pole}.  Furthermore, the relation  \eqref{eq:st-pole} implies that $\widehat{\cC}^{(M, M|i, i)}_{N,p} (\tau, \bar \tau)\big{\vert}_{\rm pole}$ satisfies the following Laplace equation 
\ie \label{eq:source1}
\Delta_{\tau} \widehat{\cC}^{(M, M|i, i)}_{N,p} (\tau, \bar \tau) \big{\vert}_{\rm pole} = -4 \sum_{(m,n) \neq (0,0) } \int^{\infty}_0 e^{- t Y_{m,n} (\tau, \bar \tau)}  {B}^{(0,0)}_{N, 1}( t) \,  dt \, ,
\fe
where the right-hand side is precisely $-4\, \cC_{N,1}^{(0, 0)}(\tau, \bar{\tau})$, with the $(m,n)=(0,0)$ excluded in the lattice sum. 

\subsubsection{The $\tau$-independent term}
\label{sec:C-const}

We now consider the $\tau$-independent part of the integrated correlators. As shown in \eqref{eq:const} in appendix \ref{app:lattice}, it consists of two contributions in the large-$p$ expansion, which we quote below
\ie \label{eq:const1}
\widehat{\cC}^{(M,M'|i,i')}_{N, p}\big {\vert}_{\rm const} &=
2\int_0^{\infty}   \widehat{B}^{(M, M'|i,i')}_{N,p}(t) \, dt + 2 \sum_{g=0}^{\infty} p^{-g-1}\, \underset{s\to 0}{\text{lim}} {\hat{c}^{(M,M'|i,i')}_{N;\, g;\, s} \over s }
 \, .
\fe
Since it is independent of $\tau$, the Laplace-difference equation \eqref{eq:gen-LDu} for $\widehat{\cC}^{(M,M'|i,i')}_{N, p}\big {\vert}_{\rm const}$ becomes
\ie \label{eq:recursionx}
(a+1)(\delta+1)\, \widehat{\cC}^{(M,M'|i,i')}_{N, p}\big {\vert}_{\rm const}  &=\, \left(p+1+ \delta  \right)\left(p+ a+1 \right) \, \widehat{\cC}^{(M,M'|i,i')}_{N, p+1}\big {\vert}_{\rm const} -2p(p+a+\delta+1) \, \widehat{\cC}^{(M,M'|i,i')}_{N, p}\big {\vert}_{\rm const} \cr
& + p\left(p+ a +\delta \right) \, \widehat{\cC}^{(M,M'|i,i')}_{N, p-1}\big {\vert}_{\rm const}  -4 \, \delta_{M,M'} \delta_{i, i'}\, \cC_{N,1}^{(0,0)}\big {\vert}_{\rm const}  \, ,
\fe
where $a$ is given in \eqref{eq:para} and $\cC_{N,1}^{(0,0)}\big {\vert}_{\rm const}={N(N-1)}/{4}$. 
Solving the recursion relation, we obtain 
\ie \label{eq:solCI}
\widehat{\cC}^{(M,M'|i,i')}_{N, p}\big {\vert}_{\rm const}&= \widehat{\cC}^{(M,M'|i,i')}_{N, 0}\big {\vert}_{\rm const} + \beta \frac{(a+1)
   (\delta+1)}{a\,\delta}\left(1-\frac{(a+p+1)_\delta}{\binom{a+\delta}{\delta} (p+1)_\delta} \right) \cr 
   &+ \delta_{M,M'} \delta_{i,i'} N(N-1) \left[ (a+1)H_{p+a}-H_p \right] \,,
\fe
where $H_{n}=\sum_{k=1}^n \frac{1}{k}$ are the harmonic numbers, with the analytic continuation $H_{n}=\gamma + \psi(n+1)$, where $\psi$ is the digamma function and $\gamma$ is the Euler-Mascheroni constant. And $\widehat{\cC}^{(M,M'|i,i')}_{N, 0}\big {\vert}_{\rm const}$ is the initial condition and $\beta$ is $p$-independent. When $M=M'$, the limit $\delta \rightarrow 0$ should be understood.

Using the results of $\widehat{B}^{(M, M'|i,i')}_{N,p}(t)$, we can evaluate the first contribution in \eqref{eq:const1} straightforwardly. We find
\ie \label{eq:intB}
2\int_0^{\infty}   \widehat{B}^{(M, M'|i,i')}_{N,p}(t) \, dt=2\int_0^{\infty}   \widehat{B}^{(M, M|i,i)}_{N,0}(t) \, dt + \delta_{M, M'} \, \delta_{i,i'} \, N(N-1) \left(H_{p+a}-H_{a}\right) \, .
\fe
 It is easy to see that \eqref{eq:intB} is a solution to the recursion relation \eqref{eq:recursionx}.
Interestingly,  for $M\neq M'$ or $i\neq i'$, $\int_0^{\infty}   \widehat{B}^{(M, M'|i,i')}_{N,p}(t) \, dt$ is independent of $p$, which is a rather non-trivial fact from the expressions of $\widehat{B}^{(M, M'|i,i')}_{N,p}(t)$. However, this is expected from the viewpoint of the recursion relation \eqref{eq:recursionx}.  Below we give some examples of the initial conditions, i.e. $\widehat{B}^{(M, M'|i,i')}_{N,0}(t)$.

When $M=M'=0$, we have $\widehat{B}^{(M, M'|i,i')}_{N,0}(t)=0$, and for $M=M'=3,4,5$ we find
\ie \label{eq:exampleC0}
\int_0^{\infty} \widehat{B}^{(3,3)}_{N,0}(t) &= \frac{N (2
   N+3)}{(N+1)
   (N+2)} \,  , \cr 
\int_0^{\infty} \widehat{B}^{(4,4)}_{N,0}(t) &= \frac{3 N^6+12 N^5+15
   N^4+17 N^3+16
   N^2-5 N-6}{N
   (N+1) (N+2) (N+3)
   \left(N^2+1\right)} \, , \cr
 \int_0^{\infty} \widehat{B}^{(5,5)}_{N,0}(t) &=  \frac{4 N^7+30 N^6+95
   N^5+245 N^4+475
   N^3+325 N^2-130
   N-120}{N (N+1)
   (N+2) (N+3) (N+4)
   \left(N^2+5\right)} \, .
\fe
As we commented earlier, when $M\neq M'$ or $i\neq i'$, $\int_0^{\infty}   \widehat{B}^{(M, M'|i,i')}_{N,p}(t) \, dt$ is independent of $p$. For example, for $(M,M')=(4,0)$ and $(M,M')=(5,3)$, we find
\ie
\int_0^{\infty}   \widehat{B}^{(4,0)}_{N,p}(t) \, dt = -\frac{4}{(N+3)(N+2)} \,  , \qquad \quad \int_0^{\infty}   \widehat{B}^{(5,3)}_{N,p}(t) \, dt  = - \frac{12}{(N+4)(N+3)} \, . 
\fe
Note that because there is no degeneracy, we have omitted the indices $i,i'$ in all these examples.

 We now consider the second term in \eqref{eq:const1}, which we find to be
\ie \label{eq:solCII}
2 \sum_{g=0}^{\infty} p^{-g-1}\, \underset{s\to 0}{\text{lim}} {\hat{c}^{(M,M'|i,i')}_{N;\, g;\, s} \over s } = \rho\,   \frac{   (p+1)_{\delta} -(a+p+1)_{\delta} }{ \delta\,  (p+1)_\delta} \, , 
\fe
where $\rho$ is independent of $p$. When $M=M', i=i'$, we find $\rho=N(N-1)/2$ and the above expression reduces to
\ie
2 \sum_{g=0}^{\infty} p^{-g-1}\, \underset{s\to 0}{\text{lim}} {\hat{c}^{(M,M|i,i)}_{N;\, g;\, s} \over s }= -{ N(N-1) \over 2} \left(H_{p+a} -H_p\right) \, .
\fe
When $M\neq M'$, for example for $(M,M')=(4,0)$ and $(M,M')=(5,3)$, we find
\ie
2 \sum_{g=0}^{\infty} p^{-g-1}\, \underset{s\to 0}{\text{lim}} {\hat{c}^{(4,0)}_{N;\, g;\, s} \over s } &= \frac{2 \left(N^2+1\right) \left(N^2+4 p+7\right)}{(N+2) (N+3) (p+1) (p+2)} \,,
\cr
2 \sum_{g=0}^{\infty} p^{-g-1}\, \underset{s\to 0}{\text{lim}} {\hat{c}^{(5,3)}_{N;\, g;\, s} \over s } &= \frac{12 \left(N^2+5\right)}{(N+3) (N+4) (p+1)} \,.
\fe

In summary, in the large-$p$ expansion, the $\tau$-independent part of the integrated correlators is given by \eqref{eq:const1}, with each contribution defined in \eqref{eq:solCI} and \eqref{eq:solCII}. When $M=M', i=i'$,  we find
\ie \label{eq:no-tau}
\widehat{\cC}^{(M,M|i,i)}_{N, p}\big {\vert}_{\rm const} &= 2\int_0^{\infty}   \widehat{B}^{(M, M|i,i)}_{N,0}(t) \, dt  + {N(N-1)} \left[ \log(p)+ \gamma - H_{a} + \frac{a+1}{2 p}-\frac{3a^2+3a+1}{12 p^2} \right. \cr
   &+ 
 \left.    \frac{a(a+1)(2a+1)}{12
   p^3}  -\frac{15 a^4+30 a^3+15 a^2-1}{120
   p^4} +O\left(p^{-5}\right) \right] \, ,
   \fe
and a few examples for the initial conditions $\int_0^{\infty}   \widehat{B}^{(M, M|i,i)}_{N,0}(t) \, dt$ may be found in \eqref{eq:exampleC0}. Whereas for the examples of $M\neq M'$, we have
\ie
\widehat{\cC}^{(4,0)}_{N, p}\big {\vert}_{\rm const} & =\frac{2 \left(N^4+4 N^2 (p+2)-4 p (p+2)-1\right)}{(N+2) (N+3) (p+1) (p+2)} \, , \qquad
\widehat{\cC}^{(5,3)}_{N, p}\big {\vert}_{\rm const}&=\frac{12 \left(N^2-2 p+3\right)}{(N+3) (N+4) (p+1)} \, .
\fe

\subsection{Summary}

Putting everything together, we find that the large-$p$ expansion of all the integrated correlators can be summarised  in the following form,\footnote{We have expanded the integrated correlators $\widehat{\cC}^{(M, M'|i, i')}_{N, p}(\tau; \bar{\tau})$ in $1/p$. As we commented earlier, the total charge of the operator $\mathcal{O}_{p|M}^{(i)}$ is  $2p+M$ rather than $p$. One could expand in $1/(2p+M)$ instead of $1/p$ as we do here. This of course can be obtained from our results by a simple change of variables. The general form presented in \eqref{eq:uni} will not change, although the precise coefficients in the expansion do change due to this shifting.} 
\ie \label{eq:uni}
& \widehat{\cC}^{(M, M'|i, i')}_{N, p}(\tau; \bar{\tau}) =\delta_{M, M'} \delta_{i,i'}\, \widehat{\cC}^{(M, M|i, i)}_{N, p}(\tau; \bar{\tau}) \big{\vert}_{\rm pole} +  \delta_{M, M'} \delta_{i,i'}\, b^{(M, M|i,i)}_N \log(p) + \sum_{k=0}^{\infty} b_{N; k}^{(M, M'|i, i')} \, p^{-k}  \cr 
&+p^{-1/2} \sum_{r=1}^{\infty} p^{-r} \sum_{m=1}^{r} a^{(M, M'|i, i')}_{N; r,m} E\left( m +  {1\over 2}; \tau, \bar \tau \right) \pm i\,  p^{n_1} \sum_{r=0}^{\infty} \sum_{m=0}^{2r} p^{-r/2} d^{(M, M'|i, i')}_{N; r,m} D_p \left(m{-}{3r\over 2} {+}\, n_2; \tau, \bar \tau \right)\, ,
\fe
where the first line is the result of subsection \ref{sec:rem-part} and the second comes from subsection \ref{sec:dispart}. The coefficients  (up to some overall factors of $\pi, \sqrt{2}$) in each contribution, as well as $n_1, n_2$, are rational numbers that may depend on the parameters $(M,M'|i,i')$ and $N$.

The results, especially the second line in the above equation, are remarkably similar to the large-$N$ expansion of the integrated correlator $\cC^{(0,0)}_{N,1}(\tau, \bar{\tau})$ that was recently obtained in \cite{Dorigoni:2022cua}. However, as we have emphasised, there are important differences. In particular, when $N$ is even, the number of non-holomorphic Eisenstein series that appear in the large-charge expansion do not grow indefinitely as we consider higher orders in the $1/p$ expansion. More precisely, 
\ie \label{eq:trancation}
a^{(M, M'|i, i')}_{N; r,m}=0\, ,  \qquad {\rm when} \qquad m>(N^2+M+M')/2-2 \, .
\fe
 This is very different from what has been seen in the large-$N$ expansion of  integrated correlators \cite{Chester:2019jas, Dorigoni:2021guq, Dorigoni:2022cua}. Furthermore, when considering the generalised 't Hooft genus expansion by introducing $\lambda = p\, g^2_{_{YM}}$,  the zero mode of the Eisenstein series give power series terms in the $1/\sqrt{\lambda}$ expansion, whereas the functions $D_p \left(s; \tau, \bar \tau \right)$ lead to exponentially decayed terms $e^{-2n \sqrt{\lambda}}$.  The property \eqref{eq:trancation} implies that in the 't Hooft expansion, at a given order of $1/p$, the power series in the $1/\sqrt{\lambda}$ expansion also truncates. Often the power series terms and exponentially decayed terms are related by resurgence \cite{Dorigoni:2021guq, Collier:2022emf, Hatsuda:2022enx, Dorigoni:2022cua}.  The truncation therefore has interesting consequences in understanding the connection between power series terms and exponentially decayed terms from the point of view of resurgence. Indeed this phenomenon, that certain physical observables have truncating perturbative expansions but still possess non-perturbative objects, has appeared in the literature of resurgence, under the name of ``Cheshire cat" resurgence \cite{Dunne:2016jsr, Kozcaz:2016wvy, Dorigoni:2017smz, Dorigoni:2019kux, Fujimori:2022qij, Dorigoni:2019yoq, Dorigoni:2020oon, Dorigoni:2022bcx}. The general idea is that one may introduce some deformation parameter such that the truncated perturbative series becomes an asymptotic series with such a deformation parameter. One can then perform more standard resurgence analysis, and turn off the deformation parameter at the end of the calculation to obtain the non-perturbative terms. In our case, presumably the rank of gauge group $N$ naturally plays the role of such a deformation parameter.

 Finally, we would like to stress that, because it is completed with the exponentially decayed terms,  the large-$p$ transseries \eqref{eq:uni} should be understood as the full result of $\widehat{\cC}^{(M, M'|i, i')}_{N, p}(\tau; \bar{\tau})$. The Borel resummation of the transseries provides a well-defined analytic continuation for all values of $p$, and should reproduce the starting-point expression in terms of generating functions. This is analogous to the large-$N$ expansion of $\cC^{(0,0)}_{N,1}(\tau, \bar{\tau})$ that was recently obtained in \cite{Dorigoni:2022cua}. In the next section, we will consider specific examples and compute explicitly the coefficients.

\section{Large-charge expansion: Examples}
\label{sec:large-p-ex}

\subsection{$SU(2)$ gauge group}
We begin by considering the simplest case, the integrated correlators with gauge group $SU(2)$. As we commented earlier, for this special case, the only independent operators are $(T_{2})^{p}$, or equivalently $\mathcal{O}_{p|0}$, which have dimension $2p$. So for this particular gauge group, the only relevant integrated correlators are $\widehat{\cC}^{(0,0)}_{2,p}( \tau, \bar \tau)$. The generating functions of the integrated correlators can be expressed as
\ie
\widehat{\cC}^{(0,0)}_{2}(w; \tau, \bar \tau) =
\sum_{(m,n) \in \mathbb{Z}^2 } \int^{\infty}_0  e^{- t Y_{m,n} (\tau, \bar \tau)}  \widehat{B}^{(0,0)}_2(w; t) \, dt \, , 
\fe
with $\widehat{B}^{(0,0)}_2(w; t)$  given in \eqref{eq:BtzN2}. The integrated correlator is then given by
\ie \label{eq:lattSU2}
\widehat{\mathcal{C}}^{(0,0)}_{2,p}( \tau, \bar \tau) =
\sum_{(m,n) \in \mathbb{Z}^2 } \int^{\infty}_0  e^{- t Y_{m,n} (\tau, \bar \tau)}  \widehat{B}^{(0,0)}_{2,p} (t) \, dt \, .
\fe
The generating functions with $SU(2)$ gauge group are special in that they do not have any branch cuts. From \eqref{eq:BtzN2}, we find
 \ie \label{eq:SU2B}
\widehat{B}^{(0,0)}_{2,p} (t)  ={1\over 2\pi i} \oint_C {dz\over w^{p+1}} \widehat{B}^{(0,0)}_{2}(w; t) =\frac{(t^2 -6t+1)}{ (t+1)^3} - \frac{ [t^2- 2 (4p+3)t+1]}{ (t+1)^3} \left(\frac{(t+1)^2}{(t-1)^2}\right)^{-p} \, .
\fe

We now consider the large-$p$ expansion of the integrated correlator following the general discussion in the previous section.
First of all,  the first term in \eqref{eq:SU2B} arises from the $\widehat{B}^{(0,0)}_{2}(w; t) \big{\vert}_{\rm pole}$, which is independent of $p$ and leads to
\ie
\widehat{\cC}^{(0,0)}_{2,p}(\tau, \bar \tau) \big{\vert}_{\rm pole} = \sum_{(m,n) \neq (0,0) } \int^{\infty}_0 e^{- t Y_{m,n} (\tau, \bar \tau)}  \frac{(t^2 -6t+1)}{ (t+1)^3} \,  dt \, .
\fe
 The $\tau$-independent part is given in \eqref{eq:no-tau} by setting $N=2$ and $M=M'=0$, 
\ie \label{eq:no-tauN2}
\widehat{\cC}^{(0,0)}_{2,p}\big{\vert}_{\rm const} &=   2\log(p)+ 2\gamma -2 H_{1/2} +   \frac{3}{2 p}-\frac{13}{24 p^2}+\frac{1}{4
   p^3} - \frac{119}{960
   p^4} +O\left(p^{-5}\right) \, .
   \fe
 We will now focus on the remaining part of the large-$p$ expansion of the integrated correlators, namely the second line of \eqref{eq:uni}, which is given by
\ie
\widehat{\cC}_{2, p}^{(0,0), {\rm NP}}( \tau, \bar \tau) = -\sum_{(m,n) \neq (0,0) } \int^{\infty}_0 e^{- t Y_{m,n} (\tau, \bar \tau)}  \frac{ [t^2- 2 (4p+3)t+1]}{(t+1)^3} \left(\frac{(t+1)^2}{(t-1)^2}\right)^{-p} dt \, .
 \fe
Firstly, we note that $\widehat{\cC}_{2, p}^{(0,0), {\rm NP}}( \tau, \bar \tau)$ obeys the following Laplace-difference equation, 
\ie \label{eq:lacSU2}
\left(\Delta_{\tau} + {3\over 2} \right) \widehat{\cC}_{2, p}^{(0,0), {\rm NP}}( \tau, \bar \tau) &= (p+1) \left(p+ \frac{3}{2} \right) \widehat{\cC}_{2, p+1}^{(0,0), {\rm NP}}( \tau, \bar \tau) \cr
&- 2p \left( p+ {3\over 2} \right) \widehat{\cC}_{2, p}^{(0,0), {\rm NP}}( \tau, \bar \tau) + p \left(p+ {1\over 2} \right) \widehat{\cC}_{2, p-1}^{(0,0), {\rm NP}}( \tau, \bar \tau)\, , 
\fe
where the ``source term", $-4\cC_{2, 1}^{(0,0)}( \tau, \bar \tau)$ in \eqref{eq:gen-LDu}, cancels out  because of the relation \eqref{eq:source1}. The superscript `NP' is to indicate that this term is only exponentially decayed in the large-$p$ limit; this is consistent with \eqref{eq:trancation}, as for this special case there are no $1/p$ power series terms with a sum of non-holomorphic Eisenstein series as the coefficients. We will study the large-$p$ behaviour of $\widehat{\cC}_{2, p}^{(0,0), {\rm NP}}( \tau, \bar \tau)$  by a saddle-point analysis as we outlined in the previous section. We write $\widehat{\cC}_{2, p}^{(0,0), {\rm NP}}( \tau, \bar \tau)$ as, 
\ie
\widehat{\cC}_{2, p}^{(0,0), {\rm NP}}( \tau, \bar \tau) = -2\sum_{(m,n) \neq (0,0) } \int^{\infty}_1 e^{-S_p(t)}  \frac{ [t^2- 2 (4p+3)t+1]}{ (t+1)^3} dt \, ,
 \fe
 with $S_p(t)$ given in \eqref{eq:action}. The result of  the saddle-point analysis can be written as, 
 \ie \label{eq:non-perSU2}
\widehat{\cC}_{2, p}^{(0,0), {\rm NP}}( \tau, \bar \tau) =\sqrt{2\pi}\, p^{1/4} \sum_{r=0}^{\infty} \sum_{m=0}^{2r} p^{-r/2} d^{(0,0)}_{2;r,m} D_p \left(m{-}{3r\over 2} {-} \frac{1}{4}; \tau, \bar \tau \right) \, ,
\fe
where the modular function $D_p(s; \tau, \bar \tau)$ is given in \eqref{eq:dD}, and some examples of the coefficients $d^{(0,0)}_{2; r,m}$ are given
\begin{align}
d^{(0,0)}_{2;0, 0} &=4\, , \quad d^{(0,0)}_{2;1, 0} =-{2 \over 3} \, , \quad d^{(0,0)}_{2;1, 1} =-6 \, , \quad d^{(0,0)}_{2;1, 2}=- {5 \over 2^3} \, , \\
d^{(0,0)}_{2;2, 0} &={1 \over 18} \, , \quad d^{(0,0)}_{2;2, 1} = 1 \, , \quad d^{(0,0)}_{2;2, 2} = {233 \over 48} \, , \quad d^{(0,0)}_{2;2, 3} = {27 \over 2^4} \, , \quad d^{(0,0)}_{2;2, 4} = {1 \over 2^{9} } \, .  \nn
\end{align} 
Not surprisingly, all the $d^{(0,0)}_{2; r, m}$ are determined by the Laplace-difference equation \eqref{eq:lacSU2}, except  $d^{(0,0)}_{2; r, 2r}$, which plays the role of the initial condition for the recursion relation. We find that $d^{(0,0)}_{2;r, 2r}$ is given by
\ie
d^{(0,0)}_{2;r, 2r} = (-1)^{r+1}\frac{  \left(4r^2-8r-1 \right) \Gamma
   \left(r-\frac{1}{2}\right)^2}{\pi \, 2^{3 r} \Gamma (r+1)} \, .
 \fe
 With this initial data, one may  determine all the coefficients $d^{(0,0)}_{2;r,m}$ using the Laplace-difference equation \eqref{eq:lacSU2}.  
 %We note the large-charge expansion of $\widehat{\cC}_{2, p}^{(0,0)}$ does not contain any power series in $1/p$ with a sum of non-holomorphic Eisenstein series as the coefficients. This is a special property for $SU(2)$ gauge group. As we will see shortly for more general gauge groups $SU(N)$ with $N>2$, the integrated correlators do contain such terms. 

Finally, it is instructive to consider the generalised 't Hooft limit by introducing $\lambda = p\, g^2_{_{YM}}$ in the region $1 \ll \lambda \ll p$.  We begin by considering  the small-$\lambda$ perturbative expansion,  for which we find  
\ie \label{eq:pertl}
 \widehat{\cC}_{2, p}^{(0,0)}( \lambda) \big{\vert}_{\rm pert} = \sum_{s=2}^{\infty} \frac{8 (-1)^s 
 \Gamma\left(s+\frac{1}{2}\right) \zeta (2 s-1) }{\pi^{2 s-3/2}\, \Gamma (s)} \lambda ^{s-1} + {1\over p} \sum_{s=3}^{\infty} \frac{6 (-1)^s 
   \zeta (2 s-1) \Gamma
   \left(s+\frac{1}{2}\right)}{
   \pi ^{2 s-3/2} \Gamma
   (s-1)}  \lambda ^{s-1} +O(p^{-2}) \, .
   \fe
It is easy to see that the perturbative expansion is convergent with a finite radius $|\lambda|<\pi^2$. This is identical to the small-$\lambda$ perturbative expansion in the large-$N$ expansion of integrated correlators with $p=2$ \cite{Dorigoni:2021guq}. However, we find the large-$\lambda$ expansion is trivial. This can be seen easily from the fact that the coefficients of $\lambda^{s-1}$ in \eqref{eq:pertl} have vanishing residues for negative half-integer $s$. This peculiar property of the large-$\lambda$ expansion is also related to the fact that the large-$p$ expansion of the integrated correlators with  finite $\tau$ only contains exponentially decayed terms, as we discussed earlier. The large-$\lambda$ expansion of the integrated correlators can be obtained from the zero mode of the non-holomorphic Eisenstein series, and for this particular case there is no $1/p$ power series with non-holomorphic Eisenstein series. 

In the 't Hooft limit, the zero-mode part (i.e. the zero-instanton sector) of $ \widehat{\cC}_{2, p}^{(0), {\rm NP}}( \tau, \bar \tau)$ leads to exponentially decayed terms in the large-$\lambda$ limit, 
\ie \label{eq:exp00}
 \widehat{\cC}_{2, p}^{(0,0), {\rm NP}}( \lambda)  &=\sqrt{\pi} \sum_{n=1}^{\infty} e^{-2n \sqrt{\lambda }} \left[ 8 n^{1\over 2} \lambda^{1\over 4}  -\frac{5 }{2\, n^{1\over 2} \lambda^{1\over 4}  }   + 
\frac{1 }{2^6\, n^{3 \over 2} \lambda ^{3 \over 4} } + \frac{33}{2^{10} \,  n^{5 \over 2} \lambda ^{5\over 4} } + \ldots \right] \cr 
&+ p^{-1} e^{-2n \sqrt{\lambda }} \left[ -6\lambda ^{3 \over 4} n^{3 \over 2}+\frac{27}{2^3} \lambda^{1\over 4}  n^{1\over 2} + \frac{117}{2^{8}  \lambda^{1\over 4}  n^{1\over 2} } -\frac{135}{2^{12} \lambda ^{3 \over 4} n^{3 \over 2}} + \ldots \right]  +O(p^{-2}) \, .
   \fe
This is obtained from the Fourier zero mode of the modular function $D_p(s; \tau, \bar \tau)$, which contains  exponentially decayed terms of the form $e^{-2m \sqrt{\lambda }}$. This arises from $D_p(s; \tau, \bar \tau)$ by setting $n=0$ in $D_p(s; \tau, \bar \tau)$ and using the definition of $Y_{m,n}(\tau, \bar \tau)$ as given by \eqref{eq:Ymn}.\footnote{See \cite{Dorigoni:2022cua} for more detailed analysis of the Fourier mode decomposition of $D_p(s; \tau, \bar \tau)$.} From the zero mode of $D_p(s; \tau, \bar \tau)$, one may also obtain terms that behave as $e^{-2m \sqrt{\tilde{\lambda}}}$ for $\tilde{\lambda}=4\pi p/\lambda$ in a different region where $1 \ll \tilde{\lambda} \ll p$. As emphasised in \cite{Dorigoni:2022cua}, all these types of exponentially decayed terms in terms of the 't Hooft coupling are contained in our $SL(2, \mathbb{Z})$-invariant expression, as given in \eqref{eq:non-perSU2}. 

We can also consider the finite $\lambda$ or $\tilde{\lambda}$ region, generalising the results of \cite{Hatsuda:2022enx, Dorigoni:2022cua}. In these regions, $ \widehat{\cC}_{2, p}^{(0,0), {\rm NP}}( \lambda)$ behaves as
\ie
\exp \left(-p\, A \left(n \pi/\sqrt{\lambda} \right) \right)
\qquad {\rm or} \qquad  
\exp \left(-p\, A \left(n \pi/\sqrt{\tilde{\lambda}} \right) \right) \, ,
\fe
where the function $A(x)$ is defined as $A(x)=4\left( x \sqrt{x^2+1} + \sinh ^{-1}(x)\right)$. 
All these behaviours are very similar to what have been found in large-$N$ expansion of $\cC_{N, 1}^{(0,0)}( \tau, \bar \tau)$ in the 't Hooft limit (for which $\lambda=N g^2_{_{YM}}$), so we will not consider all the regions in detail. 

In the large-$N$ expansion of $\cC_{N, 1}^{(0,0)}( \tau, \bar \tau)$, besides the exponentially decayed terms $e^{-2n\sqrt{\lambda}}$, at each order in $1/N$ expansion, $\cC_{N, 1}^{(0,0)}( \tau, \bar \tau)$ also contains power series terms $1/\sqrt{\lambda}$. In fact, the exponentially decayed terms may be considered as the non-perturbative completion for the power series terms through a resurgence procedure \cite{Dorigoni:2021guq,Collier:2022emf, Hatsuda:2022enx}, because of the fact that the power series terms are not Borel summable. In comparison, for the case we are considering, there are no power series terms at all, therefore applying the standard resurgence procedure to obtain the above exponentially decayed terms given in \eqref{eq:exp00} may not be straightforward.  As we will show later for higher-rank gauge groups $SU(N)$, there will be both power series terms and exponentially decayed terms. However when $N$ is even,  the power series terms are not an infinite series as we commented on earlier, and once again relating the power series terms and exponentially decayed terms through resurgence procedure is not as straightforward.  The resurgence analysis may be done through the so-called ``Cheshire cat" resurgence as we commented earlier. We will leave such analysis for future work.

\subsection{$M=M'=0$ with $SU(3)$ gauge group}

We now consider the large-charge expansion of the integrated correlator $\widehat{\cC}^{(0,0)}_{3,p}(\tau, \bar \tau)$.  As described in section \ref{sec:rem-part}, the $p$-independent and $\tau$-independent contributions are relatively simple, so we will focus on the more interesting part  from the branch cuts \eqref{eq:B30cuts}, 
\ie
\widehat{\cC}^{(0,0)}_{3, p}(\tau, \bar \tau) \big{\vert}_{\rm disc} =&\, {1\over 2\pi i} \sum_{(m,n) \neq (0,0)} \int_0^\infty e^{-t Y_{m n}(\tau, \bar{\tau})}  \int^{w_1}_1  {dw \over w^{p+1}} {\rm disc} \widehat{B}^{(0,0)}_3(w;t) \,dt\cr
=&\, \widehat{\cC}^{(0,0), {\rm P}}_{3, p}(\tau, \bar \tau) + \widehat{\cC}^{(0,0), {\rm NP}}_{3, p}(\tau, \bar \tau)\, ,
\fe
where the perturbative and non-perturbative terms are given by \eqref{eq:integral-form}, which we quote again below: 
\ie
\widehat{\cC}^{(0,0), {\rm P}}_{3, p}(\tau, \bar \tau) &= {1\over 2\pi i} \sum_{(m,n) \neq (0,0)} \int_0^\infty e^{-t\, Y_{m n}(\tau, \bar{\tau})}  \int^{\infty\pm i\epsilon}_1  {dw \over w^{p+1}} {\rm disc} \widehat{B}^{(0,0)}_3(w;t)\,dt \, ,  \cr
\widehat{\cC}^{(0,0), {\rm NP}}_{3, p}(\tau, \bar \tau) &= -{1\over 2\pi i} \sum_{(m,n) \neq (0,0)} \int_0^{\infty} e^{-t\, Y_{m n}(\tau, \bar{\tau})}  \int^{\infty\pm i \epsilon}_{w_1}  {dw \over w^{p+1}} {\rm disc} \widehat{B}^{(0,0)}_3(w;t) \,dt \, . 
\fe
Using the substitution $w=e^{\frac{\mu}{p}}$  for $\widehat{\cC}^{(0,0), {\rm P}}_{3, p}(\tau, \bar \tau)$, we find
\ie
 \frac{1}{p}\int_0^\infty & {\rm disc} \widehat{B}_3^{(0,0)}(e^\frac{\mu}{p}; t)\, e^{-\mu} \, \frac{d\mu}{2 \pi i} = \frac{27  \left(t^2+1\right)}{4 \sqrt{\pi }\, 
   t^{3/2} p^{3/2}}-\frac{27  \left(13 t^4+24 t^3+24
   t+13\right)}{32 \sqrt{\pi } \, t^{5/2} p^{5/2} } \cr
   & +\frac{135  \left(59
   t^6+832 t^5+857 t^4+857 t^2+832 t+59\right)}{2048 \sqrt{\pi } \, t^{7/2} p^{7/2}} + O(p^{-9/2}) \, .
\fe
From the definition \eqref{eq:EE}  of the non-holomorphic Eisensetein series, we find 
\ie \label{eq:SU3T2}
 \widehat{\cC}^{(0,0), {\rm P}}_{3, p}(\tau, \bar \tau) = \sum_{r=1}^{\infty} p^{-r-1/2} \sum_{m=1}^{r} a^{(0,0)}_{3; r,m} E(m+1/2; \tau, \bar \tau) \, , 
\fe
where the first few examples of the coefficients are given as 
\ie
a^{(0,0)}_{3;1,1} &=\frac{27}{4} \, , \qquad  a^{(0,0)}_{3;2,1}=-\frac{81}{4} \, ,  \qquad  a^{(0,0)}_{3;2,2}=-\frac{1053}{2^6} \, , \cr 
a^{(0,0)}_{3;3,1} &=\frac{115695}{2^{11}} \, , \quad  a^{(0,0)}_{3;3,2}=\frac{5265}{2^6} \, ,  \quad  a^{(0,0)}_{3;3,3}=\frac{119475}{2^{13}}\, . 
\fe
For the non-perturbative part $\widehat{\cC}^{(0,0), {\rm NP}}_{3, p}(\tau, \bar \tau)$, we make the change of variables $w = w_1 e^{\frac{\mu}{p}}$ to give, as in \eqref{eq:NP-sad},
\ie
\widehat{\cC}^{(0,0), {\rm NP}}_{3, p}(\tau, \bar \tau) = 2\sum_{(m,n) \neq (0,0)} \int_1^\infty h^{(0,0)}_{3,p}(t) e^{-S_p(t)} \, dt \, , 
\fe
with $S_p(t)$ defined in \eqref{eq:action}, and, as in \eqref{eq:h}, with
\ie
h^{(0,0)}_{3,p}(t)  = -\frac{1}{p} \int_0^{\infty \pm i \epsilon} {\rm disc} \widehat{B}^{(0,0)}_3(z_1 e^{\frac{\mu}{p}}, t) \, e^{-\mu} \, \frac{d\mu}{2 \pi i} \, ,
\fe
which is only polynomial in $p$. Performing the analysis described in the above leads to the final result, which is given by 
 \ie \label{eq:N3NP}
\widehat{\cC}^{(0,0), {\rm NP}}_{3, p}(\tau, \bar \tau) = \pm\, i \sum_{r=0}^{\infty} \sum_{m=0}^{2r} p^{-r/2} d^{(0,0)}_{3;r,m} D_p \left(m{-}{3r\over 2}; \tau, \bar \tau \right) \, ,
\fe
with $D_p(s;  \tau, \bar \tau )$ defined in \eqref{eq:dD}, and the coefficients $d^{(0,0)}_{3;r,m}$ (some examples) given by
\begin{align}
d^{(0,0)}_{3;0, 0} &=-24\, , \quad d^{(0,0)}_{3;1, 0} =4 \, , \quad d^{(0,0)}_{3;1, 1} =96 \, , \quad d^{(0,0)}_{3;1, 2} = -51 \, , \\
d^{(0,0)}_{3;2, 0} &=-{1 \over 3} \, , \quad d^{(0,0)}_{3;2, 1} = -16 \, , \quad d^{(0,0)}_{3;2, 2} = -{367 \over 2} \, , \quad d^{(0,0)}_{3;2, 3} = 204 \, , \quad d^{(0,0)}_{3;2, 4} = -{1839 \over 2^4 } \, . \nn
\end{align}
It is straightforward to verify that the results \eqref{eq:SU3T2} and \eqref{eq:N3NP} obey the Laplace-difference equation \eqref{eq:gen-LDu} with the ``source term" $-4 {\cC}^{(0,0)}_{3, 1}(\tau, \bar \tau)$ on the RHS removed. It is important to note the $\pm i$ ambiguity in the final expression, which is due to the branch cuts of the ${\rm disc} \widehat{B}_N^{(M,M'|i,i')}(w;t)$ along $(w_1\,  , \infty)$ for any odd $N$, as we explained earlier.

As before, from the $SL(2, \mathbb{Z})$-invariant results, we can extract the expression in the 't Hooft limit. In particular, from \eqref{eq:SU3T2}, we find  that the perturbative terms become
\ie \label{eq:SU3llam}
\widehat{\cC}^{(0,0), {\rm P}}_{3, p}(\lambda) =&\,
\left[ \frac{108 \zeta (3)}{\lambda ^{3/2}}-\frac{1053 \zeta (5)}{\lambda
   ^{5/2}}+\frac{119475 \zeta (7)}{2^5 \lambda ^{7/2}}+\frac{959175
   \zeta (9)}{2^6 \lambda ^{9/2}}+ O(\lambda ^{-11/2}) \right] \cr
   -&\, {1\over p} \left[\frac{324  \zeta (3)}{\lambda ^{3/2}}-\frac{5265 \zeta (5)}{\lambda
   ^{5/2}}+\frac{836325 \zeta (7)}{2^5 \lambda ^{7/2}}+\frac{8632575
   \zeta (9)}{2^6 \lambda ^{9/2}} + O(\lambda ^{-11/2}) \right] + O(p^{-2})\, ,
\fe
and the non-perturbative terms lead to 
\begin{align}
\widehat{\cC}^{(0,0), {\rm NP}}_{3, p}(\lambda) &=
\pm\, i \sum_{n=1}^{\infty}  e^{-2n \sqrt{\lambda }} \left[ \left(-48 -  {204 \over n \, \lambda^{1/2} } -{1839 \over 2 \, n^2 \, \lambda}-\frac{20211 }{2^3 \, n^3\, \lambda^{3/2}  } -{578817 \over 2^{7} \, n^4 \, \lambda^2}   + \ldots \right) \right. \\
&+ \left. {1  \over p}\left(96  n \lambda^{1/2} + 408+  {2043 \over n \, \lambda^{1/2} } +{27567 \over 2^2\, n^2 \, \lambda}+\frac{1063881}{2^6\, n^3\, \lambda^{3/2}  }   + \ldots \right) \right]+ O(p^{-2})\, ,
\end{align}
which can be expressed in terms of polylogarithm's, 
\begin{align}
\widehat{\cC}^{(0,0), {\rm NP}}_{3, p}(\lambda)  &= \pm\, i \left[ \left(-48\text{Li}_0\left(x\right)- {204 \, \text{Li}_1\left(x\right) \over \lambda^{1/2}  }- {1839 \, \text{Li}_2\left(x\right) \over 2\, \lambda  } - {20211 \, \text{Li}_3\left(x\right) \over 2^3 \, \lambda^{3/2}  } + \ldots \right) \right. \\
& \left. + {1\over p}\left(96 \lambda^{1/2} \, \text{Li}_{-1}\left(x\right) + {408 \, \text{Li}_0\left(x\right) } + {2043 \, \text{Li}_1\left(x\right) \over  \lambda^{1/2}  } + {27567 \, \text{Li}_2\left(x\right) \over 2^2 \, \lambda  } + \ldots \right) \right] + O(p^{-2})\, .  \nonumber
\end{align}
where the argument $x=e^{-2\sqrt{\lambda }}$. 
Importantly, unlike the even-$N$ cases, $\widehat{\cC}^{(0,0), {\rm P}}_{3, p}(\lambda)$ contains infinite numbers of terms at each order in $1/p$ expansion. This property can also be seen from small-$\lambda$ expansion, which is given by 
\ie \label{eq:pertSU3}
\widehat{\cC}^{(0,0)}_{3, p}(\lambda) \big{\vert}_{{\rm pert}}=&\sum_{s=2}^{\infty}\frac{48 (-1)^s 
   \left(s^2-s+6\right) \zeta
   (2 s-1) \Gamma \left(s+\frac{1}{2}\right)^2}{\pi ^{2 s-1} \Gamma
   (s) \Gamma (s+3)} \lambda ^{s-1} 
   \cr
   & +{1\over p} \sum_{s=3}^{\infty} \frac{96 (-1)^s 
   \left(s^2-s+6\right)  \zeta
   (2 s-1) \Gamma
   \left(s+\frac{1}{2}\right)^2}{ \pi ^{2s-1} \Gamma
   (s-1) \Gamma (s+3)} \lambda ^{s-1} + O(p^{-2})\, . 
   \fe
   We see the coefficients of $\lambda^{s-1}$ have non-trivial residues at $s=-1/2-n$ for all $n=0, 1,\ldots$. Using the contour integral representation described in appendix \ref{app:lattice}, one can precisely rederive \eqref{eq:SU3llam} from \eqref{eq:pertSU3} by computing the residues at $s=-1/2-n$. As we commented earlier, $\widehat{\cC}^{(0,0), {\rm NP}}_{3, p}(\lambda)$ can be understood as the non-perturbative completion of $\widehat{\cC}^{(0,0), {\rm P}}_{3, p}(\lambda)$ by the process of resurgence.  In particular, the large-$\lambda$ expansion of the integrated correlator at each order in $1/p$ expansion is asymptotic and not Borel summable,  and $\widehat{\cC}^{(0,0), {\rm NP}}_{3, p}(\lambda)$ gives the non-perturbative completion. This analysis was performed recently in \cite{Paul:2023rka}; we will not repeat the analysis here. Once again, in our analysis, all these results in the 't Hooft limit are simply the zero Fourier mode of the complete $SL(2, \mathbb{Z})$ invariant expression.

\subsection{$M=M'=0$ with $SU(4)$ gauge group}

The computation of the large-$p$ expansion for $N=4, M=M'=0$ is very similar to those we described in the previous subsections. We will only list the results in this section. Let us  begin with the power series terms. We once again find that they are expressed in terms of non-holomorphic Eisenstein series. However, as we already emphasised in the general expression \eqref{eq:uni}, due to \eqref{eq:trancation}, it turns out not all the Eisenstein series appear in the $1/p$ expansion of the integrated correlators. In fact, we find it terminates at $E(13/2;  \tau, \bar \tau)$ for the integrated correlators $ \widehat{\cC}^{(0,0), {\rm P}}_{4, p}(\tau, \bar \tau)$.  Explicitly, using the result of the discontinuity given in \eqref{eq:N4-disc}, we find that the power series terms in the large-$p$ expansion take the following form, 
\ie \label{eq:SU4T2}
 \widehat{\cC}^{(0,0), {\rm P}}_{4, p}(\tau, \bar \tau) = \pi \sum_{r=1}^{\infty} p^{-r-1/2} \sum_{m=1}^{r} a^{(0,0)}_{4; r,m} E(m+1/2; \tau, \bar \tau) \, . 
\fe
where the coefficients $a^{(0,0)}_{4;r,m}$ are rational numbers. For the first few examples, they are given by
\begin{align}
a^{(0,0)}_{4;1,1} &= \frac{4279275}{2^{18}}\, , \qquad a^{(0,0)}_{4;2,1} =- \frac{192567375}{2^{21}}\, , \qquad a^{(0,0)}_{4;2,2} =- \frac{364459095}{2^{21}} \, , \\ 
a^{(0,0)}_{4;3,1} &= \frac{17052910875}{2^{25}}\, , \qquad a^{(0,0)}_{4;3,2} = \frac{27334432125}{2^{24}}\, , \qquad a^{(0,0)}_{4; 3,3} = \frac{55307377125}{2^{25}}\, , \nonumber
%\cr
%a^{(0)}_{4;4,1} &=- \frac{779576923125}{2^{28}}\, , \,\,\, a^{(0)}_{4; 4,2} = -\frac{3201773149575}{2^{28}}\, , \,\,\, a^{(0)}_{4;4,3} =- \frac{5807274598125}{2^{28}}\, , \,\,\, a^{(0)}_{4; 4,4} =- \frac{1765488099375}{2^{27}} \, . \nonumber
\end{align}
and $a^{(0,0)}_{4; r,m}=0$ for all $m>6$, which implies that $E(s; \tau, \bar \tau)$ with $s>13/2$ do not appear in the expansion of $ \widehat{\cC}^{(0,0), {\rm P}}_{4, p}(\tau, \bar \tau)$.  As we explained before, this happens because the factor $(t+1)^{2b} \left(1-{w / w_1} \right)^{b}$ in ${\rm disc} \widehat{B}^{(M,M|i,i)}_N(w;t)$ given in \eqref{eq:dis} is a polynomial when $N$ is even, in which case $b$ is an integer. Therefore the degree of $t$ does not increase indefinitely in the large-$p$ expansion. 

We now move to the non-perturbative terms. For the same reason we discussed above (i.e. the factor $(t+1)^{2b} \left(1-{w / w_1} \right)^{b}$ is polynomial when $N$ is even), the ${\rm disc} \widehat{B}^{(0,0)}_{4}(w;t)$ does not have branch cuts on the $w$-plane, therefore there is no ambiguity in the choice of contour as in the $SU(3)$ case (or more general odd $N$ cases). We find the non-perturbative terms are given by
%\ie 
%C^{(0), {\rm NP}}_{4, p}(\tau, \bar \tau) =p^{-3/4} \sum_{r=0}^{\infty} \sum_{m=0}^{2r} p^{-r/2} d_{r,m} D_p \left({r\over 2} {+} \frac{3}{4}-m; \tau, \bar \tau \right) \, , 
%\fe
\ie \label{eq:NP-SU4}
\widehat{\cC}^{(0,0), {\rm NP}}_{4, p}(\tau, \bar \tau) ={i\over \sqrt{2}}  p^{-3/4} \sum_{r=0}^{\infty}  \sum_{m=0}^{2r} p^{-r/2} d^{(0,0)}_{4; r,m} D_p \left(m{-}{3r\over 2} {+} \frac{3}{4}; \tau, \bar \tau \right) \, , 
\fe
and below are a few examples of the rational coefficients $d^{(0,0)}_{4;r,m}$, 
\begin{align}
    d^{(0,0)}_{4;0,0}&={45045 \over 2^{7}} \, , \qquad 
d^{(0,0)}_{4;1,0}=-{15015 \over 2^{8}}\, , \qquad 
d^{(0,0)}_{4;1,1}= -{675675 \over 2^{8}}\, , \qquad 
d^{(0,0)}_{4;1,2}= {14909895 \over 2^{12}} \\
d^{(0,0)}_{4;2,0} &={5005 \over 2^{10}} \, , \quad 
d^{(0,0)}_{4;2,1}={225225 \over 2^{9}}\, , \quad 
d^{(0,0)}_{4;2,2}={75570495\over 2^{13}}\, , \quad 
d^{(0,0)}_{4;2,3}= -{3476347875\over 2^{16}}\, , \quad 
d^{(0,0)}_{4;2,4}= {6121840725 \over 2^{18}} \, . \nonumber
\end{align}

Once again, the complete finite-$\tau$ results allow us to obtain the  't Hooft limit of the integrated correlators expressed in terms of $\lambda$. We find 
\begin{align} \label{eq:SU4l}
 & \widehat{\cC}^{(0,0), {\rm P}}_{4, p}(\lambda) = {45045\pi \over 2^{14}}\left[ \frac{95 \zeta (3)}{\lambda ^{3/2}}-\frac{8091 \zeta (5)}{2 \lambda
   ^{5/2}}+\frac{1227825 \zeta (7)}{2^3 \lambda
   ^{7/2}}-\frac{39193875 \zeta (9)}{2^8 \lambda
   ^{9/2}}+\frac{29268894375 \zeta (11)}{2^8 \lambda
   ^{11/2}} \right. 
   \cr &-\, \left. \frac{738779042925 \zeta (13)}{2^9 \lambda ^{13/2}} \right] - {10135125 \pi \over 2^{17} p} \left[ \frac{19 \zeta (3)}{\lambda ^{3/2}}-\frac{2697 \zeta (5)}{2 \lambda
   ^{5/2}}+\frac{572985 \zeta (7)}{2^3 \lambda ^{7/2}}-\frac{23516325
   \zeta (9)}{2^3 \lambda ^{9/2}} \right. \cr
   & \left. +\, \frac{21463855875 \zeta (11)}{2^8
   \lambda ^{11/2}}  -\frac{640275170535 \zeta (13)}{2^9 \lambda
   ^{13/2}} \right]  + O(p^{-2})  %- {3378375  \over 2^{23} p^2} \left[ \frac{75715 \zeta (3)}{\lambda ^{3/2}}-\frac{14215887 \zeta (5)}{2
  % \lambda ^{5/2}}+\frac{1882255725 \zeta (7)}{2^2 \lambda
 %  ^{7/2}}\right. \cr &- \, \left. \frac{185059367955 \zeta (9)}{2^3 \lambda
 %  ^{9/2}}  + \frac{196716239094375 \zeta (11)}{2^8 \lambda
 %  ^{11/2}}-\frac{6694076907943425 \zeta (13)}{2^9 \lambda ^{13/2}} \right]\, .
 \end{align}
It is notable that at each order in $1/p$ expansion the large-$\lambda$ expansion terminates at $O(\lambda^{-13/2})$. This is related to the fact that in \eqref{eq:SU4T2}, $a^{(0,0)}_{4; r,m}=0$ for $m>6$, namely Eisenstein series with indices larger than $13/2$ do not appear. This interesting property can also be seen from the small-$\lambda$ expansion  
\begin{align}
\widehat{\cC}^{(0,0)}_{4, p}(\lambda)\big{\vert}_{\rm pert} &=  \sum_{s=2}^{\infty}\frac{90090\, (-1)^s 
   \left(s^2-s+4\right)
   \left(s^2-s+18\right) 
   \zeta (2 s-1) \Gamma
   \left(s+\frac{1}{2}\right)}{\pi ^{2 s-3/2}(2 s+1) (2
   s+3) (2 s+5) (2 s+7) (2 s+9) (2 s+11)
   \Gamma (s)} \lambda ^{s-1} \\
 &  + {1\over p} \sum_{s=3}^{\infty} \frac{675675\, (-1)^s 
   (s^2-s+4) (s^2-s+18) 
   \zeta (2 s-1) \Gamma
   \left(s+\frac{3}{2}\right)}{\pi ^{2 s-3/2} (2 s+1)^2
   (2 s+3) (2 s+5) (2 s+7) (2 s+9) (2 s+11)
   \Gamma (s-1)} \lambda ^{s-1} + O(p^{-2}) \, . \nonumber
\end{align}
We see that the coefficients of $\lambda ^{s-1}$ at each order in $1/p$ have non-trivial residues at $s=-1/2, \ldots, -11/2$, and these residues precisely lead to \eqref{eq:SU4l}. 

Even though there are only finite number of power series terms, our results show that at each order in the $1/p$ expansion, there are associated exponentially decayed terms. They can be obtained from \eqref{eq:NP-SU4} and are given by 
\begin{align}
 \widehat{\cC}^{(0,0), {\rm NP}}_{4, p}(\lambda) &=  i \,\sum_{n=1}^{\infty} e^{-2n \sqrt{\lambda}} \left[ {45045 \over 2^5 \, n^{3/2}\, \lambda ^{3/4}  } \left( 1+\frac{331}{2^4 \, n \, \lambda^{1/2}}+\frac{135905}{2^9
   \, n^2\, \lambda}+\frac{20197489}{2^{13}\, n^3 \, \lambda^{3/2}}+\ldots \right) \right. \cr
  & \left. -\, { 675675 \over 2^{7}\,n^{1/2}\, \lambda^{1/4}\, p}  \left( 1+\frac{343}{2^4 \, n \, \lambda^{1/2} }+\frac{149145}{2^9
   \, n^2\, \lambda }+\frac{24002829 }{2^{13}\, n^3 \, \lambda^{3/2}} +\ldots \right) + O(p^{-2}) \right] \, . 
   %\\
%  & +{ 225225 \,n^{3/2}\, \lambda^{3/4} \over 2^{13}\, p^2} e^{-2n \sqrt{\lambda}}\left[-1+\frac{5033}{2^4 \, n \, \lambda^{1/2}}+\frac{3807847}{2^9
%   \, n^2\, \lambda }+\frac{899010803}{2^{13}\, n^3 \, \lambda^{3/2}}+\ldots \right] + O(p^{-3})\, . \nonumber
 \end{align}
Unlike the $SU(3)$ case, there is no ``ambiguity" of choosing the contours for  $\widehat{\cC}^{(0,0), {\rm NP}}_{4, p}(\lambda)$. From a resurgence point of view, this is related to the fact the perturbative series $\widehat{\cC}^{(0,0), {\rm P}}_{4, p}(\lambda)$ as shown in \eqref{eq:SU4l} has only a finite number of terms at each order in $1/p$ expansion, and so there are no issues of resummation of the perturbave series.

\subsection{Operators in other identical towers}

We now consider integrated correlators involving  operators in other identical towers, i.e. $M=M' > 0$. The computation is very similar. We begin with  the operators with $M=M'=3$, for which the discontinuity is given in \eqref{eq:dsB333}. Because this tower of operators involve $T_3$, we must have $N \geq 3$ for the gauge group $SU(N)$. Let us take the simplest $N=3$ case as an example; the computation and the results for higher $N$ are very similar. The perturbative  terms in this case are given by
\ie
 \widehat{\cC}^{(3,3), {\rm P}}_{3, p}(\tau, \bar \tau) = \sum_{r=1}^{\infty} p^{-r-1/2} \sum_{m=1}^{r} a^{(3,3)}_{3;r,m} E(m+1/2; \tau, \bar \tau) \, , 
\fe
where a few examples of $a^{(3,3)}_{3;r,m}$ are given by
\ie
a^{(3,3)}_{3;1,1} &=27\,, \qquad 
a^{(3,3)}_{3;2,1} =- {567 \over 2^2}\,, \qquad 
a^{(3,3)}_{3;2,2} =- {5427 \over 2^4} \, , \cr
a^{(3,3)}_{3;3,1} &= {373275 \over 2^9}\,, \qquad 
a^{(3,3)}_{3;3,2} ={189945 \over 2^6}\,, \qquad 
a^{(3,3)}_{3;3,3} = {6461775 \over 2^{11}}\, .
\fe
The non-perturbative terms are
\ie
 \widehat{\cC}^{(3,3), {\rm NP}}_{3, p}(\tau, \bar \tau)  = \pm i \sum_{r=0}^{\infty} \sum_{m=0}^{2r} p^{-(r+1)/2} d^{(3,3)}_{3;r,m} D_p \left(m{-}{3r\over 2} + \frac{1}{2}; \tau, \bar \tau \right) \, ,
\fe
and the some examples of the coefficients $d^{(3,3)}_{3;r,m}$ are given
\begin{align}
d^{(3,3)}_{3; 0, 0} &=-216\, , \quad d^{(3,3)}_{3;1, 0} =36 \, , \quad d^{(3,3)}_{3;1, 1} =1512 \, , \quad d^{(3,3)}_{3;1, 2} =-1917 \, , \\
d^{(3,3)}_{3;2, 0} &=-3 \, , \quad d^{(3,3)}_{3;2, 1} = -252 \, , \quad d^{(3,3)}_{3;2, 2} = -{9891 \over 2} \, , \quad d^{(3,3)}_{3;2, 3} = 13797 \, , \quad d^{(3,3)}_{3;2, 4} = -{180207 \over 2^4 } \,  .  \nn
\end{align}
In the 't Hooft limit and expressed in terms of $\lambda = p\, g^2_{_{YM}}$, we find
\begin{align}
 \widehat{\cC}^{(3,3), {\rm P}}_{3, p}(\lambda) &=27 \left[ \frac{16 \zeta (3)}{\lambda ^{3/2}}-\frac{804
   \zeta (5)}{\lambda ^{5/2}}+\frac{239325 \zeta
   (7)}{2^3 \lambda ^{7/2}}-\frac{13260625 \zeta
   (9)}{2^4 \lambda ^{9/2}} + O(\lambda ^{-11/2}) \right] \cr
 &-\frac{567}{p}\left[\frac{4 \zeta (3)}{\lambda ^{3/2}}-\frac{335
   \zeta (5)}{\lambda ^{5/2}}+\frac{558425 \zeta
   (7)}{2^5 \lambda ^{7/2}}-\frac{39781875 \zeta
   (9)}{2^6 \lambda ^{9/2}} + O(\lambda ^{-11/2})\right]+ O(p^{-2}) \, ,
 \end{align}
 and from the non-perturbative term, we have, 
\ie \label{eq:exp33}
 \widehat{\cC}^{(3,3), {\rm NP}}_{3, p}(\lambda) &=\pm i\,\sum_{n=1}^{\infty} e^{-2n \sqrt{\lambda }} \left[ -9\left( \frac{96}{n \lambda^{1/2}}   + 
\frac{1704}{n^2 \lambda} + \frac{20023}{n^{3} \lambda ^{3/2} }+ \frac{685361}{2^2 n^{4} \lambda ^{2} } + \ldots \right) \right. \cr 
& \left. + \, {36 \over p}  \left(84 + \frac{1533}{n \lambda^{1/ 2}} +\frac{152089}{2^3 n^2 \lambda ^{} } + \frac{5638493 }{2^5 n^3 \lambda ^{3/2}} + \ldots \right) \right]  +O(p^{-2})\, .
   \fe
We have also analysed the integrated correlators with higher ranks of gauge group, once again the results are all given in the general structure \eqref{eq:uni}.

We will move on to consider the integrated correlators $ \widehat{\cC}^{(4,4), {\rm P}}_{N, p}(\tau, \bar \tau)$. For this case, we require $N\geq 4$, and here we will consider the case with $SU(4)$ gauge group. The perturbative terms are given by
\ie \label{eq:per44}
 \widehat{\cC}^{(4,4), {\rm P}}_{4, p}(\tau, \bar \tau) = \pi \sum_{r=1}^{\infty} p^{-r-1/2} \sum_{m=1}^{r} a^{(4,4)}_{4;r,m}\, E(m+1/2; \tau, \bar \tau) \, , 
\fe
with 
\ie
a^{(4,4)}_{4;1,1} &= \frac{726424904205}{2^{34}}\, , \quad a^{(4,4)}_{4;2,1}=-\frac{50123318390145 
}{2^{37}}\, , \quad a^{(4,4)}_{4;2,2}=-\frac{168450013576845 
}{2^{37}} \cr 
a^{(4,4)}_{4;3,1} &= \frac{6948254208720825}{2^{41}} \, , \quad a^{(4,4)}_{4;3,2}=\frac{19371751561337175  
}{2^{40}}\, , \quad a^{(4,4)}_{4;3,3}=\frac{35448187717317375 
}{2^{40}} \, , 
\fe
and the non-perturbative terms take the following form, 
\ie
\widehat{\cC}^{(4,4), {\rm NP}}_{4, p}(\tau, \bar \tau) =i \sqrt{\pi\over 2}\, p^{-3/4} \sum_{r=0}^{\infty}  \sum_{m=0}^{2r} p^{-r/2} d^{(4,4)}_{4; r,m} D_p \left(m{-}{3r\over 2} {+} \frac{3}{4}; \tau, \bar \tau \right) \, , 
\fe
which has no $\pm i$ ambiguity and coefficients $d^{(4,4)}_{4; r,m}$ are given by
\begin{align}
d^{(4,4)}_{4;0, 0} &=\frac{21396375}{2^{15}}\, , \quad d^{(4,4)}_{4;1, 0} =-\frac{7132125}{2^{16}} \, , \quad d^{(4,4)}_{4;1, 1} =-\frac{492116625}{2^{16}} \, , \quad d^{(4,4)}_{4; 1, 2} =\frac{18721828125}{2^{20}} \, , \\
d^{(4,4)}_{4;2, 0} &=\frac{2377375}{2^{18}} \, , \qquad d^{(4,4)}_{4;2, 1} = \frac{164038875}{2^{17}} \, , \qquad d^{(4,4)}_{4;2, 2} = {84052093125 \over 2^{21}} \, , \cr d^{(4,4)}_{4;2, 3} &= -\frac{436507446375}{2^{21}} \, , \qquad d^{(4,4)}_{4;2, 4} = {21765002874615 \over 2^{26} } \, .  \nn
\end{align}

Importantly, we also find that the coefficients of the perturbative terms obey the condition that all $a^{(4,4)}_{4; r,m}=0$ for $m>10$. Once again, this implies that the Eisenstein series  $E(s; \tau, \bar \tau)$ with $s>21/2$ do not appear in the expansion. In terms of 't Hooft coupling $\lambda$, this means that there are only a finite number of terms in the large-$\lambda$ expansion for a given order in $1/p$. 
In the 't Hooft limit, we find
\begin{align}
 & \widehat{\cC}^{(4,4), {\rm P}}_{4, p}(\lambda) = {855855 \pi \over 2^{30}}\left[  \frac{848771 \zeta (3)}{\lambda
   ^{3/2}}-\frac{196820739 \zeta (5)}{2 \lambda
   ^{5/2}}+\frac{41418450225 \zeta (7)}{2^2 \lambda
   ^{7/2}} +\dots \right] \cr  
&-\frac{59053995 \pi }{2^{33}\,p}\left[\frac{848771 \zeta (3)}{\lambda
   ^{3/2}}-\frac{328034565 \zeta (5)}{2 \lambda
   ^{5/2}}+\frac{96643050525 \zeta (7)}{2^2 \lambda
   ^{7/2}} +\dots \right]+ O(p^{-2}) \, ,
 \end{align}
where for each order in $1/p$, the power series terminates at order $O(\lambda^{-21/2})$. The non-perturbative terms are given by
\ie 
\widehat{\cC}_{4, p}^{(4,4), {\rm NP}}( \lambda)  =& \, i \sqrt{\pi} \sum_{n=1}^{\infty} e^{-2n \sqrt{\lambda }} \left[\frac{21396375}{2^{13} n^{3/2} \lambda^{3/4}} \left( 1+ \frac{875}{16 n \lambda^{1/2}}   + 
\frac{25430713}{12800 n^2 \lambda} +  \ldots \right) \right. \cr 
&\left. - {1\over p} \frac{492116625}{2^{15} n^{1/2} \lambda^{1/4}} \left( 1 + \frac{887}{16 n \lambda^{1\over 2}} +\frac{26305713}{12800 n^2 \lambda ^{} } + \ldots \right) +O(p^{-2}) \right]  \, .
   \fe

\subsection{Operators in different towers}

We now consider integrated correlators involving operators in different towers, i.e. $M \neq M'$. The analysis is very similar. We begin with the case $M=4, M'=0$, with the simplest case of $SU(4)$. Using the result of \eqref{eq:dB404} for the discontinuity, we find that the perturbative terms are given by,
\ie
 \widehat{\cC}^{(4,0), {\rm P}}_{4, p}(\tau, \bar \tau) = \pi \sum_{r=1}^{\infty} p^{-r-1/2} \sum_{m=1}^{r} a^{(4,0)}_{4;r,m} E(m+1/2; \tau, \bar \tau) \, , 
\fe
with
\ie
a^{(4,0)}_{4;1,1} &= \frac{15205905}{2^{23}}\, , \quad a^{(4,0)}_{4;2,1}=-\frac{15205905 
}{2^{26}}\, , \quad a^{(4,0)}_{4;2,2}=-\frac{5524994475 
}{2^{26}} \cr 
a^{(4,0)}_{4;3,1} &= -\frac{12240753525}{2^{30}} \, , \quad a^{(4,0)}_{4;3,2}=\frac{259674740325  
}{2^{29}}\, , \quad a^{(4,0)}_{4;3,3}=\frac{4788998589375 
}{2^{31}} \, , 
\fe
and $a^{(4,0)}_{4;r,m} = 0$ for all $m>8$, and so that $ E(s; \tau, \bar \tau)$ with $s> \frac{17}{2}$ does not appear in the expansion of $\widehat{\cC}^{(4,0), {\rm P}}_{4, p}(\tau, \bar \tau)$. The non-perturbative terms take the following form,  
\ie
\widehat{\cC}^{(4,0), {\rm NP}}_{4, p}(\tau, \bar \tau) =i \sqrt{\pi\over 2}\, p^{-3/4} \sum_{r=0}^{\infty}  \sum_{m=0}^{2r} p^{-r/2} d^{(4,0)}_{4; r,m} D_p \left(m{-}{3r\over 2} {+} \frac{3}{4}; \tau, \bar \tau \right) \, , 
\fe
with  
\begin{align}
d^{(4,0)}_{4;0, 0} &=-\frac{182325}{2^{10}}\, , \quad d^{(4,0)}_{4;1, 0} =\frac{60775}{2^{11}} \, , \quad d^{(4,0)}_{4;1, 1} =\frac{4193475}{2^{11}} \, , \quad d^{(4,0)}_{4; 1, 2} =-\frac{109941975}{2^{15}} \, , \\
d^{(4,0)}_{4;2, 0} &=-\frac{60775}{2^{13}\,3} \, , \qquad d^{(4,0)}_{4;2, 1} =-\frac{1397825}{2^{12}} \, , \qquad d^{(4,0)}_{4;2, 2} = -{732764175 \over 2^{16}} \, , \cr d^{(4,0)}_{4;2, 3} &= \frac{2479802325}{2^{16}} \, , \qquad d^{(4,0)}_{4;2, 4} = -{77604703605 \over 2^{21} } \, .  \nn
\end{align}

In the 't Hooft limit, we find
\begin{align}
 & \widehat{\cC}^{(4,0), {\rm P}}_{4, p}(\lambda) = {109395 \pi \over 2^{27}}\left[  \frac{35584 \zeta (3)}{\lambda
   ^{3/2}}-\frac{6464640 \zeta (5)}{\lambda
   ^{5/2}}+\frac{700434000 \zeta (7)}{\lambda
   ^{7/2}} +\dots \right] \cr  
&-\frac{109395 \pi }{2^{26}\,p}\left[\frac{2224 \zeta (3)}{\lambda
   ^{3/2}}-\frac{18989880 \zeta (5)}{\lambda
   ^{5/2}}+\frac{4071272625 \zeta (7)}{\lambda
   ^{7/2}} +\dots \right]+ O(p^{-2}) \, ,
 \end{align}
where for each order in $1/p$, the power series terminates at order $O(\lambda^{-17/2})$. The non-perturbative terms are given by
\ie 
\widehat{\cC}_{4, p}^{(4,0), {\rm NP}}( \lambda)  =& \, i \sqrt{\pi} \sum_{n=1}^{\infty} e^{-2n \sqrt{\lambda }} \left[\frac{182325}{128 n^{3/2} \lambda^{3/4}} \left( 1+ \frac{603}{16 n \lambda^{1/2}}   + 
\frac{2128197}{2560 n^2 \lambda} +  \ldots \right) \right. \cr 
&\left. - {1\over p} \frac{4193475}{512 n^{1/2} \lambda^{1/4}} \left( 1 + \frac{13601}{368 n \lambda^{1\over 2}} +\frac{48442011}{58880 n^2 \lambda ^{} } + \ldots \right) +O(p^{-2}) \right]  \, .
\fe

Finally, we will study the case $M=5, M'=3$, with the gauge group $SU(5)$. With the expression for the discontinuity \eqref{eq:dB535}, we find that the perturbative terms in the large-$p$ expansion are given by
\ie
 \widehat{\cC}^{(5,3), {\rm P}}_{5, p}(\tau, \bar \tau) = \sum_{r=1}^{\infty} p^{-r-1/2} \sum_{m=1}^{r} a^{(5,3)}_{5;r,m} E(m+1/2; \tau, \bar \tau) \, , 
\fe
with
\ie
a^{(5,3)}_{5;1,1} &= \frac{133888}{22287}\, , \quad a^{(5,3)}_{5;2,1}=-\frac{234304 
}{7429}\, , \quad a^{(5,3)}_{5;2,2}=\frac{4047408 
}{7429} \cr 
a^{(5,3)}_{5;3,1} &= \frac{10054675}{44574} \, , \quad a^{(5,3)}_{5;3,2}=-\frac{55651860}{7429}\, , \quad a^{(5,3)}_{5;3,3}=-\frac{5811253875 
}{59432} \, , 
\fe
and the non-perturbative terms take the following form, 
\ie
\widehat{\cC}^{(5,3), {\rm NP}}_{5, p}(\tau, \bar \tau) = \pm i \, p^{-5/2} \sum_{r=0}^{\infty}  \sum_{m=0}^{2r} p^{-r/2} d^{(5,3)}_{5; r,m} D_p \left(m {-} {3r\over 2} {+} \frac{5}{2}; \tau, \bar \tau \right) \, , 
\fe
with   
\begin{align}
d^{(5,3)}_{3;0, 0} &=-2702700\, , \quad d^{(5,3)}_{5;1, 0} =450450 \, , \quad d^{(5,3)}_{5;1, 1} =45945900 \, , \quad d^{(5,3)}_{5; 1, 2} =-\frac{291215925}{2} \, , \\
d^{(5,3)}_{5;2, 0} &=-\frac{75075}{2} \, , \qquad d^{(5,3)}_{5;2, 1} =-7657650 \, , \qquad d^{(5,3)}_{5;2, 2} = -{1458331875 \over 4} \, , \cr d^{(5,3)}_{5;2, 3} &= \frac{5024994975}{2} \, , \qquad d^{(5,3)}_{5;2, 4} = -{134111352375 \over 32 } \, .  \nn
\end{align}

In the 't Hooft limit in terms of $\lambda$, we find the integrated correlators can be expressed as
\begin{align}
 & \widehat{\cC}^{(5,3), {\rm P}}_{5, p}(\lambda) = {32 \over 22287}\left[  \frac{66944 \zeta (3)}{\lambda
   ^{3/2}}+\frac{24284448 \zeta (5)}{\lambda
   ^{5/2}}-\frac{17433761625 \zeta (7)}{\lambda
   ^{7/2}} +\dots \right] \cr  
&-\frac{8}{7429\,p}\left[\frac{468608 \zeta (3)}{\lambda
   ^{3/2}}+\frac{445214880 \zeta (5)}{\lambda
   ^{5/2}}-\frac{517201594875 \zeta (7)}{\lambda
   ^{7/2}} +\dots \right]+ O(p^{-2}) \, ,
 \end{align}
The non-perturbative terms are given by
\ie 
\widehat{\cC}_{5, p}^{(5,3), {\rm NP}}( \lambda)  =& \pm i \sum_{n=1}^{\infty} e^{-2n \sqrt{\lambda }} \left[\frac{345945600}{n^{5} \lambda^{5/2}} \left( 1+ \frac{431}{4 n \lambda^{1/2}} + \frac{198485}{32 n^2 \lambda} +  \ldots \right) \right. \cr 
&\left. - {1\over p} \frac{2940537600}{n^{4} \lambda^{2}} \left( 1 + \frac{7437}{68 n \lambda^{1\over 2}} +\frac{3498373}{544 n^2 \lambda ^{} } + \ldots \right) +O(p^{-2}) \right]  \, .
\fe

\section{Conclusion and discussion}
\label{sec:confut}

In this paper we proposed a lattice-sum representation for integrated correlators that are associated with four-point correlation functions in $\mathcal{N}=4$ SYM of the form $\langle \mathcal{O}_2 \mathcal{O}_2 \mathcal{O}^{(i)}_p \mathcal{O}^{(j)}_p \rangle$, where $\mathcal{O}^{(i)}_p$ is a charge-$p$ (or dimension-$p$) half-BPS operator. This generalised the earlier proposal for the simplest integrated correlator with $p=2$ \cite{Dorigoni:2021bvj, Dorigoni:2021guq}. The formulation makes manifest $SL(2, \mathbb{Z})$ invariance of the correlators, and allows us to introduce generating functions for the integrated correlators which sum over the dependence on charge $p$ and/or gauge group rank $N$. As we reviewed in this paper, we recently proved \cite{recp} that all the integrated correlators for any $N$ are governed by a universal Laplace-difference equation that relates integrated correlators of operators with different charges. The Laplace-difference equation is a recursion relation that completely determines all the integrated correlators, once the initial conditions are given. In this paper, we utilised the Laplace-difference equation to explicitly determine the generating functions for the integrated correlators, in terms of initial data. We found that the generating functions, and especially their singularity structures, which are what is relevant for understanding the charge dependence of the integrated correlators, take a universal form for all the integrated correlators. Using the generating functions (and the singularity structures), we determined the transseries for the integrated correlators in the large-charge limit. The Laplace-difference equation of \cite{recp} is satisfied only after reorganising the operators in a particular way (as we reviewed in subsection \ref{sec:OLDE}), so accordingly we have taken the large-charge limit in a specific manner. 
 
In particular, for a fixed $N$, the large-charge expansion of integrated correlators universally contains three parts. The first part is independent of the coupling $\tau$ and behaves as a power series in $1/p$ as well as a $\log(p)$ term when $i=j$ (i.e. when two higher-dimensional operators are identical). The second part is a power series in $1/p$, where the coefficient of each term is an $SL(2, \mathbb{Z})$ invariant function given by a sum of finite numbers of non-holomorphic Eisenstein series with half-integer indices and rational pre-factors. The third part is exponentially decayed in the large-$p$ limit. It can be organised as a ``power series" in  $1/p$, and the coefficient of each term is a sum of the new $SL(2, \mathbb{Z})$ invariant functions  $D_p(s; \tau, \bar \tau)$, which behave as $e^{-c(\tau, \bar \tau) \sqrt{p}}$. Finally, when $i=j$, there is an additional modular function of $\tau$ that is independent of $p$ and is determined in terms of the integrated correlator with $p=2$.  These results (especially the second and third parts) are remarkably similar to those found  recently in \cite{Dorigoni:2022cua} for the integrated correlator with $p=2$ in the large-$N$ expansion, by exchanging $p \leftrightarrow N$. Furthermore, in appendix \ref{app:largeN} we also studied the large-$N$ expansion of the integrated correlator with a fixed $p=3$, and once again a similar structure is found. 

There are however some interesting differences between the large-$N$ expansion of integrated correlators with a fixed $p$ and the large-$p$ expansion with a fixed $N$. In particular, the precise form of the large-$p$ expansion of integrated correlators depends on whether $N$ is even or odd.  In the case of odd $N$, the number of Eisenstein series for a given order in the large-charge expansion grows as the order of the $1/p$ expansion increases. This is the same as the large-$N$ expansion of the integrated correlators \cite{Dorigoni:2021guq, Dorigoni:2022cua}.
In terms of the generalised 't Hooft coupling $\lambda=p\, g^2_{_{YM}}$, the power series terms in $1/p$ become asymptotic series in $1/\sqrt{\lambda}$, which are not Borel summable. As has been seen in the large-$N$ expansion \cite{Dorigoni:2021guq, Collier:2022emf, Hatsuda:2022enx}, through resurgence this is tightly related to the exponentially decayed terms $D_p(s; \tau, \bar \tau)$, which behave as $e^{-2n\sqrt{\lambda}}$ in the 't Hooft coupling. For even $N$, on the other hand, the number of Eisenstein series that appear in the large-$p$ expansion does not grow indefinitely.  In terms of 't Hooft coupling, this implies at a given order in $1/p$ expansion, the $1/\sqrt{\lambda}$ expansion terminates at a finite order, and there is in principle no ambiguity of defining such polynomials of $1/\sqrt{\lambda}$. Nevertheless, we find that there are still always exponentially decayed terms $e^{-2n\sqrt{\lambda}}$ for even $N$. This imposes interesting questions of relating these perturbative terms and exponentially decayed terms through the resurgence analysis, as in the case of odd $N$. As we commented earlier, this phenomenon is not new and has appeared in the literature of resurgence, see e.g. \cite{Dunne:2016jsr, Kozcaz:2016wvy, Dorigoni:2017smz, Dorigoni:2019kux, Fujimori:2022qij, Dorigoni:2019yoq, Dorigoni:2020oon, Dorigoni:2022bcx}. The general procedure of understanding such observables from the viewpoint of resurgence is to introduce some deformation parameters so that the finite perturbative series depends on the parameters and becomes asymptotic. This is often called ``Cheshire cat" resurgence. For the integrated correlators, a natural parameter here is clearly $N$. The power series in the large-$\lambda$ expansion truncates when we tune the parameter $N$ to be even. We will leave the systematic study of the resurgence analysis for the integrated correlators, especially for even $N$, for future work. 

As we commented earlier, we have studied the large-charge expansion of the integrated correlators $\widehat{C}^{(M,M'|i,i')}_{N,p}(\tau, \bar{\tau})$ by considering the large-$p$ limit with fixed $M, M'$; this allows us to utilise the Laplace-difference equation. It will be very interesting to understand the large-$M$ (and large-$M'$) behaviour of the integrated correlators with a fixed $p$ (even for $p=0$). Another research direction is to study the large-$N$ limit along with the large-charge limit. It will be of particular interest to understand this double scaling limit, and its connections with the dual string theory on AdS$_5 \times$S$^5$ along the line of \cite{Aprile:2020luw}. The double scaling limit has been considered for $\widehat{C}^{(M,M'|i,i')}_{N,p}(\tau, \bar{\tau})$ in the recent work \cite{Paul:2023rka}, but mostly only for the special cases of $M=M'=0$. The Laplace-difference equation and the concept of generating functions should provide powerful pathways to extend the analysis of \cite{Paul:2023rka}.  

In this paper (and our previous paper \cite{recp}) we have only considered the integrated correlators with $SU(N)$ gauge group. It has been shown in  \cite{Dorigoni:2022zcr} (see also \cite{Alday:2021vfb} on related work) that for the simplest integrated correlator that is associated with $\langle \mathcal{O}_2 \mathcal{O}_2 \mathcal{O}_2 \mathcal{O}_2\rangle$ (i.e. $\widehat{C}^{(0,0)}_{N,1}(\tau, \bar{\tau})$ in our notation), more general classical gauge groups also exhibit many remarkable properties such as the lattice-sum representation, simplicity in the large-$N$ expansion \cite{Dorigoni:2022cua}, and the Goddard-Nuyts-Olive duality \cite{Goddard:1976qe}. It is clearly of interest to extend our analysis for these more general integrated correlators $\widehat{C}^{(M,M'|i,i')}_{N,p}(\tau, \bar{\tau})$ to other gauge groups. 

\section*{Acknowledgements}

The authors would like to thank Daniele Dorigoni, Michael Green and Rodolfo Russo for insightful discussions. 
CW is supported by Royal Society University Research Fellowships No.~UF160350 and URF$\backslash$R$\backslash$221015. AB is supported by a Royal Society funding No.~RF$\backslash$ERE$\backslash$210067.  
%%%%%%%%%%%%%%%%%%%%%%%%%%%%%%%%%%%%%%%%%%%%%%%%%%%%%

\appendix

\section{More examples of generating functions and the singularity structures}
\label{app:genfun}

In this appendix, we give some more examples of generating functions and their singularity structures. In particular, they are relevant for the computation of large-charge expansion of the integrated correlators in the main text.  

\subsection{$M=3, M'=3, N=3$}

In the main text, we gave the singularities for the generating function $\widehat{B}_{3}^{(3,3)}(w; t)$, which is relevant for the large-charge expansion of the itnegrated correlator. For the completeness, here we present the generating function itself, 
\begin{align} \label{eq:B33}
& \widehat{B}_{3}^{(3,3)}(w; t)= \frac{1}{(w-1)} \left[\frac{3 t \left(t^4+4 t^3+17 t^2+10 t+12\right)}{(t+1)^5} \right. \\
& \left. + \frac{f^{(3,3)}_3(w,t)}{(t+1)^6\sqrt{1-w/w_1} \left(\sqrt{1-w}+\sqrt{1-w/w_1}\right)^5}  -\frac{ \, g^{(3,3)}_3(w,t)\,\sqrt{1-w}}{(t+1)^7  \sqrt{1-w/w_1} \left(\sqrt{1-w}+\sqrt{1-w/w_1}\right)^6} \right] \, , \nonumber
\end{align}
where
\begin{align}
& f^{(3,3)}_3(w,t) = 96\left(t^6 (w-1)^3+2 t^5 (w-1)^2 (5 w-8)-3 t^4 (w-1) \left(10 w^2-10 w-3\right) \right. \\
& \left. +\, 2t^3 (10 w-11) \left(w^2-w+1\right) +t^2(w-1) \left(20 w^2-22 w+17\right)-30tw(w-1)^2+11 (w-1)^3 \right) \, ,  \nonumber
\end{align} 
and
\begin{align}
& g^{(3,3)}_3(w,t)=  192  \left(t^6 (w-1)^2 (12 w-11)-t^5 (w-1) \left(46 w^2-51 w+10\right)+t^4 \left(85 w^3-164 w^2+94 w-10\right) \right.
\cr
& \left. -\, 5 t^3 \left(20 w^3-35 w^2+16
w-2\right)+t^2 (w-1) \left(85 w^2-79 w+10\right)-5 t (w-1)^2 (9 w-2)+11 (w-1)^3\right) \, . 
\end{align} 

\subsection{$M=5, M'=3, N=5$}

Here we give another example that is the case with $M\neq M'$, i.e. $M=5, M'=3$ with $SU(5)$ gauge group. We will only present the singularity structures of generating functions. As we commented previously that the generating functions for all the cases with $M\neq M'$ or $i \neq i'$ do not have the pole at $w=1$. But they do have branch cuts along $(1, w_1)$. For this particular case of $M=5, M'=3, N=5$, we find that the discontinuity is given by
\begin{align} \label{eq:dB535}
& {\rm disc}\, \widehat{B}^{(5,3)}_{5}(w;t) =-i\frac{ (t+1)^7\left(1-{w}/{w_1}\right)^{7/2}  \sqrt{w-1} }{(t\, w)^{15}} P^{(5,3)}_5(w,t) \, ,
\end{align}
where $P^{(5,3)}_5(w,t)$ is given by
\begin{align} 
& P^{(5,3)}_5(w,t) = \frac{ 1}{89148} \left[744952\tilde{t}^{22}_{0,10}-\tilde{t}^{22}_{1,9}(6044741w+4543487)\right. \\
&+ \tilde{t}^{22}_{2,8} \left(20823446 w^2+31072748 w+11074349\right)-\tilde{t}^{22}_{3,7} \left(39158777 w^3+89446063 w^2+61990132 w+13409786\right) \cr
&+7 \tilde{t}^{22}_{4,6} \left(6100411 w^4+19749968 w^3+20911974 w^2+8157878 w+1137253\right)\cr
&-3 \tilde{t}^{22}_{5,5} \left(8612009 w^5+39521965 w^4+62784928 w^3+34204774 w^2+6660697 w+579475\right)\cr
&+3 \tilde{t}^{22}_{6,4} \left(2167263 w^6+16208346 w^5+45386438 w^4+36991154 w^3+4160024 w^2-339412 w-41574\right) \cr
&+6 \tilde{t}^{22}_{7,3} \left(169560 w^7-363930 w^6-6592600 w^5-15265824 w^4+639744 w^3+2076207 w^2-1750 w+5226
\right) \cr
&-3 \tilde{t}^{22}_{8,2} \left(207523 w^8+962492 w^7+4008725 w^6-14486528 w^5 \right.\cr
&\left. -6539106 w^4+11812290 w^3-2227372 w^2+23528 w-19006\right) \cr
&-2 \tilde{t}^{22}_{9,1} \left(237328 w^9-430521 w^8+12675 w^7-13939422 w^6+38168298 w^5 \right.\cr
&\left. -32891229 w^4+9568083 w^3-324744 w^2-48426 w-7436\right) \cr
&+\tilde{t}^{22}_{10,0} \left(132769 w^{10}-353990 w^9+4937166 w^8-18083610 w^7+19406856 w^6+5960934 w^5\right.\cr
&\left.-25339020 w^4+17720010
   w^3-4901967 w^2+541348 w-37232\right) \cr
&+14 {t}^{11} \left(30511 w^9-102141 w^8+528255 w^7-2217315 w^6+4291569 w^5\right.\cr
&\left.\left.-4099383 w^4+1978413 w^3-457983 w^2+50220 w-4238\right) \right] \,. \nonumber
\end{align}

%%%%%%%%%%%%%
\section{Lattice-sum representation and $SL(2, \mathbb{Z})$ spectral decomposition}
\label{app:lattice}
%%%%%%%%%%%%

The lattice-sum representation \eqref{eq:proNp} of integrated correlators implies that they can be formally expressed as an formal infinite sum of non-holomorphic Eisenstein series,\cite{Dorigoni:2021bvj,Dorigoni:2021guq} 
\ie \label{eq:proNpE}
\widehat{\cC}^{(M, M'|i, i')}_{N, p}(\tau; \bar{\tau}) =
\int_0^{\infty}   \widehat{B}^{(M, M'|i,i')}_{N,p}(t) \, dt + \sum_{s=2}^{\infty}  \hat{c}_{N, p; s}^{(M, M'| i, i')}\,\Gamma(s)\, E(s; \tau, \bar \tau) \, ,
\fe
where the coefficients $\hat {c}_{N, p;s}^{(M, M'|i, i')}$ are related to $\widehat{B}^{(M, M'|i,i')}_{N,p}(t)$ through 
\ie
\widehat{B}^{(M, M'|i,i')}_{N,p}(t)  = \sum_{s=2}^{\infty} \hat{c}_{N, p; s}^{(M, M'| i, i')} \,  t^{s-1}   \, .
\fe
In arriving the expression \eqref{eq:proNpE}, we have used the definition of the Eisenstein series, 
\ie 
E(s; \tau, \bar \tau) ={1\over \Gamma(s)}\sum_{(m,n) \neq (0,0)} \int_0^\infty e^{-t\, Y_{m n}(\tau, \bar{\tau})} t^{s-1} \, dt \, .
\fe
Importantly, the formal sum can be re-cast into a contour integral along imaginary axis \cite{Dorigoni:2022zcr},
\ie \label{eq:spec}
\widehat{\cC}^{(M, M'|i, i')}_{N, p}(\tau; \bar{\tau}) =
2\int_0^{\infty}   \widehat{B}^{(M, M'|i,i')}_{N,p}(t) \, dt + {1\over 2\pi i} \int_{1/2-i \infty}^{1/2+i \infty} ds {\pi (-1)^s \over \sin(\pi s)}  \hat{c}_{N, p; s}^{(M, M'| i, i')}  \,\Gamma(s)\, E(s; \tau, \bar \tau) \, ,
\fe
which leads to the $SL(2, \mathbb{Z})$ spectral representation of $\widehat{\cC}^{(M, M'|i, i')}_{N, p}(\tau; \bar{\tau})$ \cite{Collier:2022emf, Paul:2022piq}, where the first term was denoted as an average $\langle \widehat{\cC}^{(M, M'|i, i')}_{N, p} \rangle$. In general, a  $SL(2, \mathbb{Z})$-invariant quantity can be decomposed into a continuous spectrum in terms of non-holomorphic Eisenstein series and a discrete spectrum in terms cusp forms. It is remarkable that $SL(2, \mathbb{Z})$ spectral decomposition of these integrated correlators only contain the non-holomorphic Eisenstein series. 

The non-holomorphic Eisenstein series can be expressed in the Fourier mode expansion, 
\ie \label{eq:Es}
E(s; \tau, \bar \tau) &= {2\zeta(2s) \over \pi^s} \tau_2^s + {2 \Gamma(s-1/2)\zeta(2s-1) \over \pi^{s-1/2} \Gamma(s)} \tau_2^{1-s} 
\cr  &+ \sum_{k \neq 0} e^{2\pi i k \tau_1} {4 \sqrt{\tau_2} \over \Gamma(s) } |k|^{s-1/2} \sigma_{1-2s}(|k|) K_{s-1/2} \left( 2\pi |k| \tau_2 \right)  \, , 
\fe
where $\sigma_{\nu}(k) = \sum_{n|k} n^{\nu}$ is the divisor sum, and $K_{\nu}(x)$ the Bessel function. The first line in the above expression is the zero mode of the non-holomorphic Eisenstein series, which gives perturbative contributions; whereas the second line accounts for non-zero modes, which lead to non-perturbative instanton contributions. 

From the above expression of $E(s; \tau, \bar \tau)$ and \eqref{eq:proNpE}, we find the zero-mode part of integrated correlators\footnote{The overall factor of $2$ on the second line is due to the fact that the resummation of the first term in \eqref{eq:Es} gives the same result as the second term  in the large-$\tau_2$ expansion \cite{Dorigoni:2021guq}. This is related to the symmetry property, $\hat{c}_{N, p; s}^{(M,M'|i,i')}= \hat{c}_{N, p; 1-s}^{(M,M'|i,i')}$. }
\ie \label{eq:propert}
\widehat{\cC}^{(M,M'|i,i')}_{N, p}(\tau; \bar{\tau})\big {\vert}_{\rm zero-mod} &=
2\int_0^{\infty}   \widehat{B}^{(M, M'|i,i')}_{N,p}(t) \, dt \cr &+ {2\over 2\pi i} \int_{1/2-i \infty}^{1/2+i \infty} ds {\pi (-1)^s \over \sin(\pi s)}  \hat{c}_{N, p; s}^{(M, M'| i, i')}  {2 \Gamma(s-1/2)\zeta(2s-1) \over \pi^{s-1/2} } \tau_2^{1-s}  \, .
\fe
We are interested in the large-$p$ expansion, for which, using the symmetry property $\hat{c}_{N, p; s}^{(M, M'| i, i')} = \hat{c}_{N, p; 1-s}^{(M, M'| i, i')}$ and the large-$p$ expansion of 
\ie
\hat{c}_{N, p; 1-s}^{(M, M'| i, i')}=p^{-s} \sum^{\infty}_{g=0} p^{-g} \hat{c}_{N; \, g;\, 1-s}^{(M, M'| i, i')} \, , 
\fe
we have
\ie \label{eq:propert2}
\widehat{\cC}^{(M,M'|i,i')}_{N, p} & (\tau; \bar{\tau})\big {\vert}_{\rm zero-mod}  =
2\int_0^{\infty}   \widehat{B}^{(M, M'|i,i')}_{N,p}(t) \, dt  \cr 
&+ \sum^{\infty}_{g=0} p^{-g} {2\over 2\pi i} \int_{1/2-i \infty}^{1/2+i \infty} ds {\pi (-1)^s \over \sin(\pi s)}  \hat{c}_{N; \, g;\, 1-s}^{(M, M'| i, i')} p^{-s}   {2 \Gamma(s-1/2)\zeta(2s-1) \over \pi^{s-1/2} } \tau_2^{1-s}  \, ,
\fe
with the contour closing from the right in the large-$p$ limit. Besides the $1/\tau_2$ perturbative terms, as $\zeta(2s-1)$ has a singularity at $s=1$, we see that the residue at $s=1$ leads to an additional $\tau$-independent term, and the total constant part of the integrated correlators in the large-$p$ expansion is given by
\ie \label{eq:const}
\widehat{\cC}^{(M,M'|i,i')}_{N, p}\big {\vert}_{\rm const} &=
2\int_0^{\infty}   \widehat{B}^{(M, M'|i,i')}_{N,p}(t) \, dt + 2 \sum_{g=0}^{\infty} p^{-g-1}\, \underset{s\to 0}{\text{lim}} {\hat{c}^{(M,M'|i,i')}_{N;\, g;\, s} \over s }
 \, .
\fe
From point of view of generating functions, the second $\tau$-independent term arises because when performing large-$p$ expansion, the generating functions may develop $1/t$ singularity and the residue at $t=0$ leads to $Y_{m,n}(\tau, \bar{\tau})$-independent (i.e. $\tau$-independent) terms. And that the contribution at $t=0$ is essentially equivalent to $s\rightarrow 0$ limit.  

%%%%%%%%%%
\section{Large-$N$ expansion with fixed-$p$} \label{app:largeN}
%%%%%%%%%

Here we consider the large-$N$ expansion of integrated correlators with a fixed charge. The example of we will consider is $\cC^{(3,3)}_{N, 0}(\tau; \bar{\tau})$\footnote{Recall  $\cC^{(3,3)}_{N, p}(\tau; \bar{\tau}) = {R^{(3,3)}_{N, p} \over 4} \widehat{\cC}^{(3,3)}_{N, p}(\tau; \bar{\tau})$, with $R^{(3,3)}_{N, 0}= {(N^2-1)(N^2-4) \over 3N}$. }, namely the integrated correlators associated with the four-point function $\langle \mathcal{O}_2 \mathcal{O}_2 \mathcal{O}_3 \mathcal{O}_3 \rangle$.  It was noted in \cite{Paul:2022piq} that $\cC^{(3,3)}_{N, 0}(\tau; \bar{\tau})$ obeys the following recursion relation, 
\begin{align} \label{eq:23}
N(N+1) \left[ \cC^{(3,3)}_{N+1, 0}(\tau; \bar{\tau})- \cC^{(3,3)}_{N, 0}(\tau; \bar{\tau}) \right] = 2N(N-1)  \cC_{N+1,1}^{(0,0)}(\tau, \bar \tau)+2(N+1)(N+2)  \cC_{N,1}^{(0,0)}(\tau, \bar \tau) \, .
\end{align}
It is easy to solve the above recursive relation, which leads to 
\ie \label{eq:solp3}
 \cC_{N,0}^{(3,3)}(\tau, \bar \tau) = 2 \sum_{i=1}^{N-1}  \left( \cC_{i,1}^{(0,0)}(\tau, \bar \tau) + \cC_{i+1,1}^{(0,0)}(\tau, \bar \tau) \right) - {4\over N} \cC_{N,1}^{(0,0)}(\tau, \bar \tau) \, . 
\fe
The generating function that sums over the $N$ dependence is defined as $\cC_{0}^{(3,3)}(z; \tau, \bar \tau) = \sum_{N=2}^{\infty} \cC_{N,0}^{(3,3)}(\tau, \bar \tau) z^N$, which can be expressed as
\ie
\cC^{(3,3)}_0 (z; \tau, \bar \tau)  =\sum_{m,n \in \mathbb{Z}^2} \int^{\infty}_0 e^{- t Y_{mn} (\tau, \bar \tau)} B^{(3,3)}_{0}(z;t) dt \, . 
\fe
From \eqref{eq:solp3}, we see 
\ie \label{eq:B3}
B^{(3, 3)}_{0}(z;t) = 2 \sum_{N=2}^{\infty}  \sum_{i=1}^{N-1}  \left(B_{i,1}^{(0,0)}(t) + B_{i+1,1}^{(0,0)}(t) \right) z^N  -4  \sum_{N=2}^{\infty}  {1\over N}B_{N,1}^{(0,0)}(t) z^N \, ,
\fe
and the generating function  $B^{(0,0)}_{1}(z;t)  = B_{N,1}^{(0,0)}(t) z^N$ is known, as given in \eqref{eq:Bztp2} which we quote below for the convenience,
\ie
B^{(0,0)}_{1}(z;t)  = \frac{3 t z^2 \left[(t-3) (3 t-1)(t+1)^2 -
   z(t+3) (3 t+1) (t-1)^2 \right]}{2 (1-z)^{\frac{3}{2}}
   \left[(t+1)^2-(t-1)^2 z\right]^{\frac{7}{2}}} \, .
   \label{eq:gresult}
\fe
We can now relate each piece in \eqref{eq:B3} to $B^{(0,0)}_{1}(z;t) $. In particular, 
\ie
 \sum_{N=2}^{\infty}  {1\over N}B_{N,1}^{(0,0)}(t) z^N =  \int^z dx \sum_{N=2}^{\infty}  B_{N,1}^{(0,0)}(t) x^{N-1} =\int^z {dx \over x} B^{(0,0)}_{1}(x;t) \, ,
\fe
and 
\ie
\sum_{N=2}^{\infty}  \sum_{i=1}^{N-1}  \left(B_{i,1}^{(0,0)}(t) + B_{i+1,1}^{(0,0)}(t) \right) z^N =   \oint_C dx \frac{(x+1) z^2}{x^2 (z-1) (z-x)} B^{(0,0)}_{1}(x;t) \, ,
\fe
where we have used 
\ie
B_{i,1}^{(0,0)}(t) =\oint_C {dx \over x^{i+1} } B^{(0,0)}_{1}(x;t) \, ,
\fe
with contour $C$ circling the origin. 
Using the explicit expression of $B^{(0,0)}_{1}(z;t)$ as given in \eqref{eq:gresult}, we find, 
\begin{align}
& B^{(3,3)}_{0}(z;t)  =-\frac{3 \left(t^3+1\right)}{2t^2} -\frac{ 2(z+1)}{z-1} B^{(0,0)}_{1}(z;t) \\
&+ \frac{3\! \left[ (t-1)^4 \left(t^2+t+1\right)  \left(
   3 t^2+4t+3 -(t-1)^2 z\right) z^2 -(t+1)^4 \left(t^2-t+1\right) \left((3 t^2-4t+3)
   z-(t+1)^2\right)\right]}{2t^2 (1-z)^{1\over 2} \left[ (t+1)^2-(t-1)^2
   z\right]^{5 \over 2}} \, . \nn
\end{align}
The function has branch cuts along $(1, z_1)$ (here $z_1=(t+1)^2/(t-1)^2$) with the discontinuity given by
\begin{align}
& {\rm disc}B^{(3,3)}_{0}(z;t)  = -{3 \, i}  \left[ \frac{2 t z^2 (z+1) \left[(t+3) (3 t+1)
   (t-1)^2 z+(3-t) (t+1)^2 (3
   t-1)\right]}{(z-1)^{5/2}
   \left[(t+1)^2-(t-1)^2 z\right]^{7/2}} \right. \\
& \left. + \frac{ (t-1)^4 z^2 \left(t^2+t+1\right)  [(t-1)^2 z-
  (3 t^2+4t+3) ] + (t+1)^4  (t^2-t+1)  [ (3t^2-4t+3) z-(t+1)^2 ] }{t^2 \sqrt{z-1} \left[(t+1)^2-(t-1)^2
   z\right]^{5/2}} \right] \, . \nn
\end{align}

With the explicit expression of ${\rm disc}B^{(3,3)}_{0}(z;t)$, we can now study the large-$N$ behaviour of the integrated correlator following the same analysis that we described in the main text. We find that $\cC^{(3,3)}_{N,0}(\tau, \bar \tau)$ can again be separated into power series terms and exponentially decayed terms in the $1/N$ expansion,  
\ie
\cC^{(3,3)}_{N,0}(\tau, \bar \tau) = \cC^{(3,3), {\rm P}}_{N,0}(\tau, \bar \tau) + \cC^{(3,3), {\rm NP}}_{N,0}(\tau, \bar \tau) \, ,
\fe
where the power series terms are given by
\ie
 \cC^{(3,3), {\rm P}}_{N,0}(\tau, \bar \tau)  =&\, {N^3\over 3} -{5 N\over 6}- {N^{3/2} \over 2}  E\left(3/2; \tau, \bar \tau \right)+ {45\, N^{1/2}\over 2^5}  E\left(5/2; \tau, \bar \tau \right)
\cr
& +{25 \over  N^{1/2}}  \left[ {31\over 2^{10}} E\left(3/2; \tau, \bar \tau \right)-{189 \over 2^{12}} E\left(7/2; \tau, \bar \tau \right) \right] \cr
& -
{225 \over  N^{3/2}}  \left[{25 \over 2^{13}} E\left(5/2; \tau, \bar \tau \right)+ {147\over 2^{15}} E\left(9/2; \tau, \bar \tau \right) \right] + O(N^{-5/2}) \, ,
\fe
 and the non-perturbative terms take the following form, 
\ie
& \cC^{(3,3), {\rm NP}}_{N,0}(\tau, \bar \tau)  =\pm i \left[ 4 N^{5/2} D_N\left(1/2; \tau, \bar \tau \right) -{N^2 \over 3} \left( 2 D_N\left(-1; \tau, \bar \tau \right) - {57\over 2} D_N\left(1; \tau, \bar \tau \right) \right) \right. \cr 
&+{N^{3/2} \over 3^2} \left( {1\over 2} D_N\left(-{5}/{2}; \tau, \bar \tau \right) - {3\over 2^2} D_N\left(-1/2; \tau, \bar \tau \right) + {3789\over 2^5} D_N\left(3/2; \tau, \bar \tau \right) \right) \cr 
&\left. - {N \over 3^4} \left( \frac{1}{4} D_N\left(-4; \tau, \bar \tau \right)+\frac{603}{80}
   D_N\left(-2; \tau, \bar \tau \right)-\frac{43065}{2^6}
   D_N\left(0; \tau, \bar \tau \right)-\frac{244215}{2^8} D_N\left(2; \tau, \bar \tau \right) \right) \right]\cr 
   &+ O(N^{1/2}) \, . 
\fe
The modular function $D_N\left(s; \tau, \bar \tau \right)$ is defined in \eqref{eq:dD} with $p$ replaced by $N$ now. 
The perturbative terms $\cC^{(3,3), {\rm P}}_{N,0}(\tau, \bar \tau)$ were also computed in \cite{Paul:2022piq}, and we find our results agree with those in the reference.

	\bibliographystyle{ssg}
	\bibliography{genfun2}

\end{document}